\documentclass[twocolumn]{aastex63}

\usepackage{multirow}
\usepackage{booktabs}

\newcommand{\dave}{\texttt{DAVE}}
\newcommand{\tri}{\texttt{TRICERATOPS}}

\begin{document}

\title{Validation of 13 Hot and Potentially Terrestrial TESS Planets}

\suppressAffiliations

\correspondingauthor{Steven Giacalone}
\email{steven$\_$giacalone@berkeley.edu}

\author[0000-0002-8965-3969]{Steven Giacalone}
\affil{Department of Astronomy, University of California Berkeley, Berkeley, CA 94720, USA}

\author[0000-0001-8189-0233]{Courtney D. Dressing}
\affiliation{Department of Astronomy, University of California Berkeley, Berkeley, CA 94720, USA}

\author[0000-0002-3385-8391]{Christina Hedges}
\affiliation{Bay Area Environmental Research Institute, P.O. Box 25, Moffett Field, CA 94035, USA}
\affiliation{NASA Ames Research Center, Moffett Field, CA, 94035, USA}

\author[0000-0001-9786-1031]{Veselin B. Kostov}
\affiliation{NASA Goddard Space Flight Center, 8800 Greenbelt Road, Greenbelt, MD 20771, USA}
\affiliation{SETI Institute, 189 Bernardo Avenue, Suite 200, Mountain View, CA 94043, USA}

\author[0000-0001-6588-9574]{Karen A.\ Collins}
\affiliation{Center for Astrophysics \textbar \ Harvard \& Smithsonian, 60 Garden Street, Cambridge, MA 02138, USA}

\author[0000-0002-4625-7333]{Eric L. N. Jensen}
\affiliation{Department of Physics \& Astronomy, Swarthmore College, Swarthmore PA 19081, USA}

\author[0000-0003-4755-584X]{Daniel A. Yahalomi}
\affil{Center for Astrophysics \textbar \ Harvard \& Smithsonian, 60 Garden Street, Cambridge, MA 02138, USA}
\affil{Department of Astronomy, Columbia University, 550 W 120th Street, New York, NY 10027, USA}

\author[0000-0001-6637-5401]{Allyson Bieryla} 
\affiliation{Center for Astrophysics \textbar \ Harvard \& Smithsonian, 60 Garden Street, Cambridge, MA 02138, USA}

\author[0000-0002-5741-3047]{David R. Ciardi} \affiliation{NASA Exoplanet Science Institute, Caltech/IPAC, Mail Code 100-22, 1200 E. California Blvd., Pasadena, CA 91125, USA}

\author[0000-0002-2532-2853]{Steve B. Howell}
\affiliation{NASA Ames Research Center, Moffett Field, CA, 94035, USA}

% ...

\shortauthors{Giacalone et al.}

\published{in the Astronomical Journal on January 28, 2022}

\begin{abstract}
The James Webb Space Telescope (JWST) will be able to probe the atmospheres and surface properties of hot, terrestrial planets via emission spectroscopy. We identify 18 potentially terrestrial planet candidates detected by the Transiting Exoplanet Survey Satellite (TESS) that would make ideal targets for these observations. These planet candidates cover a broad range of planet radii ($R_{\rm p} \sim 0.6 - 2.0 R_\oplus$) and orbit stars of various magnitudes ($K_s = 5.78 - 10.78$, $V = 8.4 - 15.69$) and effective temperatures ($T_{\rm eff }\sim 3000 - 6000$K). We use ground-based observations collected through the TESS Follow-up Observing Program (TFOP) and two vetting tools -- \texttt{DAVE} and \texttt{TRICERATOPS} -- to assess the reliabilities of these candidates as planets. We validate 13 planets: TOI-206 b, TOI-500 b, TOI-544 b, TOI-833 b, TOI-1075 b, TOI-1411 b, TOI-1442 b, TOI-1693 b, TOI-1860 b, TOI-2260 b, TOI-2411 b, TOI-2427 b, and TOI-2445 b. Seven of these planets (TOI-206 b, TOI-500 b, TOI-1075 b, TOI-1442 b, TOI-2260 b, TOI-2411 b, and TOI-2445 b) are ultra-short-period planets. TOI-1860 is the youngest ($133 \pm 26$ Myr) solar twin with a known planet to date. TOI-2260 is a young ($321 \pm 96$ Myr) G dwarf that is among the most metal-rich ([Fe/H] = $0.22 \pm 0.06$ dex) stars to host an ultra-short-period planet. With an estimated equilibrium temperature of $\sim 2600$ K, TOI-2260 b is also the fourth hottest known planet with $R_{\rm p} < 2 \, R_\oplus$.
\end{abstract}

\section{Introduction}

Over the last two decades, the combination of planet radii ($R_{\rm p}$) and planet masses ($M_{\rm p}$) measured from transit and radial velocity (RV) observations have enabled the calculations of bulk densities for hundreds of exoplanets. With the help of theoretical models of the interior structures of planets \citep{valencia2006internal, valencia2007detailed, valencia2007radius, fortney2007planetary, seager2007mass, zeng2008computational, grasset2009study, zeng2013detailed, zeng2016mass}, the bulk densities of these planets have made it possible to identify planets with terrestrial compositions \citep[e.g.,][]{batalha2011kepler, carter2012kepler, dragomir2013most, barros2014revisiting, dressing2015mass, rogers2015most, motalebi2015harps, gillon2017seven}. As a consequence, our understanding of terrestrial planets outside of the solar system has progressed significantly in recent years. For instance, terrestrial planets with orbital periods shorter than 30 days are now known to have maximum radii between 1.5 and 2.0 $R_\oplus$ \citep[e.g.,][]{rogers2015most, buchhave20161}.

Another notable discovery resulting from these surveys is the distinct gap in occurrence rate between planets with $R_{\rm p} < 1.5 \, R_\oplus$ and planets with $R_{\rm p} > 2.0 \, R_\oplus$ \citep{fulton2017california, fulton2018california} (often referred to as the ``radius gap''), with the former regime corresponding to planets with terrestrial compositions and the latter regime corresponding to planets with volatile-rich gaseous envelopes. This feature has important implications for the formation and evolution of short-period terrestrial planets, and several theories have predicted it or put forth an explanation for its origin. Some have proposed that the gap is a natural consequence of planets forming in gas-poor and gas-rich environments \citep{lee2014make, lee2016breeding, lopez2018formation}, while others contend that the gap is a result of atmospheric loss via photoevaporation \citep{jackson2012coronal, lopez2013role, owen2013kepler, jin2014planetary, owen2017evaporation, jin2018compositional}, core-powered mass loss \citep{ginzburg2016super, ginzburg2018core}, or planetesimal collision \citep{shuvalov2009atmospheric, schlichting2015atmospheric}. This gap has also been found to depend on planet orbital period \citep{van2018asteroseismic, martinez2019spectroscopic}, stellar mass \citep{fulton2018california, wu2019mass, cloutier2020evolution}, and system age \citep{berger2020gaia, david2020small}, which indicates that the terrestrial planet formation mechanism responsible for the feature could vary from system to system.

More recently, attempts have been made to more closely characterize terrestrial planets by observing their thermal emission phase curves. These near- and mid-infrared observations can reveal whether a terrestrial planet is surrounded by a thin atmosphere or has an airless surface, as only the former is expected to produce phase curves with evidence of atmospheric heat redistribution \citep{seager2009method, selsis2011thermal, koll2016temperature, kreidberg2016prospects}. Using this method, \cite{demory2016map} found evidence of atmospheric circulation for 55 Cnc e and \cite{kreidberg2019absence} inferred the absence of an atmosphere for LHS 3844 b. In addition, \cite{kreidberg2019absence} were able to use the wavelength-dependent planet-to-star flux ratio to estimate the surface composition of LHS 3844 b, finding that it is consistent with a basaltic composition that could result from widespread volcanism.

Our ability to characterize short-period terrestrial planets will improve drastically with the launch of the James Webb Space Telescope (JWST), which will allow for the characterization of exoplanet atmospheres and surface properties via transmission spectroscopy, emission spectroscopy, and emission photometry \citep{greene2016characterizing}. For most of the known terrestrial planets, detecting atmospheric absorption features in transmission spectra would be extremely challenging \citep[the exception being those orbiting ultracool dwarfs and white dwarfs;][]{lustig2019detectability, kaltenegger2020white}, but many of these planets would make excellent targets for thermal emission measurements. With these observations, one can infer the presence or lack of atmospheres surrounding short-period terrestrial planets \citep{koll2019identifying, lustig2019detectability, mansfield2019albedo}. For planets with atmospheres, relatively low-signal-to-noise-ratio (S/N) emission photometry and/or spectroscopy will reveal modest information about atmospheric composition and identify suitable targets for further atmospheric characterization with future high-precision instruments. For planets without atmospheres, emission measurements will permit the characterization of the surfaces of planets, such as those hot enough for the existence of dayside lava oceans \citep{rouan2011orbital, samuel2014constraining, kite2016lavaworlds, essack2020lavaworlds}.

The most highly anticipated JWST instrument for these observations is the Mid-Infrared Instrument (MIRI), which can perform low-resolution spectroscopy between 5 and 12 $\mu$m. This wavelength range contains a number of features that can be used to discern planets with atmospheres from those without atmospheres. \cite{morley2017observing} and \cite{lincowski2018trappist} simulated emission spectra for several terrestrial exoplanets assuming various atmospheric compositions, finding a number of notable absorption features. Specifically, Earth-like and O$_2$-dominated outgassed atmospheres can be identified via strong H$_2$O absorption between 5 and 7 $\mu$m, whereas Venus-like atmospheres display prominent SO$_2$ absorption between 7 and 9 $\mu$m and strong CO$_2$ absorption above 10 $\mu$m. \cite{lincowski2018trappist} also modeled the case of O$_2$-dominated desiccated (water-poor) atmospheres, which may be particularly relevant for planets orbiting M dwarfs \citep{luger2015extreme}, finding that they are distinguishable by a lack of H$_2$O absorption between 5 and 7 $\mu$m and strong O$_3$ absorption at 9.6 $\mu$m. \cite{zilinskas2020atmospheric} modeled emission spectra of N$_2$-dominated atmospheres for the hot terrestrial planet 55 Cnc e, finding that C-rich atmospheres have a distinct HCN feature at 7.5 $\mu$m. \cite{hu2012spectra} considered the cases of hot planets with airless surfaces when simulating thermal emission spectra. These spectra are largely blackbody-like but feature notable SiO absorption between 7 and 13 $\mu$m, which could be abundant for planets close enough to their host stars for their surfaces to vaporize \citep{schaefer2012vaporization}. This SiO absorption is expected to vary based on the types of rocks being vaporized (e.g., basaltic versus feldspathic versus ultramafic), and can therefore reveal information about surface composition.

In anticipation of the launch of JWST, many have designed methods and frameworks for identifying good targets for thermal emission observations \citep[e.g.,][]{batalha2017pandexo, kempton2018framework}. \cite{kempton2018framework} defined the emission spectroscopy metric (ESM), a proxy for the signal-to-noise ratio attainable for a terrestrial planet being observed with emission spectroscopy, in order to determine what planets should be prioritized for these observations, drawing the threshold above which the best targets exist at 7.5. As of 2018, only seven confirmed terrestrial planets (GJ 1132 b, HD 219134 b, HD 219134 c, 55 Cnc e, HD 3167 b, K2-141 b, and GJ 9827 b) had met this criterion, and three of these (HD 219134 b, HD 219134 c, and 55 Cnc e) have host stars too bright for emission spectroscopy observations with JWST. If an extensive emission photometry/spectroscopy survey of short-period terrestrial planets is to be conducted, more of these planets must be discovered.

The Transiting Exoplanet Survey Satellite mission \citep[TESS;][]{ricker2010transiting}, an ongoing survey searching for transiting planets across nearly the entire sky, has already significantly expanded the size of this sample. Since the start of the mission in mid-2018, an additional 15 planets with $R_{\rm p} < 2 \, R_\oplus$, ${\rm ESM} > 7.5$, and a host stars amenable to JWST observations have been discovered. In addition, we have identified 18 TESS Objects of Interest \citep[TOIs;][]{guerrero2021tess}, stars that exhibit decreases in brightness consistent with the signals caused by transiting planets, that would also meet these requirements if confirmed to host planets with terrestrial compositions. Nonetheless, because some of these TOIs could end up being astrophysical false positives (FPs; such as eclipsing binaries around nearby stars contaminating the TESS aperture), the community would benefit from a vetting analysis that identifies the potentially terrestrial planet candidates that have the best chances of being bona fide planets. In this paper, we  scrutinize TESS data and follow-up observations to assess the possibility that these 18 TOIs are actual planets and argue for future characterization efforts.

In Section \ref{sec: sample}, we discuss our sample of 18 TOIs and describe how they were selected. In Section \ref{sec: procedure}, we describe our vetting analysis procedure. In Section \ref{sec: followup}, we present follow-up observations of these TOIs that are incorporated into our analysis. In Section \ref{sec: results}, we present the results of our vetting analysis and validate 13 of the TOIs. In Section \ref{sec: discussion}, we discuss the implications of our results with respect to JWST emission spectroscopy. Lastly, in Section \ref{sec: conclusion}, we provide concluding remarks.

\section{Sample}\label{sec: sample}

The goal of this paper is to identify a sample of small, hot, and likely terrestrial planets that would be favorable targets for emission spectroscopy observations with JWST.  We selected our sample by first identifying all TOIs with orbital periods ($P_{\rm orb}$) $<$ 10 days and $R_{\rm p} < 2 \, R_\oplus$, which corresponds approximately to the largest a planet can be without having a volatile-rich gaseous envelope \citep[e.g.,][]{rogers2015most, buchhave20161}. The $P_{\rm orb}$ of each TOI is gathered from ExoFOP-TESS.\footnote{ \url{https://exofop.ipac.caltech.edu/tess/index.php} -- search performed on 2021-03-11.} We estimate the $R_{\rm p}$ of each TOI using the transit depths ($\delta$) listed on ExoFOP-TESS and the stellar properties in version 8.1 of the TESS Input Catalog (TIC; \citealt{stassun2018tess}).\footnote{$P_{\rm orb}$ and $\delta$ are reported by either the SPOC pipeline, which is discussed further below.} Next, we removed all TOIs that have been flagged as FPs or false alarms (FAs) on ExoFOP-TESS under ``TFOPWG Disposition.'' FPs are typically caused by eclipsing binaries around stars close enough to the target star to contaminate the TESS aperture, while FAs are typically caused by stellar rotation or instrumental variability that produces a signal resembling a planetary transit. Because the events caused by FPs and FAs are often shallow enough to be mistaken as the transits of small planets, scrutinizing observations of small TOIs for FP and FA signatures is an important step in determining which are bona fide planets. Our procedure for further vetting TOIs that pass this condition is described in Section \ref{sec: procedure}. 

Lastly, we determined which of our planet candidates would be most amenable to thermal emission measurements with JWST. To do this, we calculated the emission spectroscopy metric (ESM) for each of the remaining TOIs. The ESM is a quantity introduced in \cite{kempton2018framework} to serve as a proxy for the S/N one should expect to obtain when observing the emission spectrum of an exoplanet with JWST. More specifically, ESM is given by the equation
\begin{equation}\label{ESM}
    {\rm ESM} = 4.26 \times 10^6 \times \frac{B_{7.5} (T_{\rm day})}{B_{7.5}  (T_{\rm eff})} \times \left( \frac{R_{\rm p}}{R_\star} \right)^2 \times 10^{-m_K / 5}
\end{equation}
where $B_{7.5}$ is Planck's function evaluated at 7.5 $\mu$m for a given temperature, $T_{\rm day}$ is the dayside temperature of the planet in Kelvin (which is assumed to be 1.1x the equilibrium temperature of the planet), $T_{\rm eff}$ is the effective temperature of the host star in Kelvin, $R_\star$ is the stellar radius, and $m_K$ is the apparent magnitude of the host star in $K$ band. When calculating equilibrium temperature (here and throughout the remainder of the paper), we assume zero bond albedo and full day-night heat redistribution.\footnote{We acknowledge that, due to these assumptions, all equilibrium temperatures discussed in this paper are only rough estimates.} \cite{kempton2018framework} recommend that terrestrial planets with ${\rm ESM} \gtrsim 7.5$ be prioritized for emission spectroscopy observations. We therefore removed TOIs with ESMs lower than this threshold. The host star and the planet properties of our final list of 18 TOIs are shown in Table \ref{tab1}. 

It is worth noting that small planets are not the only good targets for JWST emission spectroscopy. In fact, Equation \ref{ESM} shows that larger planets with thick atmospheres would produce an even higher signal through these observations. However, this paper focuses specifically on terrestrial planets.

\begin{deluxetable*}{cccccccccc}[t!]\label{tab1}
\tabletypesize{\footnotesize}
\tablewidth{\textwidth}
\tablecaption{TOI Parameters from TICv8.1 and ExoFOP}
\tablehead{
\colhead{TOI} & \colhead{$K_s$ mag} & \colhead{Parallax (mas)} & \colhead{$T_{\rm eff}$ ($K$)} & \colhead{$\log g$} & \colhead{$R_\star$ ($R_\odot$)} & \colhead{$\delta$ (ppm)} & \colhead{$R_{\rm p}$ ($R_\oplus$)} & \colhead{$P_{\rm orb}$ (days)} & \colhead{ESM}
}
\startdata 
  206.01 & $10.06 \pm 0.02$ & $20.92 \pm 0.05$  & $3380 \pm 160$ & $4.87 \pm 0.01$ & $0.35 \pm 0.01$ & $1540 \pm 230$ & $1.51 \pm 0.12$ & $0.736$ & $8.7 \pm 1.4$ \\
  500.01 & $7.73 \pm 0.03$  & $21.07 \pm 0.02$  & $4450 \pm 130$ & $4.53 \pm 0.10$ & $0.75 \pm 0.06$ & $246 \pm 27$   & $1.29 \pm 0.13$ & $0.548$ & $9.3 \pm 1.2$ \\
  539.01 & $9.23 \pm 0.02$  & $9.20 \pm 0.02$   & $4900 \pm 130$ & $4.52 \pm 0.09$ & $0.81 \pm 0.05$ & $310 \pm 40$   & $1.56 \pm 0.14$ & $0.310$ & $8.1 \pm 1.1$ \\
  544.01 & $7.80 \pm 0.02$  & $24.29 \pm 0.04$  & $4220 \pm 120$ & $4.61 \pm 0.11$ & $0.66 \pm 0.06$ & $590 \pm 6$    & $1.76 \pm 0.16$ & $1.549$ & $10.3 \pm 0.9$ \\
  731.01 & $5.78 \pm 0.02$  & $106.21 \pm 0.03$ & $3540 \pm 160$ & $4.78 \pm 0.01$ & $0.46 \pm 0.01$ & $242 \pm 20$   & $0.78 \pm 0.04$ & $0.322$ & $20.4 \pm 1.7$ \\
  833.01 & $8.15 \pm 0.03$  & $23.94 \pm 0.02$  & $3920 \pm 160$ & $4.65 \pm 0.01$ & $0.60 \pm 0.02$ & $580 \pm 60$   & $1.58 \pm 0.10$ & $1.042$ & $10.0 \pm 1.2$ \\
 1075.01 & $9.11 \pm 0.02$  & $16.24 \pm 0.03$  & $3920 \pm 160$ & $4.67 \pm 0.01$ & $0.58 \pm 0.02$ & $970 \pm 90$   & $1.97 \pm 0.10$ & $0.605$ & $14.7 \pm 1.3$ \\
 1242.01 & $9.77 \pm 0.03$  & $9.06 \pm 0.03$   & $4250 \pm 130$ & $4.56 \pm 0.11$ & $0.71 \pm 0.07$ & $578 \pm 32$   & $1.87 \pm 0.18$ & $0.381$ & $9.8 \pm 0.9$ \\
 1263.01 & $7.10 \pm 0.02$  & $21.45 \pm 0.04$  & $5100 \pm 130$ & $4.55 \pm 0.08$ & $0.82 \pm 0.05$ & $258 \pm 32$   & $1.43 \pm 0.12$ & $1.021$ & $9.9 \pm 1.3$ \\
 1411.01 & $7.25 \pm 0.02$  & $30.76 \pm 0.02$  & $4180 \pm 120$ & $4.57 \pm 0.11$ & $0.69 \pm 0.06$ & $366 \pm 21$   & $1.44 \pm 0.14$ & $1.452$ & $8.9 \pm 0.9$ \\
 1442.01 & $10.09 \pm 0.02$ & $24.26 \pm 0.04$  & $3330 \pm 160$ & $4.92 \pm 0.01$ & $0.31 \pm 0.01$ & $1350 \pm 80$  & $1.24 \pm 0.05$ & $0.409$ & $10.3 \pm 0.7$ \\
 1693.01 & $8.33 \pm 0.02$  & $32.44 \pm 0.04$  & $3470 \pm 160$ & $4.77 \pm 0.01$ & $0.46 \pm 0.01$ & $1010 \pm 120$ & $1.60 \pm 0.11$ & $1.767$ & $7.8 \pm 1.1$ \\
 1860.01 & $6.79 \pm 0.02$  & $21.78 \pm 0.03$  & $5670 \pm 100$ & $4.51 \pm 0.07$ & $0.93 \pm 0.04$ & $232 \pm 29$   & $1.54 \pm 0.12$ & $1.066$ & $11.1 \pm 1.5$ \\
 2260.01 & $8.68 \pm 0.02$  & $9.85 \pm 0.03$   & $5430 \pm 130$ & $4.51 \pm 0.08$ & $0.90 \pm 0.05$ & $313 \pm 35$   & $1.73 \pm 0.13$ & $0.352$ & $10.5 \pm 1.3$ \\
 2290.01 & $9.07 \pm 0.02$  & $17.19 \pm 0.02$  & $3860 \pm 160$ & $4.68 \pm 0.01$ & $0.57 \pm 0.02$ & $600 \pm 60$   & $1.51 \pm 0.09$ & $0.386$ & $11.8 \pm 1.2$ \\
 2411.01 & $8.53 \pm 0.02$  & $16.77 \pm 0.08$  & $4100 \pm 120$ & $4.52 \pm 0.11$ & $0.73 \pm 0.07$ & $520 \pm 50$   & $1.81 \pm 0.19$ & $0.783$ & $10.7 \pm 1.4$ \\
 2427.01 & $7.05 \pm 0.02$  & $35.04 \pm 0.03$  & $4070 \pm 120$ & $4.58 \pm 0.11$ & $0.68 \pm 0.06$ & $560 \pm 24$   & $1.75 \pm 0.17$ & $1.306$ & $15.6 \pm 1.6$ \\
 2445.01 & $10.78 \pm 0.02$ & $20.56 \pm 0.10$  & $3330 \pm 160$ & $4.96 \pm 0.01$ & $0.27 \pm 0.01$ & $2400 \pm 400$ & $1.44 \pm 0.12$ & $0.371$ & $13.0 \pm 2.0$
\enddata
\vspace{-25pt}
\end{deluxetable*}

\begin{deluxetable*}{cccccccc}[t!]\label{tab: planet_fits}
\tabletypesize{\footnotesize}
\tablewidth{\textwidth}
 \tablecaption{Best-Fit Planet Parameters}
 \tablehead{
 \colhead{TOI} & \colhead{$R_{\rm p}$ ($R_\oplus$)} & \colhead{$P_{\rm orb}$ (days)} & \colhead{$T_0$ (BJD - 2457000)} & \colhead{$b$} &  \colhead{$a$ (AU)} & \colhead{$T_{\rm eq}$} & \colhead{ESM}
 }
\startdata
206.01  & $1.30 \pm 0.05$ & $0.7363104 \pm 0.0000003$ & $1325.5431 \pm 0.0004$ & $0.66 \pm 0.03$ & $0.0112 \pm 0.0001$ &  $910 \pm 36$  &  $6.4 \pm 0.5$ \\
500.01  & $1.16 \pm 0.12$ & $0.5481579 \pm 0.0000006$ & $1468.3917 \pm 0.0006$ & $0.58 \pm 0.17$ & $0.0128 \pm 0.0011$ & $1693 \pm 105$ &  $7.3 \pm 1.2$ \\
539.01  & $1.25 \pm 0.10$ & $0.3096071 \pm 0.0000004$ & $1354.1044 \pm 0.0009$ & $0.39 \pm 0.20$ & $0.0089 \pm 0.0007$ & $2311 \pm 108$ &  $5.1 \pm 0.6$ \\
544.01  & $2.03 \pm 0.10$  & $1.5483510 \pm 0.0000015$ & $1469.7570 \pm 0.0005$ & $0.64 \pm 0.08$ & $0.0251 \pm 0.0019$ & $1082 \pm 47$  & $13.0 \pm 1.8$ \\
731.01  & $0.59 \pm 0.02$ & $0.3219659 \pm 0.0000004$ & $1543.4874 \pm 0.0006$ & $0.09 \pm 0.07$ & $0.0069 \pm 0.0001$ & $1416 \pm 65$  & $11.5 \pm 0.6$ \\
833.01  & $1.27 \pm 0.07$ & $1.0418777 \pm 0.0000324$ & $1597.2560 \pm 0.0010$ & $0.31 \pm 0.14$ & $0.0171 \pm 0.0003$ & $1118 \pm 49$  &  $6.5 \pm 0.6$ \\
1075.01 & $1.72 \pm 0.08$ & $0.6047328 \pm 0.0000032$ & $1654.2511 \pm 0.0008$ & $0.18 \pm 0.12$ & $0.0118 \pm 0.0001$ & $1336 \pm 56$  & $10.9 \pm 1.0$ \\
1242.01 & $1.65 \pm 0.23$ & $0.3814851 \pm 0.0000004$ & $1683.7103 \pm 0.0004$ & $0.40 \pm 0.22$ & $0.0097 \pm 0.0010$ & $1751 \pm 110$ &  $7.7 \pm 1.4$ \\
1263.01 & $1.36 \pm 0.11$ & $1.0213646 \pm 0.0001277$ & $1683.5569 \pm 0.0018$ & $0.37 \pm 0.19$ & $0.0185 \pm 0.0014$ & $1656 \pm 75$  &  $9.2 \pm 1.3$ \\
1411.01 & $1.36 \pm 0.16$ & $1.4520358 \pm 0.0000098$ & $1739.4762 \pm 0.0014$ & $0.32 \pm 0.20$ & $0.0230 \pm 0.0026$ & $1136 \pm 59$  &  $7.6 \pm 1.1$ \\
1442.01 & $1.17 \pm 0.06$ & $0.4090677 \pm 0.0000003$ & $1683.4523 \pm 0.0003$ & $0.33 \pm 0.13$ & $0.0071 \pm 0.0002$ & $1072 \pm 54$  &  $8.9 \pm 0.8$ \\
1693.01 & $1.41 \pm 0.10$ & $1.7666957 \pm 0.0000054$ & $1817.6827 \pm 0.0014$ & $0.30 \pm 0.14$ & $0.0226 \pm 0.0004$ &  $764 \pm 19$  &  $6.0 \pm 0.9$ \\
1860.01 & $1.31 \pm 0.04$ & $1.0662107 \pm 0.0000014$ & $1683.6041 \pm 0.0003$ & $0.69 \pm 0.02$ & $0.0204 \pm 0.0002$ & $1885 \pm 28$  &  $7.9 \pm 0.4$ \\
2260.01 & $1.62 \pm 0.13$ & $0.3524728 \pm 0.0000047$ & $1928.2390 \pm 0.0007$ & $0.77 \pm 0.04$ & $0.0097 \pm 0.0001$ & $2609 \pm 86$  &  $8.7 \pm 0.9$ \\
2290.01 & $1.17 \pm 0.07$ & $0.3862224 \pm 0.0000033$ & $1764.9871 \pm 0.0013$ & $0.27 \pm 0.15$ & $0.0086 \pm 0.0001$ & $1484 \pm 31$  &  $7.1 \pm 0.8$ \\
2411.01 & $1.68 \pm 0.11$ & $0.7826942 \pm 0.0000037$ & $2116.0139 \pm 0.0010$ & $0.39 \pm 0.14$ & $0.0144 \pm 0.0001$ & $1355 \pm 45$  &  $9.9 \pm 1.2$ \\
2427.01 & $1.80 \pm 0.12$ & $1.3060011 \pm 0.0000102$ & $2169.6202 \pm 0.0004$ & $0.87 \pm 0.02$ & $0.0202 \pm 0.0002$ & $1117 \pm 46$  & $17.2 \pm 2.1$ \\
2445.01 & $1.25 \pm 0.08$ & $0.3711281 \pm 0.0000005$ & $2144.5697 \pm 0.0004$ & $0.27 \pm 0.14$ & $0.0064 \pm 0.0001$ & $1060 \pm 54$  &  $9.6 \pm 1.2$ 
\enddata
\end{deluxetable*}

\subsection{Light Curve Generation}\label{sec: lightcurves}

All of our TOIs were identified by the NASA Science Processing Operations Center (SPOC) pipeline \citep{jenkins2016tess}, which analyzes data collected at a 2-minute or 20-second cadence. The SPOC pipeline identifies potential TOIs by conducting a search for transiting planet signatures using a wavelet-based, adaptive noise-compensating matched filter with the Transiting Planet Search \citep[TPS;][]{jenkins2002impact, jenkins2010transiting} algorithm. It then performs a limb-darkened transit model fit to the detected signatures \citep{li2019kepler} and constructs a number of diagnostic tests to help assess the planetary nature of the detected signals \citep{twicken2018kepler}, which are compiled in data validation reports. The pipeline then removes the transits of each potential signature and calls TPS to detect additional transiting planet signatures, stopping when it fails to identify additional transits or reaches a limit of eight detected signatures. The SPOC pipeline generates two light curves for each TOI: light curves extracted via simple aperture photometry \citep[SAP;][]{twicken2010photometric, morris2020kepler}, and light curves extracted via SAP with an additional presearch data conditioning step \citep[PDCSAP;][]{stumpe2012kepler, smith2012kepler, stumpe2014multiscale}. The PDC step aids in planet detection by removing background trends and flux contamination due to nearby bright stars, a process that is well established in exoplanet transit surveys \citep{stumpe2012kepler}.

While the SPOC pipeline typically generates light curves that are sufficient for analyzing transits, it is not designed to preserve out-of-transit variation originating from the system. Because we are interested in whether our planet candidates show evidence of phase curves caused by reflected light in the TESS data, we take a different approach to extracting light curves that detrends the instrument systematics and stellar rotation signal while preserving transits and potential eclipses. Firstly, using the same approach as that discussed in \cite{Hedges2021}, we build design matrices consisting of (1) an estimate of the TESS scattered light background base on the top 4 principal components of the pixels outside of the optimum pipeline aperture, estimated via singular value decomposition, (2) a basis spline with a knot spacing of 0.25 days to capture stellar variability, (3) the centroids of the image in column and row dimension, (4) the single-scale cotrending basis vectors (CBVs) from the TESS pipeline, (5) a simple BLS transit model, at a fixed period, transit midpoint, and duration, and (6) a simple eclipse model, consisting of a cosine phase curve and a simple box eclipse at phase 0.5. Using the same methods from \cite{Hedges2020}, we fit these design matrices to all sectors simultaneously, fitting a single transit and a single eclipse model for all sectors, but allowing each individual sector to have unique solutions for the background, spline, centroid and CBV matrices. By taking this approach of fitting all the sectors simultaneously, we are the most sensitive to the small signal of eclipses, because all sectors are able to contribute to our eclipse measurement. Even with this rigorous approach, we detect no eclipse with a $\ge 3 \sigma$ significance for the planet candidates in this paper.

Our light curve generation code does not subtract out contamination due to nearby stars, which is an important step for correctly determining the radius of a planet candidate. However, because the code uses the same apertures as the SPOC pipeline, we are able to remove contamination using the crowding factor (labeled as \texttt{CROWDSAP} in the PDCSAP FITS headers) for each of our targets. We perform this subtraction when fitting the photometry for the orbital and physical parameters of the planet candidates, which is further described in Section \ref{sec: transit_fits}.

\subsection{Adopted Stellar Parameters}\label{sec: adopted_params}

We adopt stellar parameters for each of our host stars using a combination of spectrum analysis and empirical relation. The tools used to calculate stellar parameters from spectra are outlined in Section \ref{sec: reconspec} and the empirical relations used to calculate stellar parameters are described below. Because different methods yield different parameters (e.g., some spectrum-based analysis methods only provide effective temperature and surface gravity, whereas others also provide estimates for stellar mass and radius), we take a curated approach for each of our stars. We describe this process in detail here and present the adopted parameters in Table \ref{tab:adopted}.

\begin{deluxetable}{ccccc}
\tabletypesize{\footnotesize}
\tablewidth{\columnwidth}
 \tablecaption{Adopted Stellar Parameters \label{tab:adopted}}
 \tablehead{ 
 \colhead{TOI} & \colhead{$T_{\rm eff}$ (K)} & \colhead{$\log g$} & \colhead{$M_\star$ ($M_\odot$)} & \colhead{$R_\star$ ($R_\odot$)}
 }
 \startdata 
 206    & $3383 \pm	157$ & $4.89 \pm 0.03$ & $0.35 \pm 0.01$ & $0.35 \pm 0.01$  \\ 
 500    & $4621 \pm 50$  & $4.63 \pm 0.10$ & $0.88 \pm 0.25$ & $0.75 \pm 0.06$  \\ 
 539    & $5031 \pm 50$  & $4.58 \pm 0.10$ & $0.91 \pm 0.24$ & $0.81 \pm 0.05$  \\ 
 544    & $4369 \pm	100$ & $4.73 \pm 0.10$ & $0.85 \pm 0.20$ & $0.66 \pm 0.02$  \\ 
 731    & $3540 \pm	160$ & $4.79 \pm 0.03$ & $0.48 \pm 0.03$ & $0.46 \pm 0.01$  \\ 
 833    & $3920	\pm 160$ & $4.67 \pm 0.04$ & $0.61 \pm 0.03$ & $0.60 \pm 0.02$  \\ 
 1075   & $3921	\pm 157$ & $4.69 \pm 0.03$ & $0.60 \pm 0.02$ & $0.58 \pm 0.02$  \\ 
 1242   & $4348 \pm 100$  & $4.69 \pm 0.10$ & $0.83 \pm 0.31$ & $0.68 \pm 0.10$  \\ 
 1263   & $5166 \pm 50$  & $4.54 \pm 0.10$ & $0.78 \pm 0.20$ & $0.82 \pm 0.05$  \\
 1411   & $4266 \pm 100$  & $4.73 \pm 0.10$ & $0.59 \pm 0.23$ & $0.66 \pm 0.10$  \\
 1442   & $3330	\pm 160$ & $4.92 \pm 0.04$ & $0.29 \pm 0.02$ & $0.31 \pm 0.01$  \\
 1693   & $3499 \pm 70$  & $4.80 \pm 0.03$ & $0.49 \pm 0.03$ & $0.46 \pm 0.01$  \\ 
 1860   & $5752 \pm 100$ & $4.58 \pm 0.10$ & $0.99 \pm 0.03$ & $0.94 \pm 0.02$  \\
 2260   & $5534 \pm 100$ & $4.62 \pm 0.10$ & $0.99 \pm 0.04$ & $0.94 \pm 0.05$  \\ 
 2290   & $3813 \pm 70$  & $4.70 \pm 0.02$ & $0.56 \pm 0.01$ & $0.57 \pm 0.02$  \\ 
 2411   & $4099	\pm 123$ & $4.59 \pm 0.03$ & $0.65 \pm 0.02$ & $0.68 \pm 0.02$  \\
 2427   & $4072	\pm 121$ & $4.62 \pm 0.03$ & $0.64 \pm 0.02$ & $0.65 \pm 0.02$  \\
 2445   & $3333	\pm 157$ & $4.97 \pm 0.04$ & $0.25 \pm 0.01$ & $0.27 \pm 0.01$ 
 \enddata
\end{deluxetable}

When available, we use spectra to estimate $T_{\rm eff}$. Where more than one spectrum-based estimate of $T_{\rm eff}$ is available, we adopt the average of the estimates. If spectra are not available, or if our stellar classification tools are unable to estimate parameters for a given star (which is sometimes the case for stars with $T_{\rm eff} \le 4500$ K), we adopt the $T_{\rm eff}$ listed in the TIC.

For stars with observed with Keck/HIRES and $T_{\rm eff} > 4250$ K, we adopt the $R_\star$ estimated from the spectrum. For all other stars with $T_{\rm eff} > 4250$ K, we adopt the $R_\star$ listed in the TIC. For all stars with $T_{\rm eff} \le 4250$ K, we estimate $R_\star$ and its uncertainties with the calibrations by \cite{Mann2015}, using the 2MASS $K_{\rm S}$-band magnitudes and Gaia/DR2 parallaxes.

For stars with observed with Keck/HIRES and $T_{\rm eff} > 4700$ K, we adopt the stellar mass ($M_\star$) estimated from the spectrum. For all other stars with observed spectra and $T_{\rm eff} > 4250$ K, we calculate $M_\star$ using the $R_\star$ listed in the TIC and the surface gravity estimated from the spectra. For all stars with $T_{\rm eff} \le 4250$ K, we estimate $M_\star$ with the near-infrared mass-luminosity calibrations in \cite{Mann2019} and \cite{Benedict2016} (adopting the average of the two), using the 2MASS $K_{\rm S}$-band magnitudes and {\it Gaia}/DR2 parallaxes. 

For stars with observed spectra and $T_{\rm eff} > 4250$ K, we adopt the surface gravity ($\log g$) estimated from the spectra.  Where more than one spectrum-based estimate of $\log g$ is available, we adopt the average of the estimates. For all other stars, $\log g$ is calculated using the values of $M_\star$ and $R_\star$ determined with the methods described above.

\subsection{Transit Fits}\label{sec: transit_fits}

\begin{figure*}[t!]
  \centering
    \includegraphics[width=\textwidth]{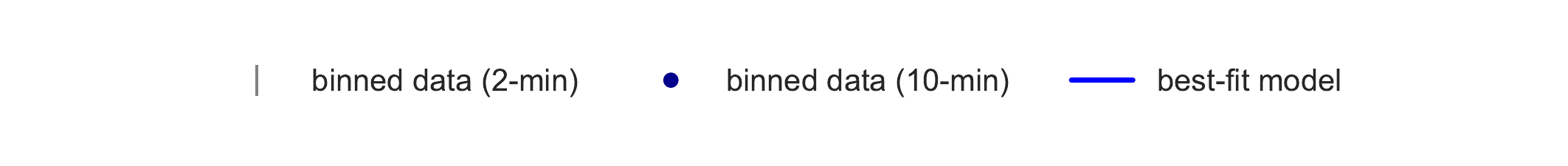}
    \includegraphics[width=\textwidth]{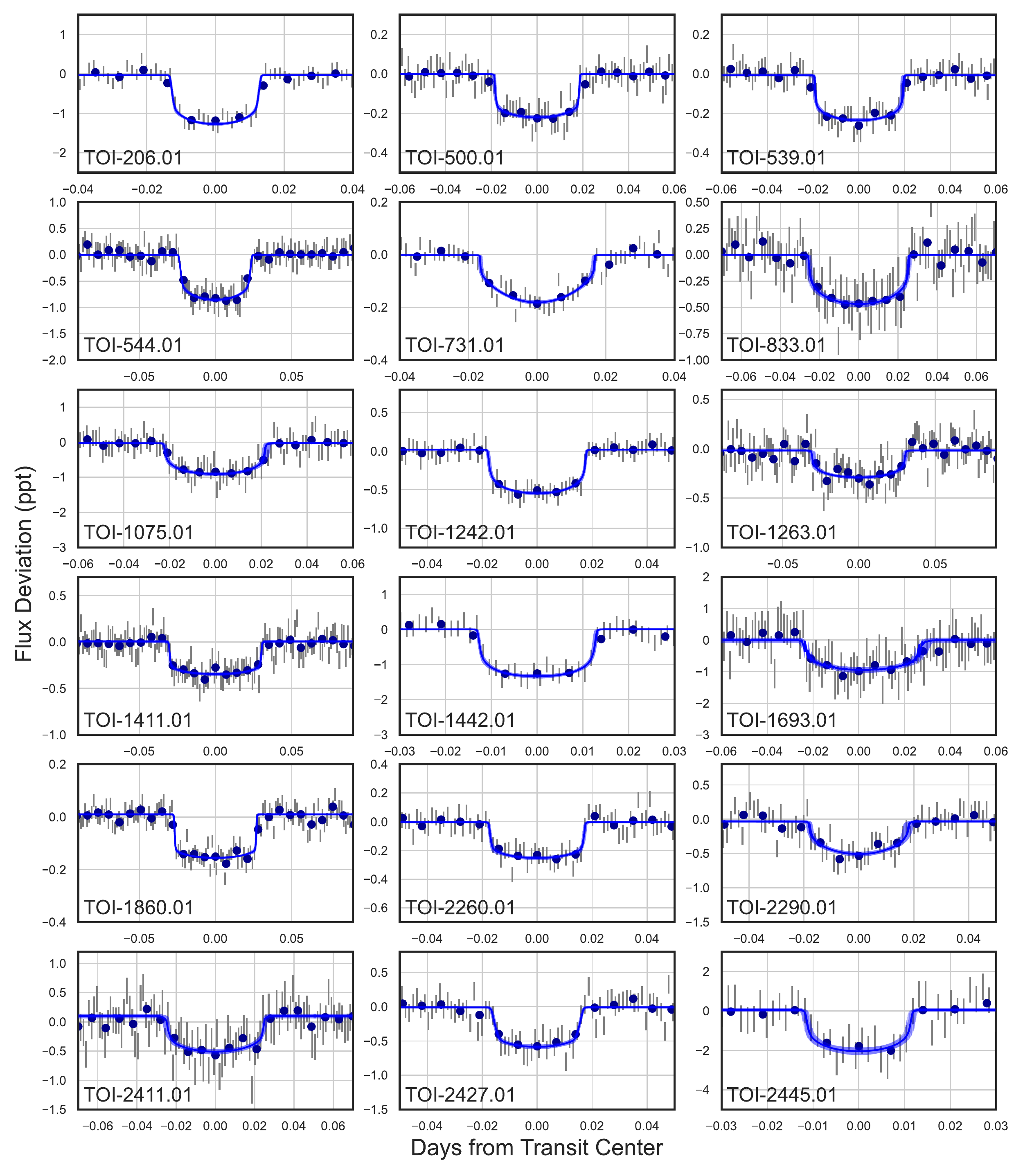}
    \caption{Phase-folded TESS data and best-fit transit models for each TOI. The parameters associated with these fits are shown in Table \ref{tab: planet_fits}. The TESS data is purged of 5$\sigma$ outliers and binned for clarity.}
    \label{fig: transits}
\end{figure*}

To estimate the orbital and planetary parameters for each of our planet candidates, we fit each of our light curves with Markov Chain Monte Carlo sampling using the \texttt{exoplanet} \citep{exoplanet:exoplanet} Python package. Our transit model assumed a circular orbit and was initialized with the following priors: (1) Gaussian priors for $M_\star$ and $R_\star$, (2) a Gaussian prior for the natural logarithm of $P_{\rm orb}$, (3) a Gaussian prior for the time of inferior conjunction ($T_0$), (4) a uniform prior for the impact parameter ($b$), (5) uniform priors for quadratic limb-darkening coefficients \citep{exoplanet:kipping13}, (6) a Gaussian prior for the natural logarithm of the transit depth, and (7) a Gaussian prior for the flux zero point of the light curve. For each TOI, we ran a 10 walker ensemble for 20000 steps and ensured that convergence was achieved, then discarded the first 10000 steps as burn-in. The best-fit parameters for each planet candidate are shown in Table \ref{tab: planet_fits} and the corresponding best-fit light curve models are shown in Figure \ref{fig: transits}. 

For most of the TOIs in this paper, these fits only included TESS data. However, transits of TOI-206.01, TOI-1075.01, TOI-1442.01, TOI-1693.01, TOI-2411.01, TOI-2411.01, TOI-2427.01, and TOI-2445.01 were also had observed by ground-based telescopes. For these targets, we perform joint fits including both the TESS data and the ground-based data. We fit for limb-darkening coefficients, transit depth, and flux zero point independently for each dataset while treating $M_\star$, $R_\star$, $P_{\rm orb}$, $T_0$, and $b$ as parameters that are shared between the datasets. The ground-based data are discussed in Section \ref{sec: SG1}. In these cases, we adopt the planet radii inferred from the TESS data.

Using these new planet properties, we recalculate the ESM for each TOI. All TOIs except for TOI-206.01, TOI-500.01, TOI-539.01, and TOI-1693.01 retained an ${\rm ESM} > 7.5$. In addition, we find that TOI-544.01 may have a radius slightly larger than $2 R_\oplus$. Even though these TOIs do not meet our initial selection criteria with their newly calculated properties, we keep them in our analysis.

\section{Vetting Procedure}\label{sec: procedure}

We examine each of the unconfirmed TOIs described in Table \ref{tab1} using both follow-up observations and analyses with the vetting tools \dave\ \citep{kostov2019discovery} and \tri\ \citep{2020ascl.soft02004G, giacalone2021triceratops}. Follow-up observations are used to search for evidence of FPs outside of the TESS data, while \dave\ and \tri\ are used to search for FP signatures within the TESS data.

We utilize three forms of follow-up observations in our vetting analysis. First, we acquired high-resolution images, such as those obtainable with adaptive optics, to search for unresolved companions (either bound or chance aligned) near the target. These companions can dilute the TESS transit, leading to an underestimation in the planet radius, or can even be the sources of FPs if the companion hosts an eclipsing binary \citep{ciardi2015understanding, furlan2017kepler, hirsch2017assessing, teske2018effects}. Second, we obtained reconnaissance spectra to search for evidence of spectroscopic binaries around the target stars. Evidence of a binary star in the form of single-line or double-line spectroscopic binaries could either indicate that the planet candidate itself is an eclipsing binary or that there is an unresolved star in the system causing us to underestimate the radius of the planet candidate. In addition, deriving stellar parameters from spectra allows us to reaffirm the photometrically derived parameters in the TIC, which is important for a correct calculation of the planet radius and equilibrium temperature. Third, we used ground-based facilities with higher spatial resolutions than TESS to obtain time-series photometry of the field of stars within $2\farcm5$ from the target during the time of transit. Because it is possible for nearby stars to contaminate the TESS aperture, transits due to nearby eclipsing binary stars can be mistakable as transits due to planet-size objects around the target star. These scenarios can be ruled out by either observing the transit on the target star, free of any contamination from nearby stars, or ruling out eclipsing binaries around all nearby stars bright enough to cause an FP. These observations are further described in Section \ref{sec: followup}.

Next, we analyze each TOI with \dave, which vets planet candidates at both the pixel and light curve levels. At the pixel level, \dave\ uses centroid offset analyses to identify evidence of FPs due to contamination from nearby stars. A similar difference image centroiding analysis is performed by the SPOC for each of its threshold-crossing events \citep[TCEs;][]{twicken2018kepler}. For all TOIs, we cross-check with the corresponding SPOC data validation report to see if an offset is detected in the SPOC analysis. At the light curve level, \dave\ searches for signatures -- such as differences in odd and even transits, secondary eclipses, and nontransit variability -- that are indicative of FPs \citep[e.g.,][]{morton2012efficient, ansdell2018scientific, shallue2018identifying}. For these analyses, we use the SAP/PDCSAP light curves generated by the SPOC pipeline.

Lastly, we analyze each TOI using \tri, which vets a planet candidate by calculating the Bayesian probability that the candidate is an astrophysical FP. The analysis begins by querying the TIC for all stars in a $2\farcm5$ radius around the target star and modeling the TESS pixel response function to determine the amount of flux contamination each contributes to the aperture. For each star that contributes enough flux to cause the observed transit, the tool simulates light curves due to transiting planets and eclipsing binaries and calculates the marginal likelihood of each transit-producing scenario. These are combined with prior probabilities to calculate the FP probability (FPP; the total probability that the observed transit is due to something other than a transiting planet around the target star) and nearby FP probability (NFPP; the total probability that the observed transit originated from a known nearby star) for the planet candidate. A planet candidate that achieves a sufficiently small FPP (${\rm FPP} < 0.01$) and NFPP (${\rm NFPP} < 10^{-3}$) can be considered validated \citep{giacalone2021triceratops}. For this analysis, we use the same light curves generated using the methodology described in Section \ref{sec: lightcurves} (without contamination due to nearby stars removed with the \texttt{CROWDSAP} factor). Because the FPPs and NFPPs returned by \tri\ have an intrinsic scatter, we run the tool 20 times on each TOI and report the means and standard deviations of these probabilities. Ultimately, we decide whether a planet is validated based on the results of this analysis.

\tri\ also has the ability to fold in follow-up observations to place tighter constraints on the chances of FP scenarios. Specifically, high-resolution images are used to constrain the area of sky around the target where unresolved companion stars can exist. Folding in these data therefore reduces the probabilities of scenarios like those involving hierarchical and background eclipsing binaries. In addition, time-series photometry allows us to remove nearby stars that have been cleared from being eclipsing binaries from the analysis. When available, we utilize these data during this step of vetting.

\section{Follow-up Observations}\label{sec: followup}

We analyzed our TOIs using observations obtained by the TFOP Working Groups.\footnote{https://tess.mit.edu/followup} The data from these observations are available for download on the ExoFOP-TESS website and are summarized below.

\subsection{High-Resolution Imaging}

\begin{figure*}[h!]
  \centering
    \includegraphics[width=1.0\textwidth]{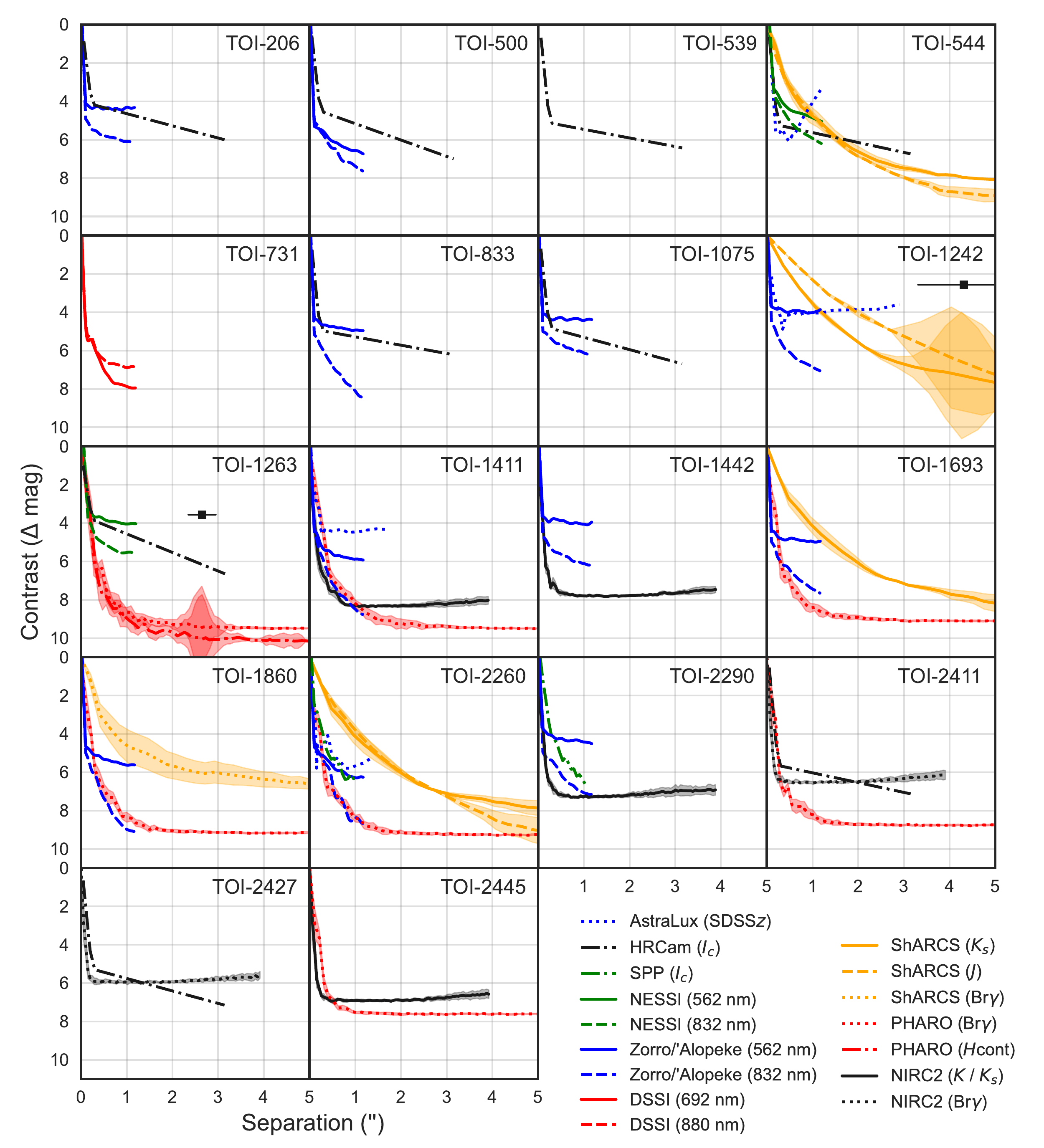}
    \caption{Contrast curves extracted from the high-resolution follow-up observations summarized in Table \ref{tab:imaging}, which allows us to rule out companions at a given separation above a certain $\Delta$mag. Curves without shading (i.e., those from lucky and speckle imaging) were constructing by taking the 5$\sigma$ upper limits of the contrasts in circular annuli around the target star. Curves with shading (i.e., those from adaptive optics imaging) were constructed by taking the mean and root-mean-square error of the contrasts in circular annuli around the target star. TOI-1242 and TOI-1263 have $< 5\arcsec$ companions, which are both known stars in the TIC. The TESS band $\Delta$mag and separations of these companions are indicated by black squares. These curves are folded into the \tri\ analysis described in Section \ref{sec: procedure}.}
    \label{fig: imaging}
\end{figure*}

We obtained high-resolution images of our TOIs using adaptive optics, speckle, and lucky imaging. In each of these observations, we search for stars within $5 \arcsec$ from the target star. In situations where companions were detected, we cross-checked the TIC to determine if these companions were previously known. These observations, which were obtained by members of TFOP Sub Group 3 (SG3), are summarized in Table \ref{tab:imaging}, are displayed in Figure \ref{fig: imaging}, and are discussed below.

\startlongtable
\begin{deluxetable*}{ccccccccccc}
\tabletypesize{\footnotesize}
\tablewidth{\columnwidth}
 \tablecaption{Summary of High-Resolution Imaging Follow-Up}
 \tablehead{
 \colhead{TOI} & \colhead{Telescope} & \colhead{Instrument} & \colhead{Filter} & \colhead{Image Type} & \colhead{Companion} & & \colhead{Contrast} & & \colhead{($\Delta$mag)} & \\ \cmidrule{7-11}
    &&&&& \colhead{($< 5\arcsec$)} & 0$\farcs$1 & 0$\farcs$5 & 1$\farcs$0 & 1$\farcs$5 & 2$\farcs$0 
 }
 \startdata
 \multirow{3}{*}{206}     & SOAR (4.1 m)  & HRCam    & $I_c$           & Speckle & - & 1.625 & 4.323 & 4.641 & 4.958 & 5.275  \\
                          & Gemini-S (8 m)& Zorro    & 562 nm          & Speckle & - & 4.115 & 4.398 & 4.309 &  -    &   -    \\
                          & Gemini-S (8 m)& Zorro    & 832 nm          & Speckle & - & 4.908 & 5.787 & 6.081 &  -    &   -    \\ \cmidrule{1-11}
 \multirow{3}{*}{500}     & SOAR (4.1 m)  & HRCam    & $I_c$           & Speckle & - & 1.721 & 4.738 & 5.164 & 5.591 & 6.017  \\
                          & Gemini-S (8 m)& Zorro    & 562 nm          & Speckle & - & 5.307 & 6.083 & 6.564 &  -    &   -    \\
                          & Gemini-S (8 m)& Zorro    & 832 nm          & Speckle & - & 5.057 & 6.441 & 7.386 &  -    &   -    \\ \cmidrule{1-11}
 539                      & SOAR (4.1 m)  & HRCam    & $I_c$           & Speckle & - & 1.660 & 5.238 & 5.462 & 5.686 & 5.910  \\ \cmidrule{1-11}
\multirow{6}{*}{544}      & CAHA (2.2 m)  & AstraLux & SDSS$z$         & Lucky   & - & 2.614 & 6.015 & 4.053 &  -    &   -    \\
                          & Shane (3 m)   & ShARCS   & $K_s$           & AO      & - & 0.588 & 3.272 & 4.774 & 5.816 & 6.625   \\
                          & Shane (3 m)   & ShARCS   & $J$             & AO      & - & 0.842 & 3.223 & 4.713 & 5.940 & 6.872   \\
                          & WIYN (3.5 m)  & NESSI    & 562 nm          & Speckle & - & 1.817 & 4.431 & 4.856 &   -   &     -   \\
                          & WIYN (3.5 m)  & NESSI    & 832 nm          & Speckle & - & 1.646 & 5.025 & 5.933 &   -   &     -   \\ 
                          & SOAR (4.1 m)  & HRCam    & $I_c$           & Speckle & - & 1.903 & 5.370 & 5.629 & 5.887 & 6.145 \\ \cmidrule{1-11}
 \multirow{2}{*}{731}     & Gemini-S (8 m)& DSSI     & 692 nm          & Speckle & - & 4.721 & 6.998 & 7.872 &   -   &    -   \\
                          & Gemini-S (8 m)& DSSI     & 880 nm          & Speckle & - & 4.498 & 6.470 & 6.889 &   -   &    -   \\ \cmidrule{1-11}
 \multirow{3}{*}{833}     & SOAR (4.1 m)  & HRCam    & $I_c$           & Speckle & - & 1.922 & 5.068 & 5.285 & 5.503 & 5.720  \\
                          & Gemini-S (8 m)& Zorro    & 562 nm          & Speckle & - & 4.319 & 4.752 & 4.932 &  -    &     -   \\
                          & Gemini-S (8 m)& Zorro    & 832 nm          & Speckle & - & 5.162 & 6.805 & 8.119 &  -    &     -   \\ \cmidrule{1-11}
 \multirow{3}{*}{1075}    & SOAR (4.1 m)  & HRCam    & $I_c$           & Speckle & - & 1.708 & 4.990 & 5.310 & 5.631 & 5.9518  \\
                          & Gemini-S (8 m)& Zorro    & 562 nm          & Speckle & - & 4.061 & 4.278 & 4.429 &  -    &   -    \\
                          & Gemini-S (8 m)& Zorro    & 832 nm          & Speckle & - & 5.009 & 5.653 & 6.126 &  -    &   -    \\ \cmidrule{1-11}
 \multirow{5}{*}{1242}    & CAHA (2.2 m)  & AstraLux & SDSS$z$         & Lucky   & - & 2.143 & 4.128 & 4.047 & 3.898 &   -    \\
                          & Shane (3 m)   & ShARCS   & $K_s$           & AO      & Y & 0.438 & 2.039 & 3.549 & 4.641 & 5.567   \\
                          & Shane (3 m)   & ShARCS   & $J$             & AO      & Y & 0.237 & 1.186 & 2.313 & 3.304 & 4.055   \\
                          & Gemini-N (8 m)& 'Alopeke & 562 nm          & Speckle & - & 3.718 & 3.980 & 4.017 &   -   &     -   \\
                          & Gemini-N (8 m)& 'Alopeke & 832 nm          & Speckle & - & 4.551 & 6.087 & 6.856 &   -   &     -   \\ \cmidrule{1-11}
 \multirow{5}{*}{1263}    & WIYN (3.5 m)  & NESSI    & 562 nm          & Speckle & - & 1.690 & 3.799 & 4.049 &   -   &     -   \\
                          & WIYN (3.5 m)  & NESSI    & 832 nm          & Speckle & - & 1.679 & 5.066 & 5.533 &   -   &     -   \\
                          & SOAR (4.1 m)  & HRCam    & $I_c$           & Speckle & Y & 1.782 & 4.081 & 4.565 & 5.049 & 5.532  \\
                          & Palomar (5 m) & PHARO    & Br$\gamma$      & AO      & Y & 1.716 & 6.869 & 8.648 & 9.145 & 9.275  \\
                          & Palomar (5 m) & PHARO    & $H$cont         & AO      & Y & 1.986 & 7.769 & 8.965 & 9.618 & 9.685  \\ \cmidrule{1-11}
 \multirow{5}{*}{1411}    & CAHA (2.2 m)  & AstraLux & SDSS$z$         & Lucky   & - & 2.368 & 4.425 & 4.461 & 4.309 &        \\
                          & Palomar (5 m) & PHARO    & Br$\gamma$      & AO      & - & 1.789 & 6.912 & 8.190 & 9.017 & 9.241  \\ 
                          & Gemini-N (8 m)& 'Alopeke & 562 nm          & Speckle & - & 4.333 & 5.609 & 5.877 &   -   &     -   \\
                          & Gemini-N (8 m)& 'Alopeke & 832 nm          & Speckle & - & 4.414 & 7.160 & 8.496 &   -   &     -   \\
                          & Keck (10 m)   & NIRC2    & $K_s$           & AO      & - & 3.892 & 7.574 & 8.308 & 8.317 & 8.312  \\ \cmidrule{1-11}
 \multirow{3}{*}{1442}    & Gemini-N (8 m)& 'Alopeke & 562 nm          & Speckle & - & 3.644 & 3.867 & 4.060 &   -    &     -   \\
                          & Gemini-N (8 m)& 'Alopeke & 832 nm          & Speckle & - & 4.703 & 5.622 & 6.118 &   -    &     -   \\
                          & Keck (10 m)   & NIRC2    & $K$             & AO      & - & 3.905 & 7.638 & 7.801 & 7.837 & 7.782  \\ \cmidrule{1-11}  
 \multirow{4}{*}{1693}    & Shane (3 m)   & ShARCS   & $K_s$           & AO      & - & 0.610 & 2.790 & 4.155 & 5.208 & 6.081   \\
                          & Palomar (5 m) & PHARO    & Br$\gamma$      & AO      & - & 2.751 & 6.982 & 8.411 & 8.847 & 8.916  \\
                          & Gemini-N (8 m)& 'Alopeke & 562 nm          & Speckle & - & 4.380 & 4.803 & 4.958 &   -    &     -   \\
                          & Gemini-N (8 m)& 'Alopeke & 832 nm          & Speckle & - & 4.979 & 6.440 & 7.443 &   -    &     -   \\ \cmidrule{1-11}
 \multirow{4}{*}{1860}    & Shane (3 m)   & ShARCS   & Br$\gamma$      & AO      & - & 0.592 & 3.287 & 4.598 & 5.096 & 5.669   \\
                          & Palomar (5 m) & PHARO    & Br$\gamma$      & AO      & - & 2.366 & 6.873 & 8.346 & 8.984 & 9.051  \\
                          & Gemini-N (8 m)& 'Alopeke & 562 nm          & Speckle & - & 4.659 & 5.327 & 5.631 &   -   &     -   \\
                          & Gemini-N (8 m)& 'Alopeke & 832 nm          & Speckle & - & 4.984 & 7.356 & 8.978 &   -   &     -   \\ \cmidrule{1-11}
 \multirow{7}{*}{2260}    & CAHA (2.2 m)  & AstraLux & SDSS$z$         & Lucky   & - & 2.456 & 5.399 & 5.666 &   -   &     -   \\
                          & SAI (2.5 m)   & SPP      & $I_c$           & Speckle & - & 2.548 & 5.293 & 6.406 &   -   &     -   \\ 
                          & Shane (3 m)   & ShARCS   & $K_s$           & AO      & - & 0.564 & 2.740 & 4.142 & 5.139 & 6.027   \\
                          & Shane (3 m)   & ShARCS   & $J$             & AO      & - & 0.547 & 2.345 & 3.799 & 5.040 & 5.968   \\
                          & Palomar (5 m) & PHARO    & Br$\gamma$      & AO      & - & 2.875 & 6.920 & 8.418 & 8.983 & 9.106   \\  
                          & Gemini-N (8 m)& 'Alopeke & 562 nm          & Speckle & - & 4.688 & 5.674 & 6.283 &   -   &     -  \\
                          & Gemini-N (8 m)& 'Alopeke & 832 nm          & Speckle & - & 4.539 & 6.577 & 8.384 &   -   &     -  \\ \cmidrule{1-11}
 \multirow{4}{*}{2290}    & SAI (2.5 m)   & SPP      & $I_c$           & Speckle & - & 1.207 & 5.176 & 6.509 &   -   &     -   \\ 
                          & Gemini-N (8 m)& 'Alopeke & 562 nm          & Speckle & - & 3.740 & 4.231 & 4.424 &   -   &     -  \\
                          & Gemini-N (8 m)& 'Alopeke & 832 nm          & Speckle & - & 4.965 & 6.128 & 7.071 &   -   &     -  \\
                          & Keck (10 m)   & NIRC2    & $K$             & AO      & - & 3.755 & 7.169 & 7.276 & 7.254 & 7.181  \\ \cmidrule{1-11}
 \multirow{3}{*}{2411}    & SOAR (4.1 m)  & HRCam    & $I_c$           & Speckle & - & 1.844 & 5.776 & 6.031 & 6.286 & 6.541   \\ 
                          & Palomar (5 m) & PHARO    & Br$\gamma$      & AO      & - & 2.566 & 7.197 & 8.199 & 8.637 & 8.712   \\ 
                          & Keck (10 m)   & NIRC2    & Br$\gamma$      & AO      & - & 3.906 & 6.505 & 6.552 & 6.476 & 6.483  \\ \cmidrule{1-11}
 \multirow{2}{*}{2427}    & SOAR (4.1 m)  & HRCam    & $I_c$           & Speckle & - & 1.955 & 5.434 & 5.758 & 6.083 & 6.408 \\
                          & Keck (10 m)   & NIRC2    & Br$\gamma$      & AO      & - & 3.949 & 5.908 & 5.972 & 5.891 & 5.922 \\ \cmidrule{1-11}
 \multirow{2}{*}{2445}    & Palomar (5 m) & PHARO    & Br$\gamma$      & AO      & - & 2.608 & 6.876 & 7.527 & 7.571 & 7.623 \\ 
                          & Keck (10 m)   & NIRC2    & $K$             & AO      & - & 3.955 & 6.939 & 6.904 & 6.912 & 6.895 
\enddata
\label{tab:imaging}
\end{deluxetable*}

\subsubsection{CAHA/AstraLux}

TOI-544, TOI-1238, TOI-1242, TOI-1411, TOI-1685, and TOI-2260 were observed with the high-spatial-resolution imaging instrument AstraLux \citep{hormuth08} mounted on the 2.2 m telescope at Calar Alto Observatory (CAHA; Almer\'ia, Spain). The instrument uses the lucky-imaging technique \citep{fried78} by combining a fast readout and a small plate scale to obtain thousands of images with exposure times below the speckle coherence time and using the Sloan Digital Sky Survey $z$ filter (SDSS$z$). We observed TOI-1411 on UT 2020-01-13, TOI-1242 on UT 2020-02-26, TOI-1238 and TOI-1685 on UT 2020-08-07, and TOI-544 and TOI-2260 on UT 2021-03-23. We used the following strategy for each target: 7\,000 frames of 10\,ms exposure time to TOI-544, 126\,500 frames of 20\,ms for TOI-1238, 12\,055 frames of 20\,ms for TOI-1242, 98\,600 frames of 10\,ms for TOI-1411, 87\,600 frames of 20\,ms for TOI-1685, and 166\,860 frames of 10\,ms for TOI-2260. The number of frames and exposure time was adapted to achieve a magnitude contrast at $1\arcsec$ separation that would allow us to discard chance aligned binaries mimicking the same transit depth as the planet candidates (see \citealt{lillo-box12,lillo-box14b}). We choose a $6\arcsec \times 6\arcsec$ field-of-view in order to be able to reduce the individual exposure time down to the 10 ms level to improve the close-by sensitivity.

The datacubes were then reduced using the observatory pipeline \citep{hormuth08}. As a compromise between magnitude sensitivity and spatial resolution, we selected the 10\% of the best frames according to their Strehl ratio \citep{strehl1902} and then aligned and combined these images to compute a final high-spatial-resolution image per target. We computed the 5$\sigma$ sensitivity curves for each of the images by using our own developed \texttt{astrasens} package\footnote{\url{https://github.com/jlillo/astrasens}} with the procedure described in \cite{lillo-box14b}. We found no stellar companions within these computed sensitivity limits. 

\subsubsection{SAI/SPP}

TOI-2260 and TOI-2290 were observed on UT 2021-02-02 and 2020-10-28, respectively, with the SPeckle Polarimeter \citep[SPP;][]{safonov2017speckle} on the 2.5 m telescope at the Caucasian Observatory of Sternberg Astronomical Institute (SAI) of Lomonosov Moscow State University. SPP uses an electron-multiplying CCD Andor iXon 897 as a detector. The atmospheric dispersion compensator allowed for observations of these relatively faint targets through the wide-band $I_c$ filter. Power spectra were estimated from 4000 frames with 30 ms exposures. The detector has a pixel scale of 20.6 mas/pixel. We did not detect any stellar companions in our observations. The 5$\sigma$ sensitivity curves are presented in Figure \ref{fig: imaging}.

\subsubsection{WIYN/NESSI}

We observed TOI-544 and TOI-1263 on UT 2019-10-12 and 2019-11-16, respectively, with the NN-Explore Exoplanet Stellar Speckle Imager \citep[NESSI;][]{2018PASP..130e4502S,2018SPIE10701E..0GS} mounted on the 3.5 m WIYN telescope at Kitt Peak. High-speed electron-multiplying CCDs were used to capture image sequences simultaneously in two passbands at 562 and 832~nm. Data were acquired and reduced following \citet{2011AJ....142...19H}, yielding the 5$\sigma$ contrast curves shown in Figure \ref{fig: imaging}. No secondary sources were detected within the reconstructed 4.6\arcsec$\times$4.6\arcsec\, images.

\subsubsection{SOAR/HRCam}

We utilize speckle interferometric observations of TOI-206, TOI-500, TOI-539, TOI-544, TOI-833, TOI-1075, TOI-1263, TOI-2411, and TOI-1427 taken with HRCam mounted on the 4.1m Southern Astrophysical Research (SOAR) telescope. These observations and their related analyses are outlined in \cite{ziegler2019soar} and \cite{ziegler2021soar}. We refer the reader to those papers for more information.

\subsubsection{Shane/ShARCS}

We observed TOI-544, TOI-1242, TOI-1693, TOI-1860, TOI-2260 using the ShARCS camera on the Shane 3 m telescope at Lick Observatory \citep{kupke2012shaneao, gavel2014shaneao} on UT 2019-09-13, 2021-03-05, 2020-12-02, 2020-12-02, and 2021-03-29, respectively. Observations were taken using the Shane adaptive optics (AO) system in natural guide star mode. We collected our observations using a 4-point dither pattern with a separation of 4$\arcsec$ between each dither position. For TOI-544, TOI-1242, and TOI-2260 we obtained observations with the $K_s$ filter $(\lambda_o =  2.150; \Delta\lambda = 0.320\mu$m) and the $J$ filter $(\lambda_o =  1.238; \Delta\lambda = 0.271\mu$m). For TOI-1242, we detected a $\sim 4\farcs3$ companion in both filters. For TOI-1693 we obtained observations with only the $K_s$ filter. For TOI-1860 we obtained observations with only the narrowband Br$\gamma$ filter $(\lambda_o =  2.167; \Delta\lambda = 0.020\mu$m). See \cite{savel2020closer} for a detailed description of the observing strategy and reduction procedure.

\subsubsection{Palomar/PHARO}
The Palomar Observatory observations of TOI-1263, TOI-1693, TOI-1860, TOI-1411, TOI-2260, TOI-2411, and TOI-2445 were made with the PHARO instrument \citep{hayward2001pharo} behind the natural guide star AO system P3K \citep{dekany2013palm} on UT 2019-06-13, 2021-09-19, 2021-06-21, 2020-01-08, 2021-03-03, 2021-08-23, and 2021-09-20, respectively, in a standard five-point quincunx dither pattern with steps of 5\arcsec. Each dither position was observed three times, offset in position from each other by 0\farcs5, for a total of 15 frames.  The camera was in the narrow-angle mode with a full field of view (FOV) of $\sim25\arcsec$ and a pixel scale of approximately $0.025\arcsec$ per pixel. Observations were made in the narrowband Br$\gamma$ filter $(\lambda_o = 2.1686; \Delta\lambda = 0.0326\mu$m) for the three targets. TOI-1263, which was detected to have a $\sim 2\farcs6$ companion, was also observed in the $H$cont $(\lambda_o = 1.668; \Delta\lambda = 0.018\mu$m) filter to enable a color-based determination of the boundedness \citep{lund2020}. 

The AO data were processed and analyzed with a custom set of IDL tools.  The science frames were flat-fielded and sky-subtracted.  The flat fields were generated from a median average of dark subtracted flats taken on sky.  The flats were normalized such that the median value of the flats is unity.  The sky frames were generated from the median average of the 15 dithered science frames; each science image was then sky-subtracted and flat-fielded.  The reduced science frames were combined into a single combined image using an intrapixel interpolation that conserves flux, shifts the individual dithered frames by the appropriate fractional pixels, and median-coadds the frames.  The final resolution of the combined dither was determined from the full-width half-maximum of the point spread function which was typically 0.1\arcsec.

\subsubsection{Gemini-N/'Alopeke, Gemini-S/Zorro, and Gemini-S/DSSI}

For TOI-206, TOI-500, TOI-833, TOI-1075, TOI-1242, TOI-1411, TOI-1442, TOI-1634, TOI-1693, TOI-1860, TOI-2260, and TOI-2290, speckle interferometric observations were performed using 'Alopeke and Zorro, dual-channel high-resolution imaging instruments mounted on the Gemini 8-m North and South telescopes, respectively \citep{2018SPIE10701E..0GS}.\footnote{\url{https://www.gemini.edu/instrumentation/current-instruments/alopeke-zorro/} } Those observations were obtained on UT 2020-12-26, 2020-03-16, 2020-01-12, 2019-09-12, 2020-02-17, 2020-06-07, 2020-06-07, 2020-12-02, 2020-02-18, 2020-06-08, 2021-06-24, and 2021-06-24 respectively.

Many thousands of 60 ms images were collected on two EMCCDs, each preceded by a narrowband filter to minimize atmospheric dispersion. The full set of observations taken in 562 nm and 832 nm was then combined in Fourier space to produce their power spectrum and autocorrelation functions. From these, interferometric fringes were detected if a companion star was present within our $\sim 1 \farcs 2$ FOV, with an inner angle at the diffraction limit of the Gemini telescope. The data reduction pipeline produces final data products that include 5$\sigma$ contrast curves and reconstructed images \citep{horch1996speckle, horch2012observations, 2011AJ....142...19H}. The contrast curves at both 562 nm and 832 nm sample the spatial region near the target star from approximately 1 au to 50-100 au (depending on the distance to the target star) yielding contrast levels of 5-8 magnitudes.

For TOI-731, speckle interferometric observations were performed using the Differential Speckle Survey Instrument \citep[DSSI;][]{horch2009}, a dual-channel, high-resolution imager that allows simultaneous observations with filters centered at 692 nm and 880 nm. DSSI can resolve binaries down to 20 mas. The observations were obtained on UT 2018-03-30 when DSSI was mounted on the Gemini-South 8-m telescope as a visiting instrument.

\subsubsection{Keck/NIRC2}
TOI-1411, TOI-1442, TOI-2290, TOI-2411, TOI-2427, and TOI-2445 were observed with the NIRC2 instrument on Keck-II behind the natural guide star AO system on UT 2020-09-09, 2020-05-28, 2021-08-28, 2021-08-28, 2021-08-24, and 2021-08-28 respectively.  The observations were made in the standard 3-point dither pattern that is used with NIRC2 to avoid the left lower quadrant of the detector which is typically noisier than the other three quadrants. The dither pattern step size was $3\arcsec$ and was repeated twice times offset from each other by 0.5\arcsec\ for a total of 9 dithered observations.  The observations for TOI-1411 were made in the $K_s$ filter $(\lambda_o = 2.146; \Delta\lambda = 0.311\mu$m) and TOI-1442 were made in the $K$ $(\lambda_o = 2.196; \Delta\lambda = 0.336\mu$m) filter.  The camera was in the narrow-angle mode with a full FOV of $\sim10\arcsec$ and a pixel scale of approximately $0.099442\arcsec$ per pixel. The Keck AO observations revealed no additional stellar companions to within a resolution $\sim 0.05\arcsec$ FWHM.  The data were processed and analyzed with the same software suite used for the Palomar PHARO observations. 

\subsection{Reconnaissance Spectroscopy}\label{sec: reconspec}

We obtained reconnaissance spectra of several of our TOIs to search for evidence of FPs and characterize the target stars. These spectra were obtained by members of TFOP Sub Group 2 (SG2). The observations and the stellar parameters extracted from the acquired spectra are summarized in Table \ref{tab:spec}. Further details on the observations and the analyses performed to search for FP signatures and characterize the stars are provided below.

\begin{deluxetable*}{ccccccccccc}
\tabletypesize{\footnotesize}
\tablewidth{\columnwidth}
 \tablecaption{Summary of Reconnaissance Spectroscopy Follow-Up and Derived Stellar Parameters \label{tab:spec}}
 \tablehead{ 
 \colhead{TOI} & \colhead{Telescope} & \colhead{Instrument} & \colhead{$N_{\rm obs}$} & \colhead{$T_{\rm eff}$ (K)} & \colhead{$\log g$} & \colhead{$M_\star$ ($M_\odot$)} & \colhead{$R_\star$ ($R_\oplus$)} & \colhead{[Fe/H]} & \colhead{[M/H]} & \colhead{$v \sin{i}$ (km/s)}
 }
 \startdata 
 \multirow{1}{*}{500}  & SMARTS (1.5 m) & CHIRON & 2 & $4621 \pm 50$  & $4.63 \pm 0.10$ &         -       &        -        &-& $-0.22 \pm 0.10$ & $2.00 \pm 0.50$  \\ \cmidrule{1-11}
 \multirow{1}{*}{539}  & SMARTS (1.5 m) & CHIRON & 2 & $5031 \pm 50$  & $4.58 \pm 0.10$ &         -       &        -        &-& $-0.14 \pm 0.10$ & $3.30 \pm 0.50$  \\ \cmidrule{1-11}
 \multirow{2}{*}{544}  & SMARTS (1.5 m) & CHIRON & 4 &       -        &       -         &         -        &        -        &-&       -         &        -         \\
                       & FLWO (1.5 m)   & TRES   & 2 & $4369 \pm 100$ &  $4.73 \pm 0.10$  &         -        &        -        &-& $-0.42 \pm 0.08$ & $1.80 \pm 0.50$ \\ \cmidrule{1-11}
 \multirow{1}{*}{731}  & SMARTS (1.5 m) & CHIRON & 2 &      -         &      -          &        -        &        -         &-&         -        &         -        \\ \cmidrule{1-11}
 \multirow{1}{*}{833}  & SMARTS (1.5 m) & CHIRON & 2 &      -         &      -         &        -         &        -       &-&         -        &         -        \\ \cmidrule{1-11}
 \multirow{1}{*}{1075} & SMARTS (1.5 m) & CHIRON & 1 &      -         &      -         &        -         &        -       &-&         -        &         -        \\ \cmidrule{1-11}
 \multirow{2}{*}{1242} & FLWO (1.5 m)   & TRES   & 2 & $4437 \pm 100$  & $4.69 \pm 0.10$ &         -       &        -        &-& $-0.13 \pm 0.08$ & $3.60 \pm 0.50$  \\ 
                       & Keck (10 m)    & HIRES  & 2 & $4259 \pm 70$  &        -        &         -       & $0.68 \pm 0.10$ & $0.00 \pm 0.09$ & - &        -        \\ \cmidrule{1-11}
 \multirow{2}{*}{1263} & FLWO (1.5 m)   & TRES   & 1 & $5160 \pm 50$  & $4.58 \pm 0.10$ &         -       &        -        &-& $0.04 \pm 0.08$  & $2.10 \pm 0.50$  \\
                       & NOT (2.56 m)   & FIES   & 4 & $5172 \pm 50$  & $4.50 \pm 0.10$ &         -       &        -        &-& $0.00 \pm 0.08$  & $0.60 \pm 0.50$  \\
 \cmidrule{1-11}
 \multirow{2}{*}{1411} & FLWO (1.5 m)   & TRES   & 2 & $4352 \pm 100$  & $4.73 \pm 0.10$ &         -       &        -        &-& $-0.37 \pm 0.08$ & $2.00 \pm 0.50$  \\
                       & Keck (10 m)    & HIRES  & 2 & $4180 \pm 70$  &        -        &         -       & $0.66 \pm 0.10$ & $0.10 \pm 0.09$ & - &        -        \\
 \cmidrule{1-11}
 \multirow{2}{*}{1693} & FLWO (1.5 m)   & TRES   & 2 &       -        &        -        &           -        &         -          & - & - &            -                 \\ 
                       & Keck (10 m)    & HIRES  & 1 & $3466 \pm 70$  &        -        &        -        & $0.44 \pm 0.10$ & $0.03 \pm 0.09$  &-&         -       \\ \cmidrule{1-11}
 \multirow{2}{*}{1860} & NOT (2.56 m)   & FIES   & 1 & $5780 \pm 50$  & $4.54 \pm 0.10$ &         -        &          -       &-& $-0.09 \pm 0.08$ & $11.10 \pm 0.50$  \\
                       & Keck (10 m)    & HIRES  & 1 & $5724 \pm 100$ & $4.61 \pm 0.10$ & $0.99 \pm 0.03$ & $0.94 \pm 0.02$ & $0.04 \pm 0.06$  &-& $10.37 \pm 1.00$  \\\cmidrule{1-11}
 \multirow{1}{*}{2260} & Keck (10 m)    & HIRES  & 1 & $5534 \pm 100$ & $4.62 \pm 0.10$ & $0.99 \pm 0.04$ & $0.94 \pm 0.05$ & $0.22 \pm 0.06$  &-& $5.05 \pm 1.00$  \\ \cmidrule{1-11}
 \multirow{2}{*}{2290} & FLWO (1.5 m)   & TRES   & 2 &       -        &       -         &         -        &        -        &-&       -         &        -         \\
                       & Keck (10 m)    & HIRES  & 1 & $3813 \pm 70$ &     -     &     -     & $0.57 \pm 0.10$ & $-0.03 \pm 0.09$  &   -         &         -       \\ \cmidrule{1-11}
 \multirow{2}{*}{2411} & SMARTS (1.5 m) & CHIRON & 3 &       -        &       -         &         -        &        -        &-&       -         &        -         \\
                       & FLWO (1.5 m)   & TRES   & 2 &       -        &       -         &         -        &        -        &-&       -         &        -         \\ \cmidrule{1-11}
 \multirow{2}{*}{2427} & SMARTS (1.5 m) & CHIRON & 2 &       -        &       -         &         -        &        -        &-&       -         &        -         \\
                       & FLWO (1.5 m)   & TRES   & 2 &       -        &       -         &         -        &        -        &-&       -         &        -         \\
 \enddata
\tablecomments{Spectrum-derived parameters for each TOI. Entries that list no stellar parameters correspond to stars too cool to have parameters extracted using data collected with the specified instrument. More details on how these parameters were derived are in Section \ref{sec: reconspec}.}
\end{deluxetable*}

\subsubsection{FLWO/TRES and NOT/FIES}\label{TRESandFIES}

We obtained reconnaissance spectra of TOI-544, TOI-1242, TOI-1263, TOI-1411, TOI-1693, TOI-2290, TOI-2411, and TOI-2427 using the Tillinghast Reflector Echelle Spectrograph \citep[TRES;][]{TRES} on the 1.5 m Tillinghast Reflector at the Fred L. Whipple Observatory (FLWO) on Mt. Hopkins, AZ. We also obtained reconnaissance spectra of TOI-1263 and TOI-1860 using the high-resolution FIbre-fed Echelle Spectrograph \citep[FIES;][]{FIES} at the 2.56 m Nordic Optical Telescope (NOT) on La Palma, Spain. We analyzed the TRES and FIES spectra in order to rule out spectroscopic binaries and to confirm that the assumed luminosity classes were correct.

The TRES and FIES reconnaissance spectroscopic observations were analyzed using the Stellar Parameter Classification tool \citep[SPC;][]{SPC}. In brief, SPC uses a correlation analysis of the observed spectra against a library of synthetic spectra calculated using Kurucz model atmospheres \citep{Kurucz}. SPC fits for the $T_{\rm eff}$, $\log g$, [M/H], and projected rotational velocity ($v \sin i$) that give the highest peak correlation value using a multidimensional fit. We ran SPC with priors from the Yonsei-Yale isochrones on the fit \citep{Yonsei-Yale}. The library of calculated spectra used by SPC covers the following ranges: 3500 K $< T_{\rm{eff}} < $ 9750 K, 0.0 $< \log g <$ 5.0 (cgs), $-2.5 < \rm{[m/H]} < +0.5$, and 0 km s$^{-1}$ $< \textit{v} \: \rm{sin} \: \textit{i} <$ 200 km s$^{-1}$ \citep{SPC}. SPC is optimized for slow-rotating Solar type stars. Because it was not designed to classify cool stars ($T_{\rm{eff}} \lesssim 4000$ K), for TOI-544, TOI-1693, TOI-2290, TOI-2411, and TOI-2427 we used empirical relations in order to estimate the stellar parameters (see Section \ref{sec: adopted_params} for more information).

\subsubsection{SMARTS/CHIRON}

We obtained reconnaissance spectra of TOI-500, TOI-539, TOI-544, TOI-731, TOI-833, TOI-1075, TOI-2411, and TOI-2427 using the CHIRON spectrograph on the 1.5 m SMARTS telescope, located at Cerro Tololo Inter-American Observatory (CTIO), Chile \citep{CHIRON}. The spectra were analyzed using a machine learning procedure based on $\sim 10,000$\ TRES spectra classified by SPC and interpolated via a gradient boosting regressor that provides estimates of $T_{\rm eff}$, $\log g$, [M/H], and $v \sin i$. These classifications therefore suffer the same limitations as SPC for the coolest stars, so we estimate parameters for TOI-544, TOI-731, TOI-833, TOI-1075, TOI-2411, and TOI-2427 using the same empirical relations described in Section \ref{sec: adopted_params}. The spectra for all four TOIs have cross-correlation profiles indicative of a single star and no significant RV variations.

\subsubsection{Keck/HIRES}

\begin{deluxetable*}{cccccccccccccccc}
\tabletypesize{\footnotesize}
\tablewidth{\columnwidth}
 \tablecaption{Elemental Abundances Derived with \texttt{KeckSpec} \label{tab:keckspec}}
 \tablehead{ 
 \colhead{TOI} & \colhead{[C/H]} & \colhead{[N/H]} & \colhead{[O/H]} & \colhead{[Na/H]} & \colhead{[Mg/H]} & \colhead{[Al/H]} & \colhead{[Si/H]} & \colhead{[Ca/H]} & \colhead{[Ti/H]} & \colhead{[V/H]} & \colhead{[Cr/H]} & \colhead{[Mn/H]} & \colhead{[Fe/H]} & \colhead{[Ni/H]} & \colhead{[Y/H]}
 }
 \startdata 
1860 &  0.04 &  -0.05 &  0.15  &  -0.05 &  -0.03 &  -0.06  &   0.00 &   0.14  &  0.09  &  0.09  &  0.09  &  0.02  &  0.09  &  -0.02  & 0.17 \\ 
2260 &  0.01 & -0.03 &  0.07  &   0.02 &   0.01 &   0.02  &   0.08 &   0.20  &  0.08  &  0.08  &  0.16  & 0.12  &  0.16  &  0.07  &  0.26 \\ \cmidrule{1-16}
Error & 0.07 & 0.09 & 0.09 & 0.07 & 0.04 & 0.08 & 0.06 & 0.07 & 0.05 & 0.07 & 0.05 & 0.07 & 0.05 & 0.05 & 0.09 
 \enddata
\tablecomments{The bottom row contains the systematic uncertainty for each abundance.}
\end{deluxetable*}

We obtained reconnaissance spectra of TOI-1242, TOI-1411, TOI-1693, TOI-1860, TOI-2260, and TOI-2290 using the High Resolution Echelle Spectrometer (HIRES) \citep{vogt1994society} mounted on the Keck-I 10 m telescope on Maunakea. Our HIRES spectra were analyzed to rule out double-lined spectroscopic binaries and confirm that the stars are not giants. To do the former, we used ReaMatch \citep{kolbl2014detection}, which can identify double-line spectroscopic binaries with contamination ratios as small as 1$\%$, to constrain the presence of unresolved binary stars near each TOI. To do the latter, we classified each star using SpecMatch Synthetic \citep{2015PhDT........82P} and SpecMatch Empirical \citep{2017ApJ...836...77Y}. SpecMatch Synthetic classifies stars by searching a multidimensional grid of synthetic spectra for that which best matches the observed spectrum. SpecMatch Empirical works similarly but instead compares the observed spectrum to a library of spectra of well-characterized stars. The former provides estimates for $T_{\rm eff}$, $\log g$, $M_\star$, $R_\star$, [Fe/H], and $v \sin i$, while the latter provides estimates for $T_{\rm eff}$, $R_\star$, and [Fe/H]. Because SpecMatch Empirical outperforms SpecMatch Synthetic for cooler stars, we adopt the SpecMatch Empirical results for stars that SpecMatch Empirical determines to have $T_{\rm eff} < 4700$ K and we adopt the SpecMatch Synthetic results for stars that SpecMatch Empirical determines to have $T_{\rm eff} > 4700$ K.

In addition, we estimated the activity levels of targets observed with HIRES by calculating their $\log R^\prime _{\rm H K}$ values \citep{linsky1979stellar}. In general, stars with higher $\log R^\prime _{\rm H K}$ values are younger and more active. Rotationally modulated starspots on active stars introduce more scatter in RV observations, making planet mass measurement more difficult \citep{hillenbrand2014empirical}. This quantity is therefore useful for planning future planet characterization efforts.

Lastly, we measured fifteen elemental abundances for two stars (TOI-1860 and TOI-2260) using the \texttt{KeckSpec} algorithm \citep{rice2020stellar} on our high-S/N HIRES spectra. This algorithm is able to reliably measure abundances for stars with $T_{\rm eff} > 4700$ K. The spectra were reduced, extracted, and calibrated following the standard approach of the California Planet Search consortium \citep{howard2010california}.  We then interpolated the spectra onto the wavelength grid required for \texttt{KeckSpec} before feeding them to the algorithm. The resulting abundances are shown in Table \ref{tab:keckspec}. Because elemental abundances are believed to influence the compositions of planet interior and atmospheres \citep[e.g.,][]{bond2010compositional, konopacky2013detection, moriarty2014chemistry}, the quantities may be useful when characterizing these planets and their systems in the future.

\subsection{Time-Series Photometry}\label{sec: SG1}

To determine whether or not the signal observed by TESS is on the presumed target star and to help eliminate FPs from blends, we compile a set of observations collected by members of TFOP Sub Group 1 (SG1). These follow-up observations were scheduled using the {\tt TESS Transit Finder}, which is a customized version of the {\tt Tapir} software package \citep{Jensen:2013}. A summary of these observations is given in Table \ref{table:timeseries} and details about the facilities used are given in Table \ref{table:observatories}. 

We search for transits around the target stars in our observations using the Bayesian Information \citep{schwarz1978estimating}, considering a transit detected if a transit model is preferred over a flat line. For several of our TOIs, transits were verified on-target using these observations. These cases are further described below. We incorporate these data into the transit fits described in Section \ref{sec: transit_fits} to obtain tighter constraints on the ephemerides of the planet candidates.

\subsubsection{LCO 1.0 m / Sinistro}

We observed full transits of TOI-206.01, TOI-1075.01, TOI-1442.01, TOI-1693.01, TOI-2411.01, and TOI-2427.01 using the Sinistro cameras on the Las Cumbres Observatory (LCO) 1.0 m telescopes. Images were calibrated by the standard LCOGT BANZAI pipeline \citep{McCully:2018} and the photometric data were extracted using the {\tt AstroImageJ} ({\tt AIJ}) software package \citep{Collins:2017}. \citep{2013PASP..125.1031B}. 

Transits of TOI-206.01 were observed with a $i'$ filter on UT 2018-11-23, 2018-12-01, and 2018-12-09 and were found to have a depth of $\sim 1.0 - 1.5$ ppt. Transit of TOI-1075.01 were observed with a $z_s$ filter on UT 2019-08-26, 2019-09-23, 2019-09-24, and 2019-09-26 and were found to have a depth of $\sim 0.5 - 1.0$ ppt. Transit of TOI-1442.01 were observed with a $i'$ filter on UT 2020-08-14, 2020-09-26, and 2020-10-21 and were found to have a depth of $\sim 1.0 - 2.0$ ppt. Transits of TOI-1693.01 were observed with a $z_s$ filter on UT 2020-02-14 and 2020-10-11 and were found to have a depth of $\sim 0.5 - 1.0$ ppt. Transits of TOI-2441.01 were observed with a $i'$ filter on UT 2021-07-10, 2021-07-25, 2021-08-27, 2021-08-29, 2021-08-30, and 2021-09-09 and were found to have a depth of $\sim 0.25 - 0.75$ ppt. Transits of TOI-2427.01 were observed with a $z_s$ filter on UT 2021-08-14 and 2021-08-17 and were found to have a depth of $\sim 0.25 - 0.75$ ppt. The data for each of these TOIs can be seen in Figures \ref{fig: 206-joint} -- \ref{fig: 2427-joint}.

\subsubsection{MEarth-South}

We observed full transits of TOI-1075.01 on UT 2019-09-22 and 2019-09-28 using the MEarth-South telescope array at the Cerro Tololo Inter-American Observatory \citep{2008PASP..120..317N, 2015csss...18..767I}. The observations were collected with an RG715 filter and were found to have a transit depth of $\sim 0.5 - 1.0$ ppt. The data can be seen in Figure \ref{fig: 1075-joint}.

\subsubsection{OMM 1.6 m/PESTO}

We observed a full transit of TOI-1442.01 on UT 2020-02-09 using the PESTO camera installed at the 1.6\,m Observatoire du Mont-Mégantic (OMM), Canada. PESTO is equipped with a $1024\times 1024$ EMCCD detector with a scale of $0\farcs466$ per pixel, providing a FOV of $7\farcm95\times7\farcm95$. The observations were collected with a $i'$ filter and with a 30\,s exposure time. Image calibrations, including bias subtraction and flat field correction, and differential aperture photometry were performed with \texttt{AstroImageJ} \citep{Collins:2017}. The events were observed with a $i'$ filter and were found to have a transit depth of $\sim 1$ ppt. The data can be seen in Figure \ref{fig: 1442-joint}.

\subsubsection{TRAPPIST-South}

We observed two full transits of TOI-2445.01 using the TRAPPIST-South telescope \citep{Jehin2011,Gillon2011,Barkaoui_2018_AJ} on UT 2021-01-08 and 2021-02-14. TRAPPIST-South is a 60 cm robotic telescope installed at  La Silla observatory in Chile since 2010, and it is equipped with a thermoelectrically 2Kx2K FLI ProLine PL3041-BB CCD camera with a FOV of 22'x22' and a pixel scale of 0.65". Data calibration and photometric measurements were performed using the \textit{PROSE}\footnote{\url{https://github.com/lgrcia/prose}} pipeline \citep{garcia2021}. Both events were observed in the $I+z$ filter and were found to have a transit depth of $\sim 2.5$ ppt. The data can be seen in Figure \ref{fig: 2445-joint}.

\subsubsection{NAOJ 188 cm/MuSCAT, TCS/MuSCAT2, and LCO 2.0 m/MuSCAT 3}

We observed transits of TOI-1442.01, TOI-1693.01, and TOI-2445.01 using the MuSCAT, MuSCAT2, and MuSCAT3 instruments \citep{2015JATIS...1d5001N, 2019JATIS...5a5001N, 2020SPIE11447E..5KN}, which collect simultaneous observations using several filters. We observed full transits of TOI-1442.01 on UT 2021-05-21, 2021-06-06, and 2021-06-17 using MuSCAT3 on the LCO 2.0 m telescope at Haleakala Observatory. Observations were collected with $g'$, $r'$, $i'$, and $z_s$ filters and measured a transit depth of $\sim 1.0-2.0$ ppt. We observed a full transit TOI-1693.01 on UT 2020-09-18 using MuSCAT2 on the Telescopio Carlos S\'{a}nchez (TCS) at Teide Observatory. Observations were collected with $g'$, $i'$, and $z_s$ filters and measured a transit depth of $\sim 0.5-1.0$ ppt. We observed a full transit of TOI-2445.01 on UT 2021-02-07 using MuSCAT on the National Astronomical Observatory of Japan (NAOJ) 188 cm telescope. Observations were collected with $g'$, $r'$, and $z_s$ filters and measured a transit depth of $\sim 1.0-5.0$ ppt. These data can be seen in Figure \ref{fig: 1442-joint}, Figure \ref{fig: 1693-joint}, and Figure \ref{fig: 2445-joint}.

\startlongtable
\begin{deluxetable}{ccccc}
\tabletypesize{\footnotesize}
\tablewidth{\columnwidth}
 \tablecaption{Summary of Time-Series Photometry Follow-Up \label{tab:timeseries}}
 \tablehead{ 
 \colhead{TOI} & \colhead{TIC ID} & \colhead{Telescope} & \colhead{Date (UT)} & \colhead{Filter(s)}
 }
\startdata 
\multirow{8}{*}{206.01} & \multirow{8}{*}{55650590}   &  LCO 1.0 m  &  2018-11-19  &  $r^\prime$ \\
                           &                          &  SLR2    &  2018-11-22  &  $V$ \\
                           &                          &  LCO 1.0 m  &  2018-11-23  &  $i^\prime$ \\
                           &                          &  CKD700  &  2018-11-30  &  $r^\prime$ \\
                           &                          &  LCO 1.0 m  &  2018-12-01  &  $r^\prime i^\prime$ \\
                           &                          &  LCO 1.0 m  &  2018-12-02  &  $i^\prime$ \\
                           &                          &  LCO 1.0 m  &  2018-12-06  &  $r^\prime$ \\
                           &                          &  LCO 1.0 m  &  2018-12-09  &  $i^\prime$ \\
\cmidrule{1-5}
\multirow{5}{*}{500.01} & \multirow{5}{*}{134200185}  &  LCO 1.0 m  &  2019-03-15  &  $r^\prime$ \\
                           &                          &  TRAPPIST-S.  &  2019-03-24  &  $B$ \\
                           &                          &  LCO 0.4 m  &  2019-03-30  &  $i^\prime$ \\
                           &                          &  PEST  &  2019-03-30  &  $R_c$ \\
                           &                          &  LCO 1.0 m  &  2019-05-02  &  $z_s$ \\
\cmidrule{1-5}
\multirow{5}{*}{539.01} & \multirow{5}{*}{238004786} &  PEST  &  2019-03-29  &  $R_c$ \\
                           &                          &  MKO CDK700  &  2019-03-31  &  $r^\prime$ \\
                           &                          &  LCO 1.0 m  &  2019-04-06  &  $i^\prime$ \\
                           &                          &  LCO 1.0 m  &  2019-04-08  &  $z_s$ \\
                           &                          &  LCO 1.0 m  &  2019-04-17  &  $i^\prime$ \\
\cmidrule{1-5}
\multirow{2}{*}{544.01} & \multirow{2}{*}{50618703}   &  LCO 1.0 m  &  2019-09-20  &  $z_s$ \\
                           &                          &  TCS &  2019-10-13  &  $g^\prime r^\prime i^\prime$ \\
\cmidrule{1-5}
\multirow{4}{*}{731.01} & \multirow{4}{*}{34068865} &  LCO 1.0 m  &  2019-06-10  &  $V$ \\
                           &                          &  MKO CDK700  &  2019-06-11  &  $i^\prime$ \\
                           &                          &  PEST  &  2020-01-05  &  $R_c$ \\
                           &                          &  LCO 1.0 m  &  2020-05-12  &  $z_s$ \\
\cmidrule{1-5}
\multirow{3}{*}{833.01} & \multirow{3}{*}{362249359}  &  LCO 1.0 m  &  2020-03-28  &  $z_s$ \\
                           &                          &  LCO 1.0 m  &  2020-05-14  &  $z_s$ \\
                           &                          &  MKO CDK700   &  2020-05-15  &  $i^\prime$ \\
\cmidrule{1-5}
\multirow{7}{*}{1075.01} & \multirow{7}{*}{351601843} &  LCO 1.0 m  &  2019-08-25  &  $z_s$ \\
                           &                          &  LCO 1.0 m  &  2019-08-26  &  $z_s$ \\
                           &                          &  MEarth-S   &  2019-09-22  &  RG715 \\
                           &                          &  LCO 1.0 m  &  2019-09-23  &  $z_s$ \\
                           &                          &  LCO 1.0 m  &  2019-09-24  &  $z_s$ \\
                           &                          &  LCO 1.0 m  &  2019-09-26  &  $z_s$ \\
                           &                          &  MEarth-S   &  2019-09-28  &  RG715 \\
\cmidrule{1-5}
\multirow{5}{*}{1242.01} & \multirow{5}{*}{198212955} &  TCS  &  2020-01-27  &  $g^\prime r^\prime i^\prime z_s$ \\
                           &                          &  TCS  &  2020-02-01  &  $g^\prime r^\prime i^\prime z_s$ \\
                           &                          &  TCS  &  2020-02-09  &  $g^\prime r^\prime i^\prime z_s$ \\
                           &                          &  ULMT  &  2020-05-18  &  $i^\prime$ \\
                           &                          &  TCS  &  2020-06-09  &  $g^\prime r^\prime i^\prime z_s$ \\
\cmidrule{1-5}
\multirow{2}{*}{1263.01} & \multirow{2}{*}{406672232} &  LCO 1.0 m  &  2020-06-15  &  $z_s$ \\
                           &                          &  LCO 1.0 m  &  2020-07-28  &  $z_s$ \\
\tablebreak
\multirow{6}{*}{1411.01} & \multirow{6}{*}{116483514} &  LCO 1.0 m  &  2020-02-28  &  $i^\prime$ \\
                           &                          &  DSW CDK500  &  2020-04-16  &  $r^\prime$ \\
                           &                          &  TCS  &  2020-04-21  &  $g^\prime r^\prime i^\prime z_s$ \\
                           &                          &  LCO 1.0 m  &  2020-04-29  &  $r^\prime$ \\
                           &                          &  ULMT  &  2020-05-02  &  $i^\prime$ \\
                           &                          &  TCS  &  2020-05-10  &  $g^\prime r^\prime i^\prime z_s$ \\
\cmidrule{1-5}
\multirow{4}{*}{1442.01} & \multirow{4}{*}{235683377} &  OMM 1.6 m  &  2020-02-09  &  $i^\prime$ \\
                           &                          &  LCO 1.0 m  &  2020-08-14  &  $i^\prime$ \\
                           &                          &  LCO 1.0 m  &  2020-08-30  &  $I_c$ \\
                           &                          &  LCO 1.0 m  &  2020-09-26  &  $i^\prime$ \\
                           &                          &  LCO 2.0 m  &  2021-05-21  &  $g^\prime r^\prime i^\prime z_s$ \\
                           &                          &  LCO 2.0 m  &  2021-06-06  &  $g^\prime r^\prime i^\prime z_s$ \\
                           &                          &  LCO 2.0 m  &  2021-06-17  &  $g^\prime r^\prime i^\prime z_s$ \\
\cmidrule{1-5}
\multirow{3}{*}{1693.01} & \multirow{3}{*}{353475866} &  LCO 1.0 m  &  2020-02-14  &  $z_s$ \\
                           &                          &  LCO 1.0 m  &  2020-10-11  &  $z_s$ \\
                           &                          &  TCS  &  2020-09-18  &  $g^\prime i^\prime z_s$ \\
\cmidrule{1-5}
\multirow{2}{*}{1860.01} & \multirow{2}{*}{202426247} &  Adams   &  2020-06-06  &  $I_c$ \\
                           &                          &  TCS  &  2020-07-20  &  $g^\prime r^\prime i^\prime z_s$ \\
\cmidrule{1-5}
\multirow{2}{*}{2260.01} & \multirow{2}{*}{232568235} &  TRAPPIST-N  &  2020-09-28  &  $z^\prime$ \\
                         &                            &  Adams       &  2021-06-26  &  $I_c$ \\
\cmidrule{1-5}
\multirow{1}{*}{2290.01} & \multirow{1}{*}{321688498} &  LCO 1.0 m  &  2020-10-15  &  $i^\prime$ \\
\cmidrule{1-5}
\multirow{8}{*}{2411.01} & \multirow{8}{*}{10837041} &  MKO CDK700  &  2021-01-13  &  $i^\prime$ \\
                         &                           &  LCO 1.0 m   &  2021-06-19  &  $r^\prime$ \\
                         &                           &  LCO 1.0 m   &  2021-07-10  &  $i^\prime$ \\
                         &                           &  LCO 1.0 m   &  2021-07-25  &  $i^\prime$ \\
                         &                           &  LCO 1.0 m   &  2021-08-27  &  $i^\prime$ \\  
                         &                           &  LCO 1.0 m   &  2021-08-29  &  $i^\prime$ \\
                         &                           &  LCO 1.0 m   &  2021-08-30  &  $i^\prime$ \\
                         &                           &  LCO 1.0 m   &  2021-09-09  &  $i^\prime$ \\
\cmidrule{1-5}
\multirow{5}{*}{2427.01} & \multirow{5}{*}{142937186} &  PEST    &  2021-01-12  &  $R_c$ \\
                         &                            &  LCO 1.0 m  &  2021-01-30  &  $z_s$ \\
                         &                            &  LCO 1.0 m  &  2021-02-22  &  $z_s$ \\
                         &                            &  LCO 1.0 m  &  2021-08-14  &  $z_s$ \\
                         &                            &  LCO 1.0 m  &  2021-08-17  &  $z_s$ \\
\cmidrule{1-5}
\multirow{4}{*}{2445.01} & \multirow{4}{*}{439867639} &  MLO  &  2021-01-10  &  $I_c$ \\
                         &                            &  TRAPPIST-S  &  2021-01-08  &  $I+z$ \\
                         &                            &  TRAPPIST-S  &  2021-01-14  &  $I+z$ \\
                         &                            &  NAOJ 188 cm &  2021-02-07  &  $g^\prime r^\prime z_s$ \\
\enddata
\label{table:timeseries}
\end{deluxetable}
\begin{deluxetable*}{llDDc}
\tablecaption{Facilities used for TFOP SG1 followup\label{table:observatories}}
\tablehead{\colhead{Observatory} & \colhead{Telescope/Instrument} & \twocolhead{Aperture} & \twocolhead{Pixel scale} & \colhead{FOV}\\[-2mm]
 & & \twocolhead{(m)} & \twocolhead{(arcsec)} & \colhead{(arcmin)}
}
\decimals
\startdata
Austin College Adams Observatory (Adams) & - & 0.61 & 0.38 & $26 \times 26$ \\
Cerro Tololo Inter-American Observatory  & MEarth-South & 0.4 & 0.84 & $29 \times 29$ \\
Deep Sky West Remote Observatory (DSW) & DSW CDK500 & 0.5 & 1.09 & $37 \times 37$ \\
Las Cumbres Observatory (LCO) & LCO 0.4 m / SBIG-6303 & 0.4 & 0.57 & $29.2 \times 19.5$ \\
Las Cumbres Observatory (LCO) & LCO 1.0 m / Sinistro & 1.0 & 0.39 & $26.5 \times 26.5$ \\
Haleakala Observatory & LCO 2.0 m / MuSCAT3 & 2.0 & 0.27 & $9.1 \times 9.1$ \\
Maury Lewin Astronomical Observatory (MLO) & - & 0.36 & 0.84 & $23 \times 17$ \\
Mt. Kent Observatory (MKO) & MKO CDK700 & 0.7 & 0.4 & $27 \times 27$ \\
Observatoire du Mont-M\'{e}gantic (OMM) & OMM 1.6 m / PESTO & 1.6 & 0.47 & $7.95 \times 7.95$ \\
National Astronomical Observatory of Japan (NAOJ) & NAOJ 188 cm / MuSCAT & 1.88 & 0.36 & $6.1 \times 6.1$ \\
Ouka\"{i}meden Observatory & TRAPPIST-North & 0.6 & 0.64 & $ 22 \times 22$ \\
South African Astronomical Observatory & SLR2 & 0.5 & 0.37 & $12 \times 12 $ \\
Teide Observatory & Telescopio Carlos S\'{a}nchez (TCS) / MuSCAT2 & 1.52 & 0.44 & $7.4 \times 7.4 $ \\
La Silla Observatory & TRAPPIST-South & 0.6 & 0.64 & $ 22 \times 22$ \\
Mt. Lemmon Observatory & Univ.\ of Louisville Manner Telescope (ULMT) & 0.61 & 0.39 & $26 \times 26$ \\
- & Perth Exoplanet Survey Telescope (PEST) & 0.3 & 1.2 & $31 \times 21$ \\
\enddata
\end{deluxetable*}

\section{Results}\label{sec: results}

Below, we provide a brief summary about each of the planet candidates analyzed in this paper. We begin with details about the target stars, including their brightnesses, distances, and the TESS sectors in which they were observed. In addition, we analyze the available data of each star to search for activity indicators and signs of system youth. Specifically, we apply a Lomb-Scargle periodogram to each individual sector of TESS photometry to constrain the level of starspot variability. We consider the detection of photometric variability to be significant if the maximum peak calculated by the periodogram across all sectors is $>0.5$. When available, we also consider the spectrum-derived $v \sin{i}$ and $\log R'_{\rm HK}$.

Next, we present information gleaned from each step of our vetting process. We also summarize this information in Table \ref{tab: results}. For TFOP SG3 high-resolution imaging observations, we refer to the TOI as ``clear'' if no stars were resolved within the detection limits stated in Figure \ref{fig: imaging}. For TFOP SG2 reconnaissance spectroscopy observations, we refer to the TOI as ``clear'' if the target star is confirmed to be on the main sequence and no evidence of a spectroscopic binary was detected. For TFOP SG1 time-series photometry observations, we identify all stars from Gaia DR2 within 2\farcm5 of the target star that are bright enough to cause the TESS transit detection based on the observed transit depth, the angular distance from the target star, and the difference in magnitude from the target star. For convenience, we refer to these as ``neighbor stars''  in the discussion below, and we describe them as ``cleared'' if our photometric follow-up observations show that they have no transit-like events of the depth that would be necessary to reproduce the TESS event when blended with the central star. 

At the end of each subsection, we decide whether the TOI is validated based on the results of the \tri\ analysis. To forecast the potential of measuring the masses of the planet candidates via precise RVs, we also estimate the semiamplitude ($K_{\rm RV}$) and planet mass ($M_{\rm p}$) of each using the probabilistic planet mass-radius relation given in \cite{chen2016probabilistic} and the adopted stellar masses listed in Table \ref{tab:adopted}. However, we stress that these  estimates are merely illustrative, and should not be quoted as the actual masses and semiamplitudes.

\movetabledown=20mm
\begin{rotatetable*}
\begin{deluxetable*}{cllllccc}\label{tab: results}
\tablecaption{Vetting Results}
 \tablehead{
 \colhead{TOI} & \colhead{High-resolution Imaging} & \colhead{Recon Spectroscopy} & \colhead{Time-series Photometry} & \colhead{\texttt{DAVE} Results} & \colhead{FPP} & \colhead{NFPP} & \colhead{Validated}
 }
 \startdata 
 \multirow{1}{*}{206.01} & Clear & No data & Verified on-target      & Potential secondary eclipse & $< 0.01$ & $< 0.001$ & Y \\ \cmidrule{1-8}
 \multirow{1}{*}{500.01} & Clear & Clear   & All neighbors cleared  & Clear but unreliable centroid analysis     & $< 0.01$ & $< 0.001$ & Y \\ \cmidrule{1-8}
 \multirow{1}{*}{539.01} & Clear & Clear   & 1 neighbor not cleared & Clear but unreliable centroid analysis     & $> 0.01$ & $< 0.001$ & N \\ \cmidrule{1-8}
 \multirow{1}{*}{544.01} & Clear & Clear   & 2 neighbors not cleared &  Clear   & $< 0.01$ & $< 0.001$ & Y \\ \cmidrule{1-8}
 \multirow{1}{*}{731.01} & Clear & Clear & 1 neighbor not cleared & Clear but unreliable centroid analysis     & $> 0.01$ & $< 0.001$ & N  \\ \cmidrule{1-8}
 \multirow{1}{*}{833.01} & Clear & Clear   & 1 neighbor not cleared & Potential centroid offset & $< 0.01$ & $< 0.001$ & Y \\ \cmidrule{1-8}
 \multirow{1}{*}{1075.01} & Clear & Clear   & Verified on-target  &  Clear             & $< 0.01$ & $< 0.001$ & Y \\ \cmidrule{1-8}
 \multirow{1}{*}{1242.01} & $4\farcs3$ companion detected & Clear & 1 neighbor not cleared & Potential centroid offset  & $< 0.01$ & $> 0.001$ & N \\ \cmidrule{1-8}
 \multirow{1}{*}{1263.01} & $2\farcs6$ companion detected & Clear & 2 neighbors not cleared & Different odd-even transit depths & $> 0.01$ & $> 0.001$ & N \\ \cmidrule{1-8}
 \multirow{1}{*}{1411.01} & Clear & Clear & All neighbors cleared  & Clear  & $< 0.01$ & $< 0.001$ & Y  \\ \cmidrule{1-8}
 \multirow{1}{*}{1442.01} & Clear & No data & Verified on-target & Clear but unreliable centroid analysis    & $< 0.01$ & $<0.001$ & Y \\ \cmidrule{1-8}
 \multirow{1}{*}{1693.01} & Clear & Clear & Verified on-target     & Clear & $< 0.01$ & $< 0.001$ & Y \\ \cmidrule{1-8}
 \multirow{1}{*}{1860.01} & Clear & Clear & 1 neighbor not cleared & No results & $< 0.01$ & $< 0.001$ & Y \\ \cmidrule{1-8}
 \multirow{1}{*}{2260.01} & Clear & Clear & All neighbors cleared &  Clear          & $< 0.01$ & $< 0.001$ & Y \\ \cmidrule{1-8}
 \multirow{1}{*}{2290.01} & Clear & Clear & All neighbors cleared &  Potential centroid offset & $> 0.01$ & $<0.001$ & N \\ \cmidrule{1-8}
 \multirow{1}{*}{2411.01} & Clear & Clear & Verified on-target &  No results        & $< 0.01$ & $< 0.001$ & Y \\ \cmidrule{1-8}
 \multirow{1}{*}{2427.01} & Clear & Clear & Verified on-target      &  Potential centroid offset  & $< 0.01$ & $< 0.001$ & Y \\ \cmidrule{1-8}
 \multirow{1}{*}{2445.01} & Clear & No data & Verified on-target & Clear but unreliable centroid analysis & $< 0.01$ & $< 0.001$ & Y 
 \enddata
 \vspace{-25pt}
\end{deluxetable*}
\end{rotatetable*}

\subsection{TOI-206.01}

TOI-206.01 is a $1.30 \pm 0.05 \, R_\oplus$ planet candidate with a 0.74 day orbital period orbiting an M dwarf (TIC 55650590) that is 47.7 pc away and has a $V$ magnitude of 14.94. A Lomb-Scargle periodogram of the photometry from each TESS sector finds a maximum peak of 0.04, indicating that the star is quiet. TOI-206 has been observed in 26 TESS sectors (1--13 and 27--39).

\begin{figure}
    \centering
    \includegraphics[width=0.45\textwidth]{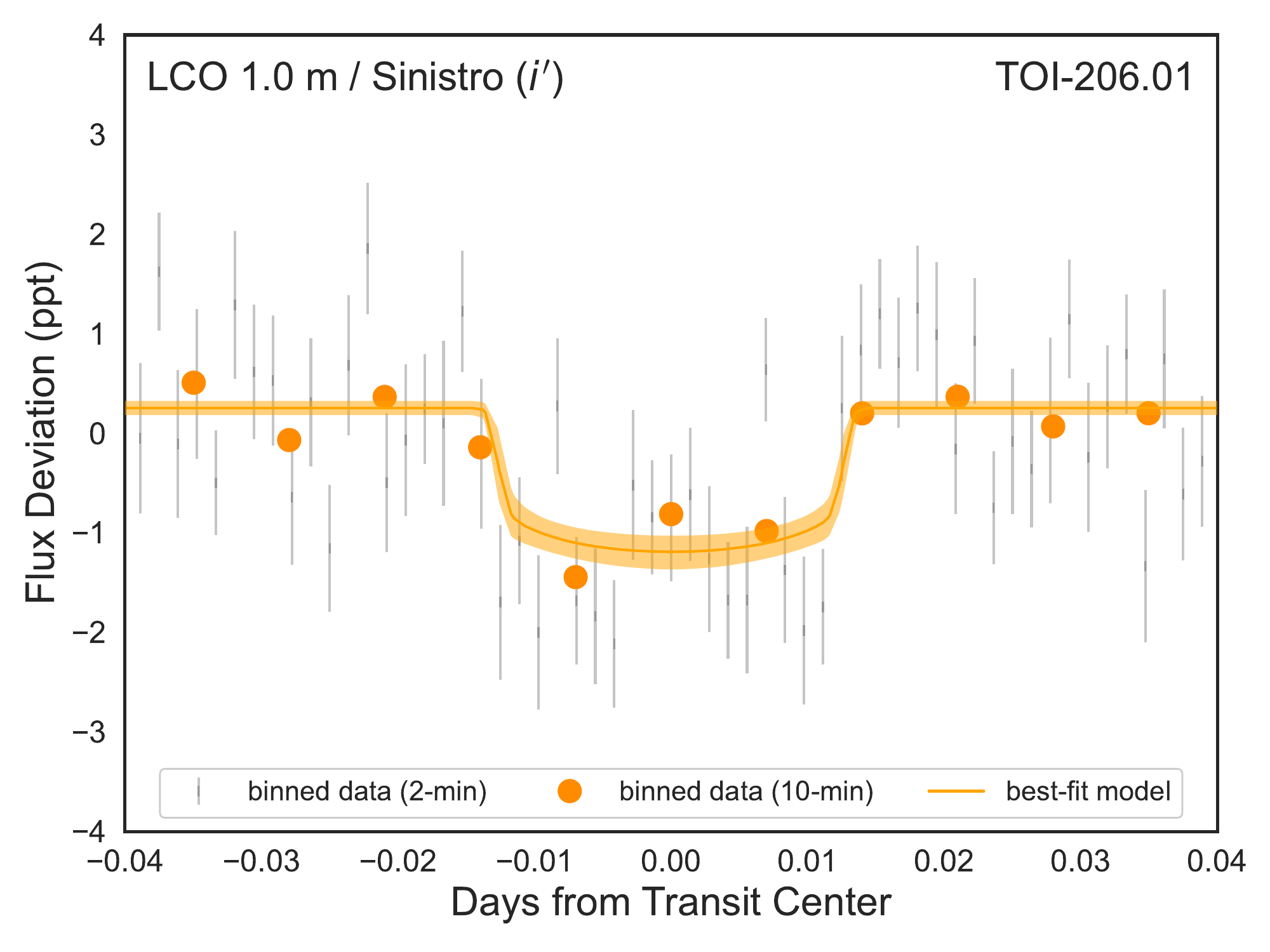}
    \caption{Phase-folded ground-based data and best-fit model of the transit of TOI-206.01. The data is detrended with a linear model and 3$\sigma$ outliers are removed.}
    \label{fig: 206-joint}
\end{figure}

Follow-up observations have found no evidence of this TOI being an FP, although no spectroscopic observations have been collected. Time-series photometric follow-up has made several detections of the transit of TOI-206.01 on TIC 55650590 (shown in Figure \ref{fig: 206-joint}).

The \dave\ analysis of this TOI detects a potential secondary eclipse in the TESS light curve, which could indicate that the transit is due to an eclipsing binary. Because follow-up observations do not detect a companion star that could dilute the radius of the transiting object and because the transit was detected on-target, this eclipsing binary would need to have a grazing transit. The morphology of the transit shown in Figure \ref{fig: transits} is inconsistent with that of a grazing eclipsing binary, meaning the feature detected in the TESS photometry is unlikely to be an actual secondary eclipse. The SPOC data validation report for this TOI reports no significant centroid offset.

The \tri\ analysis of this TOI finds ${\rm FPP} = (2.02 \pm 1.48) \times 10^{-5}$. Because all neighboring stars have been cleared, \tri\ finds ${\rm NFPP} = 0.0$. This FPP is sufficiently low to consider the planet validated. We hereafter refer to this planet as TOI-206 b. 

We estimate the semiamplitude of the RV signal for this planet to be $K_{\rm RV} = 3.1^{+2.0}_{-1.0}$ m/s, corresponding to $M_{\rm p} = 2.2^{+1.4}_{-0.7} \, M_\oplus$.

\subsection{TOI-500.01}

TOI-500.01 is a $1.16 \pm 0.12 \, R_\oplus$ planet candidate with a 0.55 day orbital period orbiting a K dwarf (TIC 134200185) that is 47.4 pc away and has a $V$ magnitude of 10.54. A Lomb-Scargle periodogram of the photometry from each TESS sector finds a maximum peak of 0.007, indicating that the star is quiet. This is corroborated by the low $v \sin{i}$ extracted from our CHIRON spectra. TOI-500 has been observed in 6 TESS sectors (6--8 and 33--35).

Follow-up observations have found no evidence of this TOI being an FP. Time-series photometric follow-up of this TOI has cleared all neighboring stars as origins of the transit but has not yet detected the 0.23 ppt event seen in the TESS data around the target star.

\dave\ finds no strong indicators that the candidate is an FP. We note, though, that the photocenter offset analysis performed by \dave\ suffers from low S/N and poor quality in most of the per-transit difference images. As a result, there is large scatter in the measured photocenters for each individual transit, making it difficult for \dave\ to detect a significant photocenter offset. The SPOC data validation report, however, does detect significant centroid offsets in sectors 8, 34, and 35. No offsets were detected in sectors 7 or 33 by SPOC, and no data validation report was generated by the SPOC pipeline for sector 6. Given that all neighboring stars have been cleared from being nearby eclipsing binaries, these offsets are unlikely to be caused by a FP originating from a nearby star. 

The \tri\ analysis of this TOI finds ${\rm FPP} = (7.12 \pm 1.13) \times 10^{-3}$. Because all neighboring stars have been cleared, \tri\ finds ${\rm NFPP} = 0.0$. This FPP is sufficiently low to consider the planet validated.  We hereafter refer to this planet as TOI-500 b. 

We estimate the semiamplitude of the RV signal for this planet to be $K_{\rm RV} = 1.4^{+1.1}_{-0.7}$ m/s, corresponding to $M_{\rm p} = 1.6^{+1.3}_{-0.7} \, M_\oplus$.

\subsection{TOI-539.01}

TOI-539.01 is a $1.25 \pm 0.10 \, R_\oplus$ planet candidate with a 0.31 day orbital period orbiting a K dwarf (TIC 238004786) that is 108.4 pc away and has a $V$ magnitude of 11.73. A Lomb-Scargle periodogram of the photometry from each TESS sector finds a maximum peak of 0.07, indicating that the star is quiet. This is corroborated by the low $v \sin{i}$ extracted from our CHIRON spectrum. TOI-539 has been observed in 11 TESS sectors (2, 6, 8, 9, 12, 29, 32--35, and 39).

Follow-up observations have found no evidence of this TOI being an FP. Time-series photometric follow-up of this TOI has cleared all neighboring stars as origins of the transit except for TIC 767067264, which is 7\farcs2 west and 7.9 magnitudes fainter in the Gaia $G_{\rm Rp}$ band. This nearby star appears not to show an event of the necessary depth but is not cleared at high confidence. The 0.31 ppt event seen in the TESS data has not been detected around the target star.

The \dave\ analysis of this TOI finds no strong indicators that the candidate is an FP. However, like TOI-500 b, the per-transit difference images used by \dave\ have very low S/N and the measured photocenters are unreliable. The SPOC data validation report for this TOI reports no significant centroid offset.

The \tri\ analysis of this TOI finds ${\rm FPP} = (3.98 \pm 0.03) \times 10^{-2}$ and ${\rm NFPP} = (7.76 \pm 0.26) \times 10^{-22}$. This $> 1 \%$ FPP comes from the scenario that the TOI is a blended eclipsing binary. While this NFPP indicates that this TOI is unlikely to originate from the nearby star TIC 767067264, the FPP is too high to validate the planet candidate.

Assuming this is a real planet, we estimate the semiamplitude of its RV signal to be $K_{\rm RV} = 1.9^{+1.6}_{-0.7}$ m/s, corresponding to $M_{\rm p} = 1.9^{+1.6}_{-0.7} \, M_\oplus$.

\subsection{TOI-544.01}

TOI-544.01 is a $2.03 \pm 0.10 \, R_\oplus$ planet candidate with a 1.55 day orbital period orbiting a K dwarf (TIC 50618703) that is 41.1 pc away and has a $V$ magnitude of 10.78. A Lomb-Scargle periodogram of the photometry from each TESS sector finds a maximum peak of 0.25, indicating that the star is quiet. This is corroborated by the low $v \sin{i}$ extracted from our TRES spectrum. TOI-544 has been observed in 2 TESS sectors (6 and 32).

Follow-up observations have found no evidence of this TOI being an FP. Time-series photometric follow-up of this TOI has cleared all neighboring stars as origins of the transit except for TIC 713009339 (located 5$\farcs$26 south-southeast and 9.5 magnitudes fainter in the TESS band) and TIC 50618707 (located 9$\farcs$18 east-southeast and 6.9 magnitudes fainter in the TESS band). TIC 713009339 is too faint to be the source of an astrophysical FP, but  TIC 50618707 is not. We would like to note that the former of these nearby stars was detected by Gaia but not by 2MASS, while the latter was detected by 2MASS but not by Gaia. The parallaxes and proper motions of these two stars are unknown, so it is possible that they are the same star observed at two different epochs. If this were the case, the star would have been within the $\sim 10\arcsec \times 10\arcsec$ FOV of the Shane/ShARCS observations obtained on UT 2019-09-13, which reach contrasts of $> 8$ mags in the $K_s$ and $J$ bands. However, no stars other than TIC 50618703 were detected in these observations. If this star (or stars, if they are indeed different sources) are really there, it (or they) would be far too faint to host eclipsing binaries mistakable for the TOI-544.01 transit. Regardless, we consider these two nearby stars in the remaining vetting steps for the sake of completeness.

The \dave\ analysis of this TOI finds no strong indicators that the candidate is an FP. The SPOC data validation report for this TOI reports no significant centroid offset.

The \tri\ analysis of this TOI finds ${\rm FPP} = (8.25 \pm 0.91) \times 10^{-3}$ and ${\rm NFPP} = (1.67 \pm 0.16) \times 10^{-16}$. This FPP and NFPP are sufficiently low to consider the planet validated. We hereafter refer to this planet as TOI-544 b.

We estimate the semiamplitude of the RV signal for this planet to be $K_{\rm RV} = 3.2^{+2.4}_{-1.4}$ m/s, corresponding to $M_{\rm p} = 5.0^{+4.0}_{-2.0} \, M_\oplus$.

\subsection{TOI-731.01}

TOI-731.01 is a $0.59 \pm 0.02 \, R_\oplus$ planet candidate with a 0.32 day orbital period orbiting a high-proper-motion ($\mu_\alpha = -462.5$ mas/yr, $\mu_\delta = -582.8$ mas/yr) M dwarf (TIC 34068865) that is 9.4 pc away and has a $V$ magnitude of 10.15. A Lomb-Scargle periodogram of the photometry from each TESS sector finds a maximum peak of 0.07, indicating that the star is quiet. TOI-731 has been observed in 3 TESS sectors (9, 35, and 36).

Follow-up observations have found no evidence of this TOI being an FP. Time-series photometric follow-up of this TOI has cleared all neighboring stars as transit sources except for TIC 34068883, which is 6.2 magnitudes fainter in Gaia $G_{\rm Rp}$ and was 6\farcs4 southwest at epoch 2020.361.\footnote{This separation is continuing to decrease and will lead to a weak microlensing event with a closest approach of 510 mas in December 2028 \citep{2018AcA....68..183B}. The event will not produce a brightening of more than 0.4 mmag, but is predicted to produce an astrometric shift of 1 mas, possibly detectable by a future astrometric mission.} However, this follow-up has not detected the 0.24 ppt transit around the target star that is seen in the TESS data.

The \dave\ analysis of this TOI finds no strong indicators that the candidate is an FP. Compared to TOI-500, the S/N of the per-transit difference images used by \dave\ is even lower and the measured centroids are unreliable. The SPOC data validation report for this TOI reports no significant centroid offset.

The \tri\ analysis of this TOI finds ${\rm FPP} = (1.89 \pm 0.46) \times 10^{-2}$ and ${\rm NFPP} = (9.21 \pm 1.48) \times 10^{-26}$. This $> 1 \%$ FPP comes from the scenario that the TOI is a blended eclipsing binary. This FPP is too high to consider the planet candidate validated.

Assuming this is a real planet, we estimate the semiamplitude of the RV signal to be $K_{\rm RV} = 0.22^{+0.11}_{-0.07}$ m/s, corresponding to $M_{\rm p} = 0.15^{+0.07}_{-0.04} \, M_\oplus$. 

\subsection{TOI-833.01}

TOI-833.01 is a $1.27 \pm 0.07 \, R_\oplus$ planet candidate with a 1.04 day orbital period orbiting a K dwarf (TIC 362249359) that is 41.7 pc away and has a $V$ magnitude of 11.72. A Lomb-Scargle periodogram of the photometry from each TESS sector finds a maximum peak of 0.20, indicating that the star is quiet. TOI-833 has been observed in 5 TESS sectors (9, 10, 11, 36, and 37).

Follow-up observations have found no evidence of this TOI being an FP. Time-series photometric follow-up of this TOI has made tentative detections of a $\sim$0.8--0.9 ppt transit on two different occasions. The field around this TOI is crowded and it is not clear if the event is on target or due to blending with TIC 847323367 (located 3\farcs1 north and 7.9 magnitudes fainter in the TESS band). 

The \dave\ analysis of this TOI detects a potential centroid offset to the north-east, but found no other indicators that this TOI is an FP. The SPOC data validation report for this TOI reports no significant centroid offset.

The \tri\ analysis of this TOI finds ${\rm FPP} = (2.32 \pm 0.23) \times 10^{-4}$ and ${\rm NFPP} = (3.89 \pm 0.11) \times 10^{-10}$. This FPP and NFPP are sufficiently low to consider the planet validated. We hereafter refer to this planet as TOI-833 b. 
We estimate the semiamplitude of the RV signal for this planet to be $K_{\rm RV} = 1.8^{+1.3}_{-0.5}$ m/s, corresponding to $M_{\rm p} = 2.0^{+1.5}_{-0.6} \, M_\oplus$.

\subsection{TOI-1075.01}

TOI-1075.01 is a $1.72 \pm 0.08 \, R_\oplus$ planet candidate with a 0.60 day orbital period orbiting a K dwarf (TIC 351601843) that is 61.5 pc away and has a $V$ magnitude of 12.62. A Lomb-Scargle periodogram of the photometry from each TESS sector finds a maximum peak of 0.02, indicating that the star is quiet. TOI-1075 has been observed in 2 TESS sectors (13 and 27).

\begin{figure*}
    \centering
    \includegraphics[width=\textwidth]{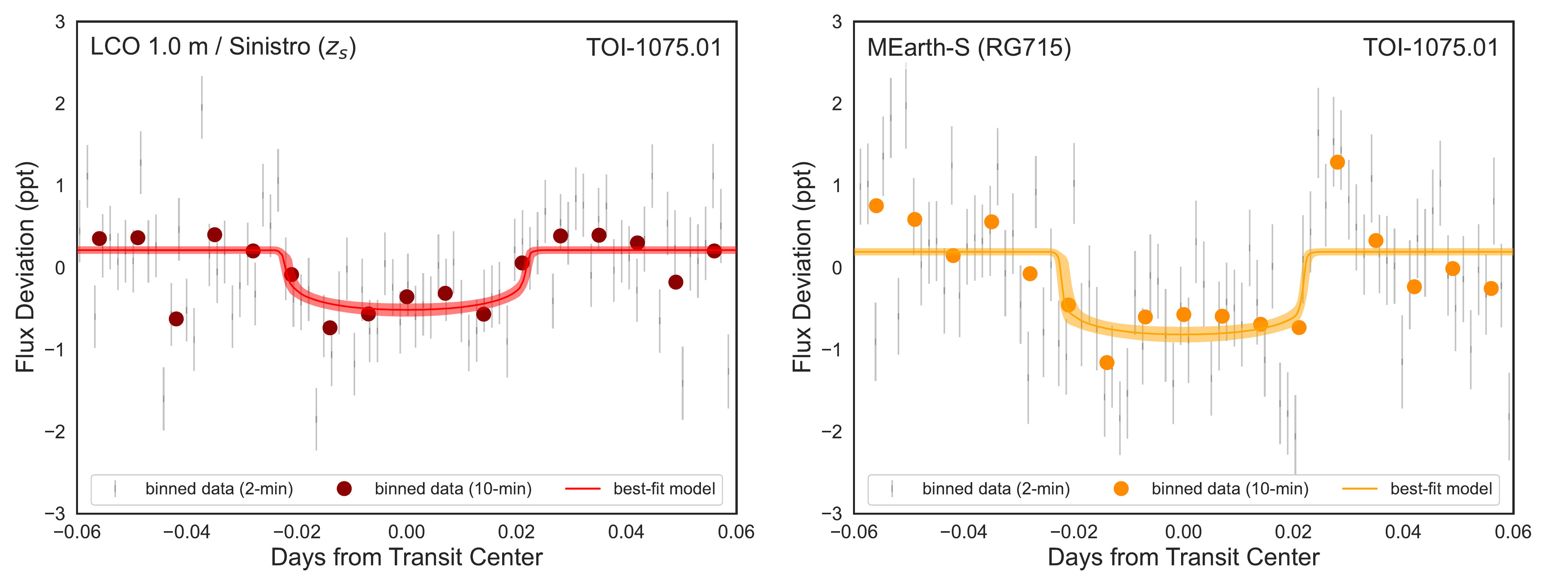}
    \caption{Phase-folded ground-based data and best-fit model of the transit of TOI-1075.01. The data is detrended with a linear model and 3$\sigma$ outliers are removed.}
    \label{fig: 1075-joint}
\end{figure*}

Follow-up observations have found no evidence of this TOI being an FP. Time-series photometric follow-up has has made several detections of the transit of TOI-1075.01 on TIC 351601843 (shown in Figure \ref{fig: 1075-joint}).

The \dave\ analysis of this TOI found no strong indicators that the candidate is an FP. The SPOC data validation report for this TOI reports no significant centroid offset.

The \tri\ analysis of this TOI finds ${\rm FPP} = (1.01 \pm 0.16) \times 10^{-3}$. Because TIC 351601843 has been verified as the host of the transit, \tri\ finds ${\rm NFPP} = 0.0$. This FPP is sufficiently low to consider the planet validated. We hereafter refer to this planet as TOI-1075 b. 

We estimate the semiamplitude of the RV signal for this planet to be $K_{\rm RV} = 4.3^{+2.9}_{-1.5}$ m/s, corresponding to $M_{\rm p} = 4.0^{+2.7}_{-1.4} \, M_\oplus$.

\subsection{TOI-1242.01}

TOI-1242.01 is a $1.65 \pm 0.23 \, R_\oplus$ planet candidate with a 0.38 day orbital period orbiting a K dwarf (TIC 198212955) that is 110 pc away and has a $V$ magnitude of 12.78. A Lomb-Scargle periodogram of the photometry from each TESS sector finds a maximum peak of 0.03, indicating that the star is quiet. This is corroborated by the low $v \sin{i}$ extracted from our TRES spectrum. TOI-1242 has been observed in 15 TESS sectors (14--26, 40, and 41) and is scheduled to be reobserved in another 8 sectors (48--55) between 2022-01-28 and 2022-09-01.

High-resolution imaging of this star detects TIC 198212956, a previously known star that is 4\farcs3 north and 2.6 magnitudes fainter in the TESS band, but finds no other unresolved stars within detection limits. TIC 198212956 is almost certainly bound to TIC 198212955 due to their similar parallaxes and proper motions as reported by Gaia DR2. Spectroscopic observations confirm that the star is on the main sequence and rule out obvious spectroscopic binaries. Time-series photometric follow-up of this TOI has cleared all neighboring stars as origins of the transit except for TIC 198212956. The 0.6 ppt event seen in the TESS data has not been detected around the target star or its companion.

The \dave\ analysis of this TOI detects a potential centroid offset to the north-east, but finds no other indicators that this TOI is an FP. The SPOC data validation report for this TOI reports no significant centroid offset.

The \tri\ analysis of this TOI find ${\rm FPP} = (3.36 \pm 0.17) \times 10^{-2}$ and ${\rm NFPP} = (2.92 \pm 0.16) \times 10^{-2}$. These $> 1 \%$ FPP and NFPP are driven by the uncertainty over whether or not the transit originates from the target star or TIC 198212956. This FPP and NFPP are too high to consider this planet candidate validated.

Assuming this is a real planet around TIC 198212955, we estimate the semiamplitude of the RV signal to be $K_{\rm RV} = 3.7^{+3.0}_{-1.7}$ m/s, corresponding to $M_{\rm p} = 3.7^{+2.9}_{-1.5} \, M_\oplus$.

\subsection{TOI-1263.01}

TOI-1263.01 is a $1.36 \pm 0.16 \, R_\oplus$ planet candidate with a 1.02 day orbital period orbiting a K dwarf (TIC 406672232) that is 46.6 pc away and has a $V$ magnitude of 9.36. A Lomb-Scargle periodogram of the photometry from each TESS sector finds a maximum peak of 0.12, indicating that the star is quiet. This is corroborated by the low $v \sin{i}$ extracted from our TRES and FIES spectra. TOI-1263 has been observed in 3 TESS sectors (14, 15, and 41) and is scheduled to be reobserved in another sector (55) between 2022-08-05 and 2022-09-01.

High-resolution imaging of this star detects TIC 1943945558, a previously known star that is 2\farcs6 southeast and 3.6 magnitudes fainter in the TESS band, but finds no other unresolved stars within detection limits. TIC 1943945558 is almost certainly bound to TIC 406672232 due to their similar parallaxes and proper motions as reported by Gaia DR2. Multiple spectroscopic observations confirm that the star is on the main sequence and rule out obvious spectroscopic binaries. Time-series photometric follow-up of this TOI has cleared all neighboring stars as origins of the transit except for TIC 1943945558 and TIC 1943945562, which is 9\farcs1 northeast and 7.4 magnitudes fainter in the TESS band. The 0.26 ppt event seen in the TESS data has not been detected around the target star.

The \dave\ analysis of this TOI detects a potential difference between the even and odd primary transits, which could be indicative of an FP in the form of an eclipsing binary. \dave\ did not report any other FP indicators for this TOI. The SPOC data validation report for this TOI reports no significant centroid offset.

The \tri\ analysis of this TOI finds ${\rm FPP} = (1.12 \pm 0.05) \times 10^{-2}$ and ${\rm NFPP} = (1.04 \pm 0.05) \times 10^{-2}$. These $> 1 \%$ FPP and NFPP are driven by the uncertainty over whether or not the transit originates from the target star or TIC 1943945562. This FPP and NFPP are too high to consider this planet candidate validated.

Assuming this is a real planet around TIC 406672232, we estimate the semiamplitude of the RV signal to be $K_{\rm RV} = 1.8^{+1.3}_{-0.7}$ m/s, corresponding to $M_{\rm p} = 2.4^{+1.7}_{-0.8} \, M_\oplus$.

\subsection{TOI-1411.01}

TOI-1411.01 is a $1.36 \pm 0.16 \, R_\oplus$ planet candidate with a 1.45 day orbital period orbiting a K dwarf (TIC 116483514) that is 32.5 pc away and has a $V$ magnitude of 10.51. A Lomb-Scargle periodogram of the photometry from each TESS sector finds a maximum peak of 0.02, indicating that the star is quiet. This is corroborated by the $\log R^\prime _{\rm H K}$ of -4.7252 extracted from our HIRES spectrum and the low $v \sin{i}$ extracted from our TRES spectrum. TOI-1411 has been observed in 3 TESS sectors (16, 23, and 24) and is scheduled to be reobserved in another 2 sectors (50 and 51) between 2022-03-26 and 2022-05-18.

Follow-up observations have found no evidence of this TOI being an FP. Time-series photometric follow-up of this TOI has cleared all neighboring stars as origins of the transit. Of note to this TOI is TIC 1101969798, a periodic variable with a semiamplitude of 0.1 mag and a period of 0.107 day, that is located 90\arcsec\ to the north-east.

The \dave\ analysis of this TOI finds no strong indicators that the candidate is an FP. The SPOC data validation report for this TOI reports no significant centroid offset.

The \tri\ analysis of this TOI finds ${\rm FPP} = (1.18 \pm 0.68) \times 10^{-4}$. Because all neighboring stars have been cleared, \tri\ finds ${\rm NFPP} = 0.0$. This FPP is sufficiently low to consider the planet validated. We hereafter refer to this planet as TOI-1411 b. 

We estimate the semiamplitude of the RV signal for this planet to be $K_{\rm RV} = 2.0^{+1.7}_{-1.0}$ m/s, corresponding to $M_{\rm p} = 2.5^{+2.0}_{-1.1} \, M_\oplus$. Vermilion et al (2022, in preparation), which detects the RV signal of this planet, reports a $K_{\rm RV}$ $5\sigma$ upper limit of 4.26 m/s (or a mass of $5.66 \, M_\oplus$), consistent with our estimate and with a terrestrial composition.

\subsection{TOI-1442.01}

TOI-1442.01 is a $1.17 \pm 0.06 \, R_\oplus$ planet candidate with a 0.41 day orbital period orbiting a M dwarf (TIC 235683377) that is 41.2 pc away and has a $V$ magnitude of 15.39. A Lomb-Scargle periodogram of the photometry from each TESS sector finds a maximum peak of 0.02, indicating that the star is quiet. TOI-1442 has been observed in 15 TESS sectors (14--26, 40, and 41) and is scheduled to be reobserved in another 9 sectors (47--55) between 2021-12-30 and 2022-09-01.

\begin{figure*}
    \centering
    \includegraphics[width=\textwidth]{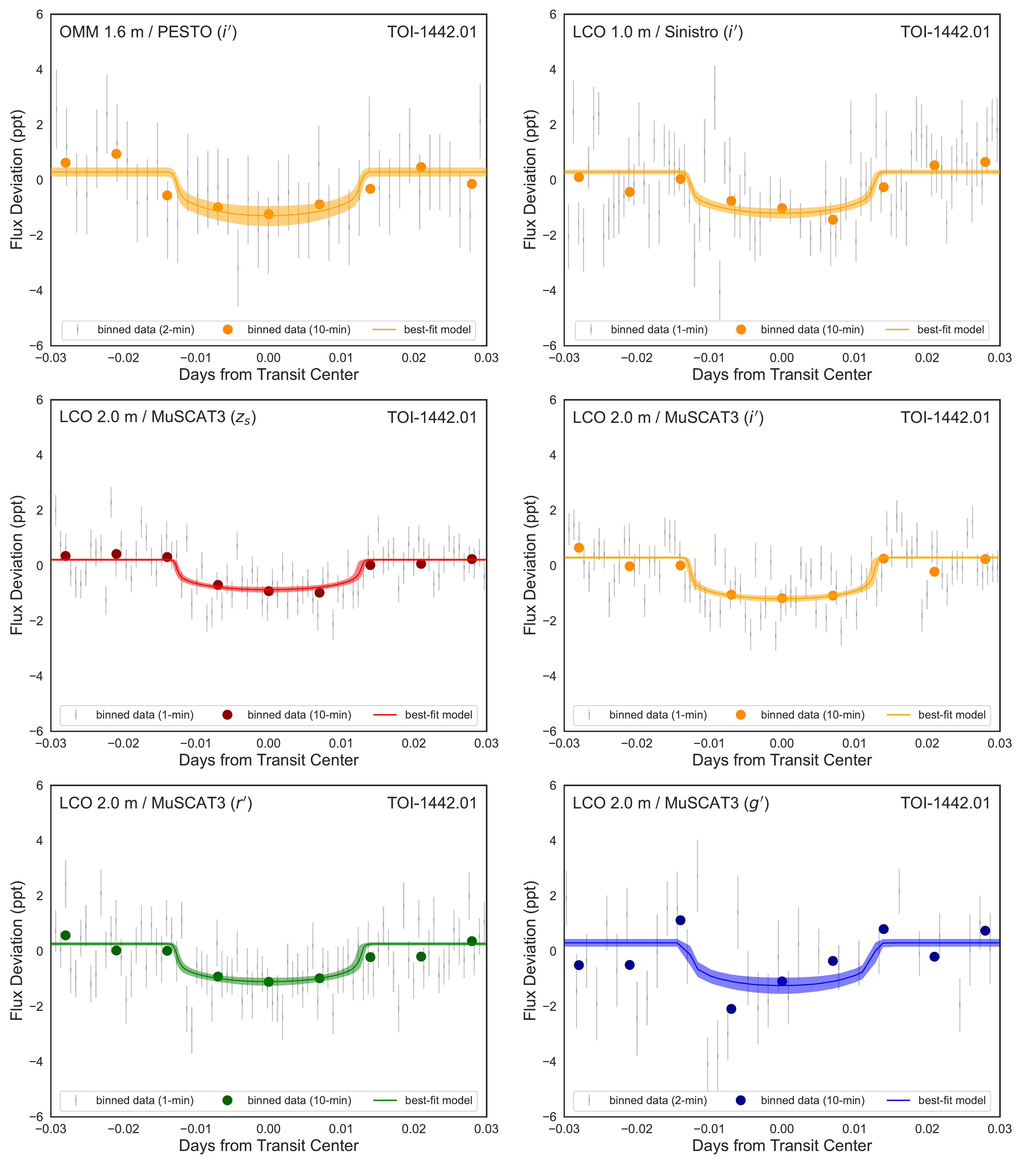}
    \caption{Phase-folded ground-based data and best-fit model of the transit of TOI-1442.01. The data is detrended with a linear model and 3$\sigma$ outliers are removed.}
    \label{fig: 1442-joint}
\end{figure*}

Follow-up observations have found no evidence of this TOI being an FP, although no spectroscopic observations of this TOI have been collected. Time-series photometric follow-up has made several detections of the transit of TOI-1442.01 on TIC 235683377 (shown in Figure \ref{fig: 1442-joint}).

The \dave\ analysis of this TOI finds no strong indicators that the candidate is an FP. However, the S/N of the per-transit difference images used by \dave\ is very low and the measured centroids are unreliable. The SPOC data validation report for this TOI reports no significant centroid offset.

The \tri\ analysis of this TOI finds ${\rm FPP} = (7.00 \pm 4.11) \times 10^{-6}$. Because the transit has been verified on-target, \tri\ finds ${\rm NFPP} = 0.0$. This FPP is sufficiently low to consider the planet validated. We hereafter refer to this planet as TOI-1442 b. 

We estimate the semiamplitude of the RV signal for this planet to be $K_{\rm RV} = 3.2^{+2.2}_{-1.0}$ m/s, corresponding to $M_{\rm p} = 1.6^{+1.1}_{-0.5} \, M_\oplus$.

\subsection{TOI-1693.01}

TOI-1693.01 is a $1.42 \pm 0.10 \, R_\oplus$ planet candidate with a 1.77 day orbital period orbiting a M dwarf (TIC 353475866) that is 30.8 pc away and has a $V$ magnitude of 12.96. A Lomb-Scargle periodogram of the photometry from each TESS sector finds a maximum peak of 0.01, indicating that the star is quiet. This is corroborated by the $\log R^\prime _{\rm H K}$ of -5.2169 extracted from our HIRES spectrum. TOI-1693 has been observed in 4 TESS sectors (19 and 43--45).

\begin{figure*}
    \centering
    \includegraphics[width=\textwidth]{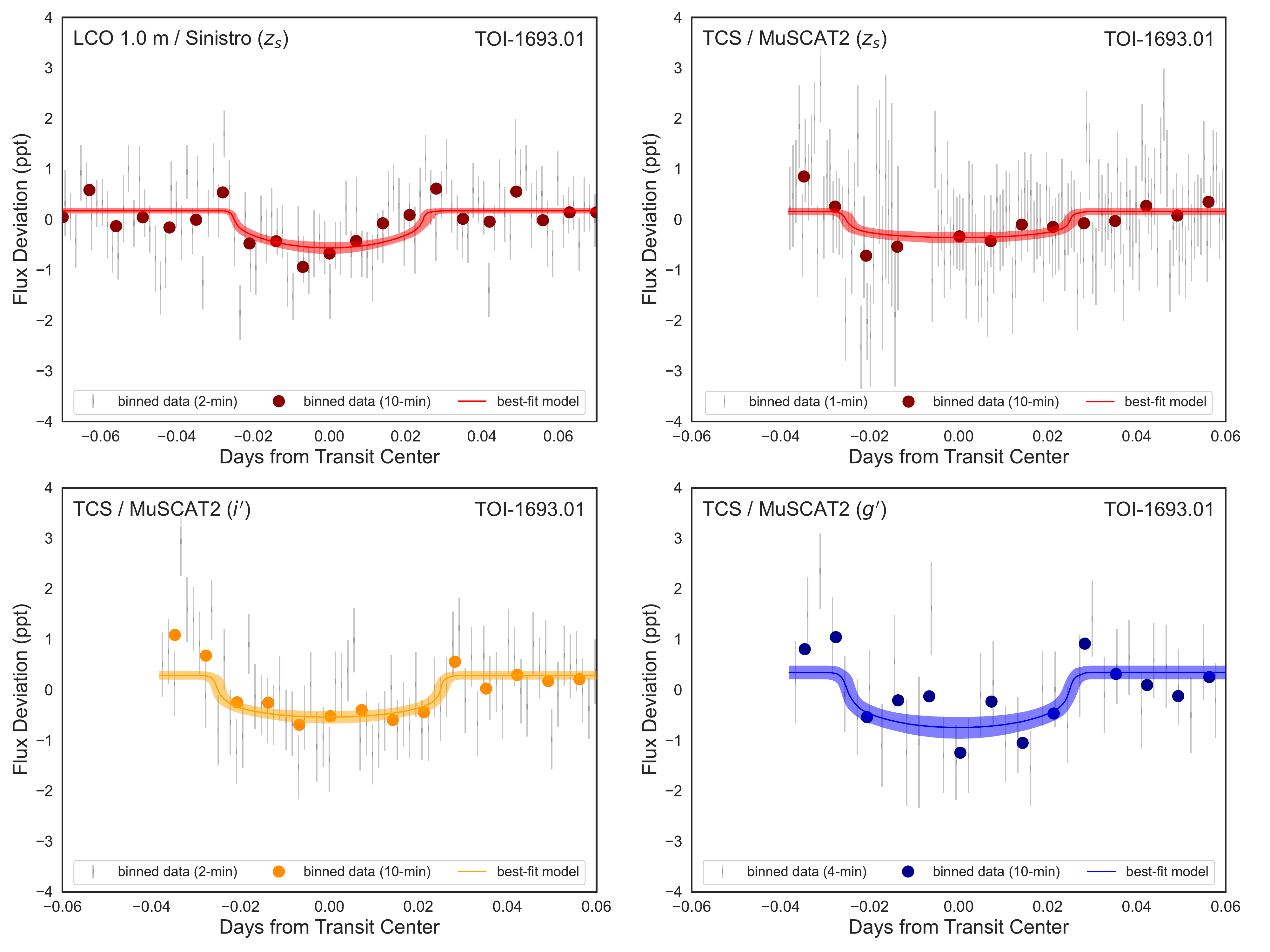}
    \caption{Phase-folded ground-based data and best-fit model of the transit of TOI-1693.01. The data is detrended with a linear model and 3$\sigma$ outliers are removed.}
    \label{fig: 1693-joint}
\end{figure*}

Follow-up observations have found no evidence of this TOI being an FP. Time-series photometric follow-up has made several detections of the transit of TOI-1693.01 on TIC 353475866 (shown in Figure \ref{fig: 1693-joint}).\footnote{Those observations are blended with TIC 723362263, which is 3\farcs75 south-west and 8.3 magnitudes fainter in the TESS band, which is marginally too faint to have caused the TESS detection.}

The \dave\ analysis of this TOI finds no strong indicators that the candidate is an FP. The SPOC data validation report for this TOI reports no significant centroid offset.

The \tri\ analysis of this TOI finds ${\rm FPP} = (1.47 \pm 0.13) \times 10^{-3}$. Because the transit has been verified on-target, \tri\ finds ${\rm NFPP} = 0.0$. This FPP is sufficiently low to consider the planet validated. We hereafter refer to this planet as TOI-1693 b. 

We estimate the semiamplitude of the RV signal for this planet to be $K_{\rm RV} = 2.4^{+1.9}_{-0.8}$ m/s, corresponding to $M_{\rm p} = 2.8^{+2.2}_{-1.0} \, M_\oplus$.

\begin{figure*}[t!]
  \centering
    \includegraphics[width=0.45\textwidth]{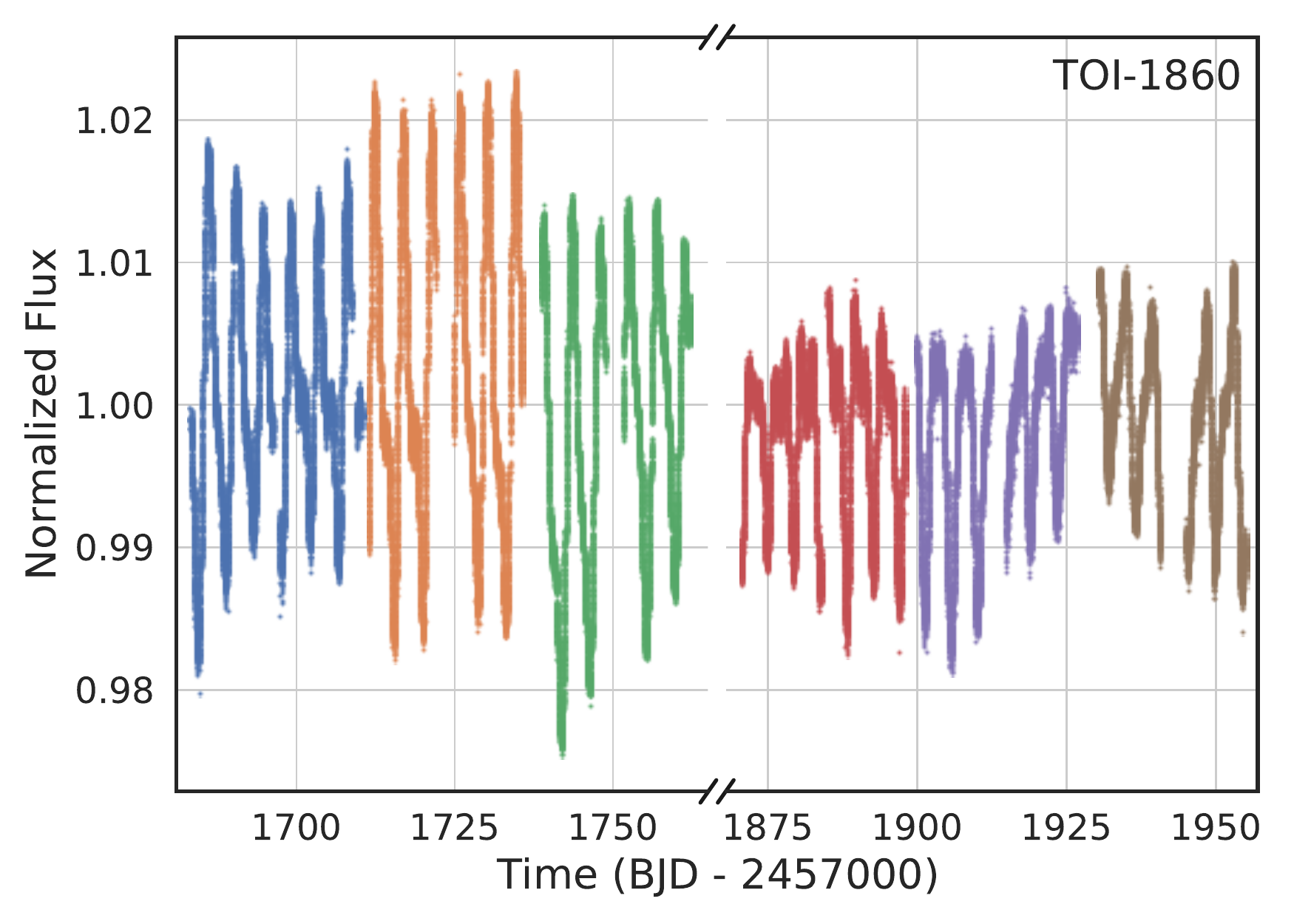}
    \includegraphics[width=0.45\textwidth]{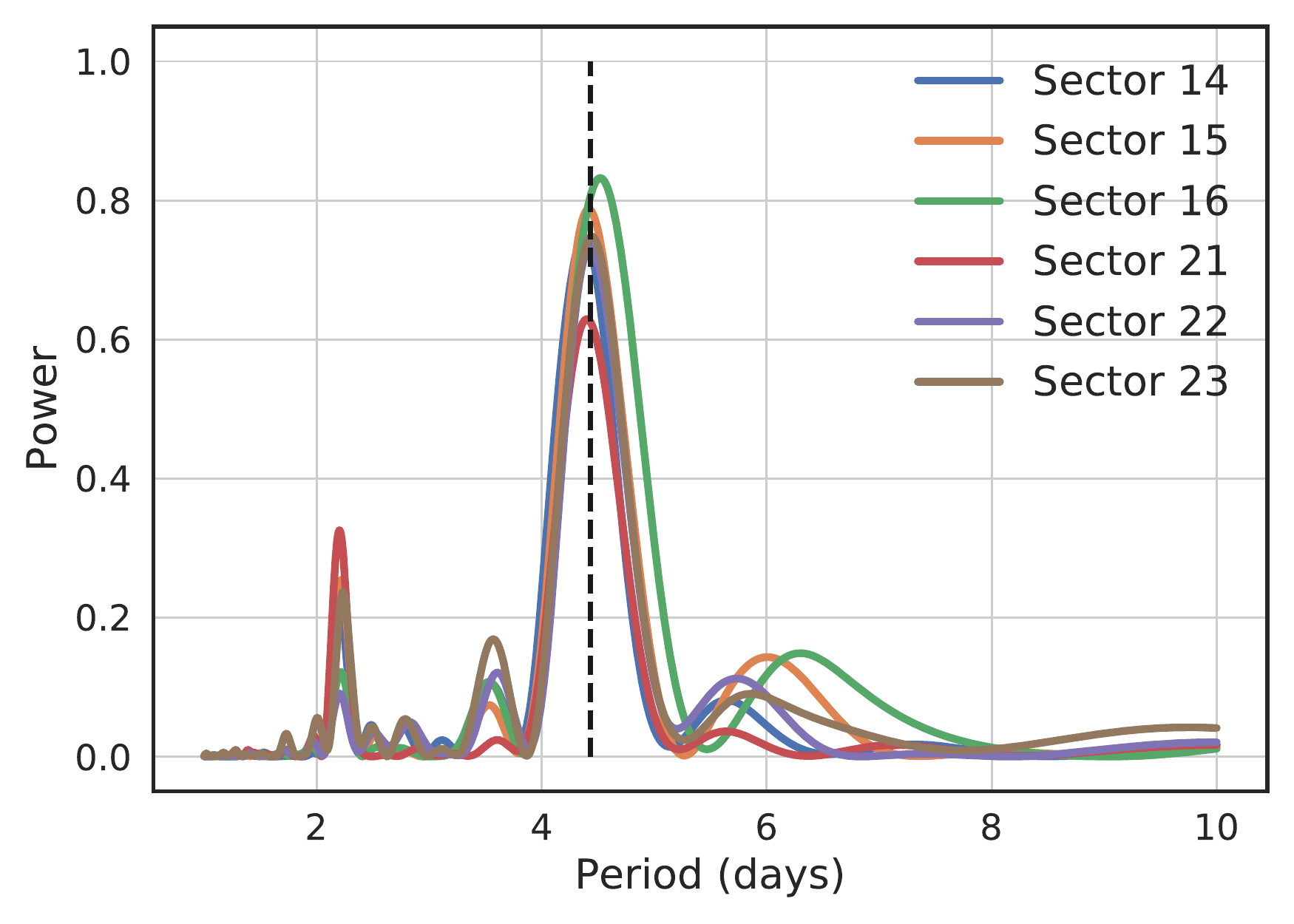}
    \includegraphics[width=0.45\textwidth]{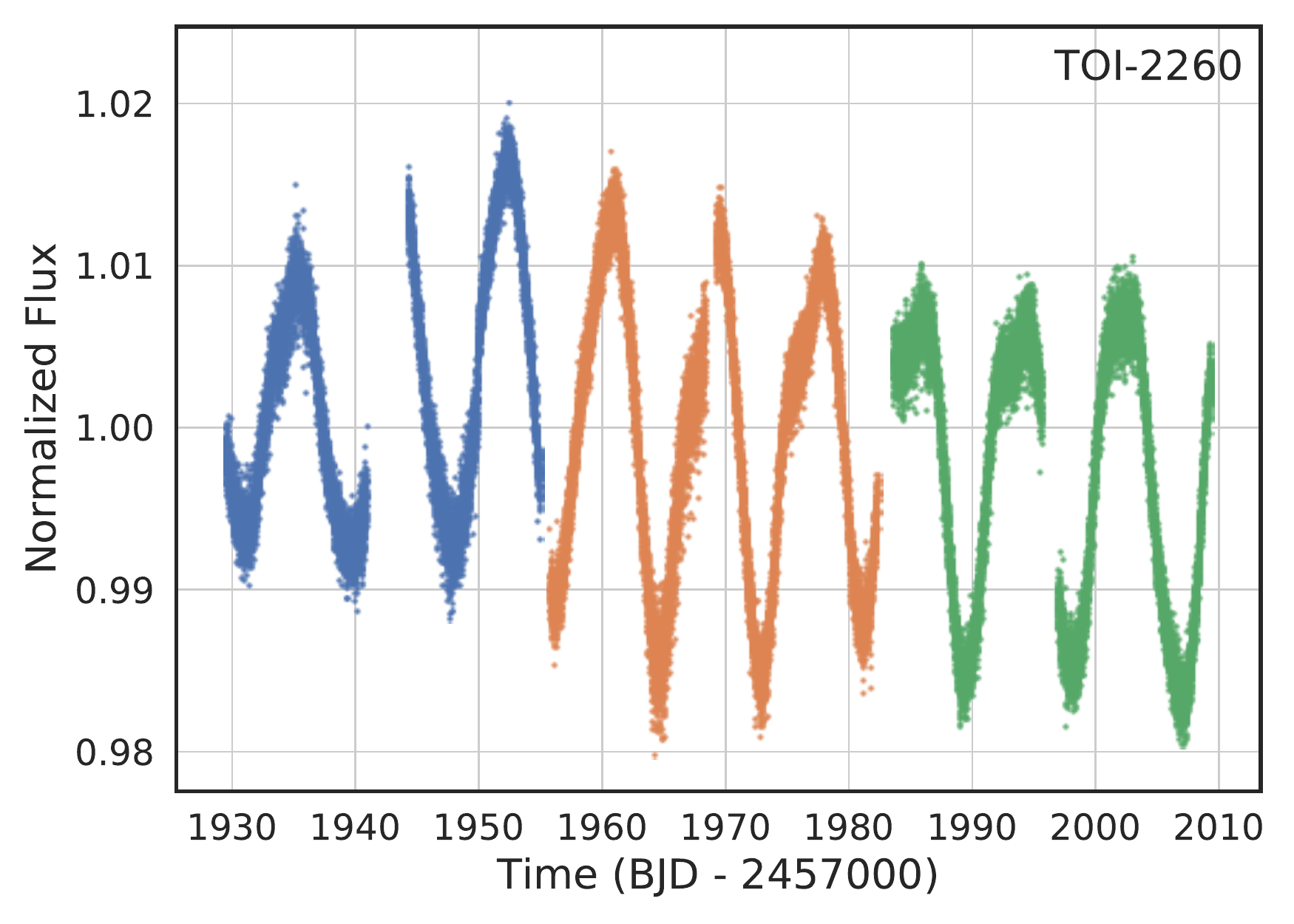}
    \includegraphics[width=0.45\textwidth]{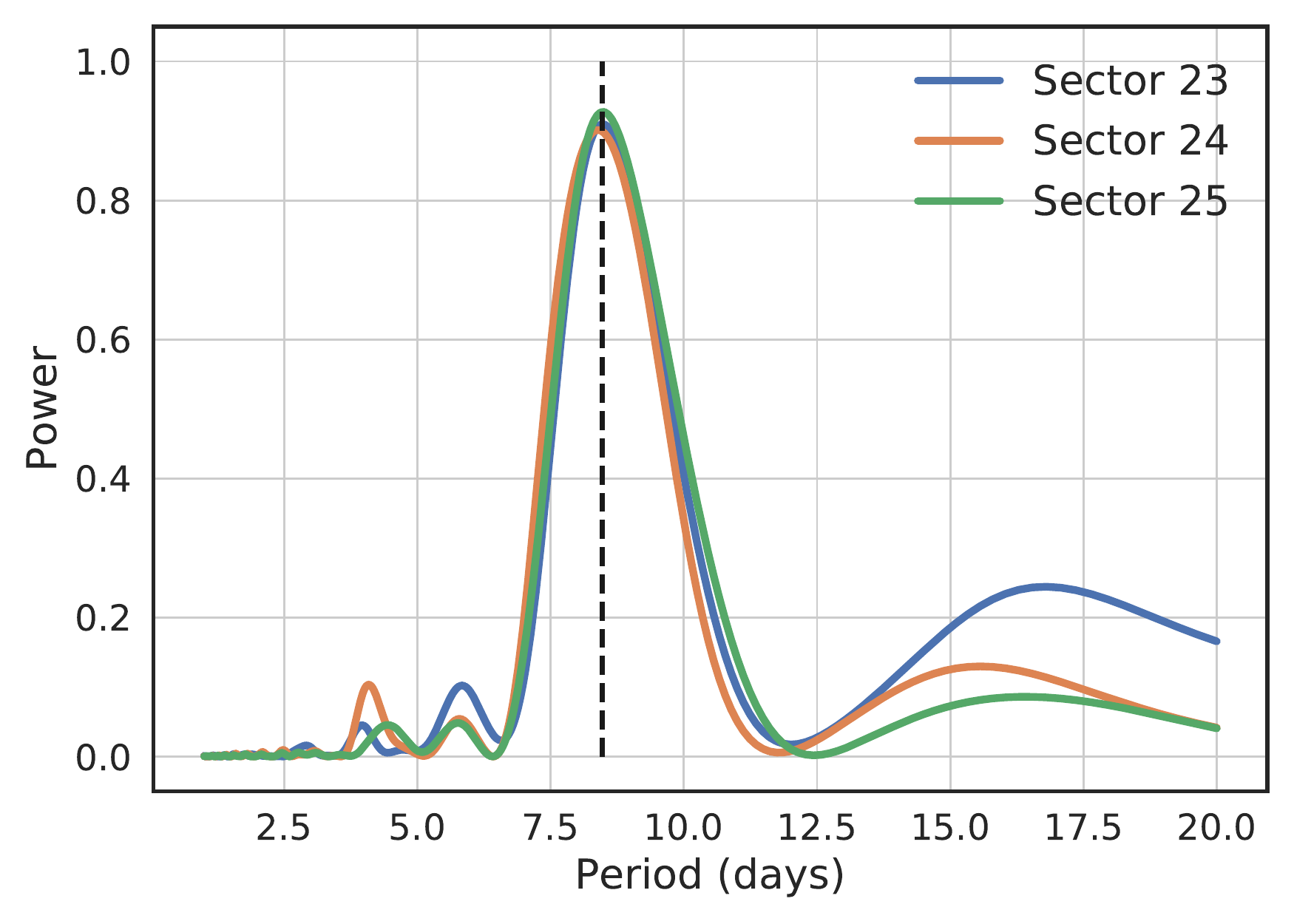}
    \caption{TESS light curves and Lomb-Scargle periodograms for TOI-1860 (top) and TOI-2260 (bottom). To estimate the rotation period of a star, we use the periodogram to calculate the peak period for each sector separately. Our estimate is then given by the mean and standard deviation of these rotation periods. For TOI-1860 and TOI-2260, we estimate a rotation period of $4.43 \pm 0.04$ days and $8.45 \pm 0.06$ days, respectively. These periods are indicated with vertical dashed lines in the right-hand panels.}
    \label{fig: rot_per}
\end{figure*}

\subsection{TOI-1860.01}\label{TOI-1860}

TOI-1860.01 is a $1.31 \pm 0.04 \, R_\oplus$ planet candidate with a 1.07 day orbital period orbiting a G dwarf (TIC 202426247) that is 45.9 pc away and has a $V$ magnitude of 8.4. A Lomb-Scargle periodogram of the photometry from each TESS sector finds a maximum peak of 0.83, indicating strong activity and a young host star. We also estimate a $\log R^\prime _{\rm H K}$ of -4.2524 from our HIRES spectrum, which indicates that the star is young and active. TOI-1860 has been observed in 7 TESS sectors (14--16, 21--23, and 41) and is scheduled to be reobserved in another 3 sectors (48--50) between 2022-01-28 and 2022-04-22.

Because this is an active star, we can use the TESS light curve to derive its rotation period. In Figure \ref{fig: rot_per}, we display the results of a Lomb-Scargle periodogram applied to each sector separately, which gives a rotation period of $4.43 \pm 0.06$ days. Using the relation defined in \cite{barnes2007ages}, we estimate the age of the star to be $133 \pm 26$ Myr. Lastly, we use BANYAN $\Sigma$ \citep{gagne2018banyan} to determine the probability that the star is a member of a nearby young association. This analysis returns a $99.9 \%$ probability that TOI-1860 is a field star.

Another interesting aspect of TOI-1860 is that it has stellar parameters and a metallicity very similar to that of the Sun, and qualifies as a solar twin according to most definitions \citep{de1996stars, ramirez2014solar}. For Solar twins, there is known to be a strong correlation between [Y/Mg] and stellar age \citep{nissen2015high, maia2016solar}. Because we obtained elemental abundances for this star using \texttt{KeckSpec} (see Table \ref{tab:keckspec}), we are able to conduct an independent check of the age of this system. Using the relation provided in \cite{maia2016solar} and [Y/Mg] = $0.196 \pm 0.090$, we estimate an age upper limit 1.93 Gyr, which is consistent with our estimation based on gyrochronology.

Follow-up observations have found no evidence of this TOI being an FP. Time-series photometric follow-up of this TOI has cleared all neighboring stars as origins of the transit except for TIC 1102367690, which is 5\farcs5 west and 5.8 magnitudes fainter in the TESS band. The 0.23 ppt event seen in the TESS data has not been detected around the target star.

\dave\ was unable to perform a vetting analysis of this TOI, due to a failure of its transit model to fit the TESS data. The SPOC data validation report for this TOI reports no significant centroid offset.

The \tri\ analysis of this TOI find ${\rm FPP} = (1.97 \pm 0.45) \times 10^{-4}$ and ${\rm NFPP} = (9.68 \pm 2.23) \times 10^{-6}$. This FPP and NFPP are sufficiently low to consider the planet validated. We hereafter refer to this planet as TOI-1860 b.

We estimate the semiamplitude of the RV signal for this planet to be $K_{\rm RV} = 1.4^{+0.8}_{-0.4}$ m/s, corresponding to $M_{\rm p} = 2.2^{+1.3}_{-0.7} \, M_\oplus$.

\subsection{TOI-2260.01}

TOI-2260.01 is a $1.62 \pm 0.13 \, R_\oplus$ planet candidate with a 0.35 day orbital period orbiting a G dwarf (TIC 232568235) that is 101.3 pc away and has a $V$ magnitude of 10.47. A Lomb-Scargle periodogram of the photometry from each TESS sector finds a maximum peak of 0.93, indicating strong activity and a young host star. We also estimate a $\log R^\prime _{\rm H K}$ of -4.438 from our HIRES spectrum, which indicates that the star is young and active. TOI-2260 has been observed in 3 TESS sectors (23--25) and is scheduled to be reobserved in another 3 sectors (50--52) between 2022-03-26 and 2022-06-13.  

Because this is an active star, we can use the TESS light curve to derive its rotation period. In Figure \ref{fig: rot_per}, we display the results of a Lomb-Scargle periodogram applied to each sector separately, which gives a rotation period of $8.45 \pm 0.03$ days. Using the relation defined in \cite{barnes2007ages}, we estimate the age of the star to be $321 \pm 96$ Myr. Lastly, we use BANYAN $\Sigma$ \citep{gagne2018banyan} to determine the probability that the star is a member of a nearby young association. This analysis returns a $99.9 \%$ probability that TOI-2260 is a field star.

Follow-up observations have found no evidence of this TOI being an FP. Time-series photometric follow-up of this TOI has cleared all neighboring stars as origins of the transit.

The \dave\ analysis of this TOI finds no strong indicators that the candidate is an FP. The SPOC data validation report for this TOI reports a significant centroid offset in sector 24, but has not conducted centroid offset analyses for sectors 23 and 25. However, given that all neighboring stars have been cleared from being nearby eclipsing binaries, this offset is unlikely to be caused by a FP coming from a nearby star.

The \tri\ analysis of this TOI finds ${\rm FPP} = (5.26 \pm 0.50) \times 10^{-3}$. Because all neighboring stars have been cleared, \tri\ finds ${\rm NFPP} = 0.0$. This FPP is sufficiently low to consider the planet validated. We hereafter refer to this planet as TOI-2260 b. 

We estimate the semiamplitude of the RV signal for this planet to be $K_{\rm RV} = 3.0^{+2.2}_{-1.1}$ m/s, corresponding to $M_{\rm p} = 3.5^{+2.5}_{-1.3} \, M_\oplus$.

\subsection{TOI-2290.01}

TOI-2290.01 is a $1.17 \pm 0.07 \, R_\oplus$ planet candidate with a 0.39 day orbital period orbiting a K dwarf (TIC 321688498) that is 58.1 pc away and has a $V$ magnitude of 12.64.  A Lomb-Scargle periodogram of the photometry from each TESS sector finds a maximum peak of 0.03, indicating that the star is quiet. However, the $\log R^\prime _{\rm H K}$ of -4.459 extracted from our HIRES spectrum suggests that the star may actually be quite active. TOI-2290 has been observed in 4 TESS sectors (17, 18, 24, and 25).

Follow-up observations have found no evidence of this TOI being an FP. Time-series photometric follow-up of this TOI has cleared all neighboring stars as origins of the transit.

The \dave\ analysis of this TOI finds a potential centroid offset, but finds no other significant FP indicators. No data validation reports have been generated by the SPOC pipeline for this TOI.

The \tri\ analysis of this TOI finds ${\rm FPP} = (4.92 \pm 0.11) \times 10^{-1}$. Because all neighboring stars have been cleared, \tri\ finds ${\rm NFPP} = 0.0$. The reason for this $> 1 \%$ FPP comes from the scenario that the TOI is a blended eclipsing binary. This FPP is too high to consider the planet validated.

Assuming this is a real planet, we estimate the semiamplitude of the RV signal to be $K_{\rm RV} = 2.1^{+1.7}_{-0.7}$ m/s, corresponding to $M_{\rm p} = 1.6^{+1.4}_{-0.6} \, M_\oplus$.

\subsection{TOI-2411.01}

TOI-2411.01 is a $1.68 \pm 0.11 \, R_\oplus$ planet candidate with a 0.78 day orbital period orbiting a K dwarf (TIC 10837041) that is 59.5 pc away and has a $V$ magnitude of 11.27. A Lomb-Scargle periodogram of the photometry from each TESS sector finds a maximum peak of 0.002, indicating that the star is quiet. TOI-2411 has been observed in 2 TESS sectors (3 and 30).

\begin{figure}
    \centering
    \includegraphics[width=0.45\textwidth]{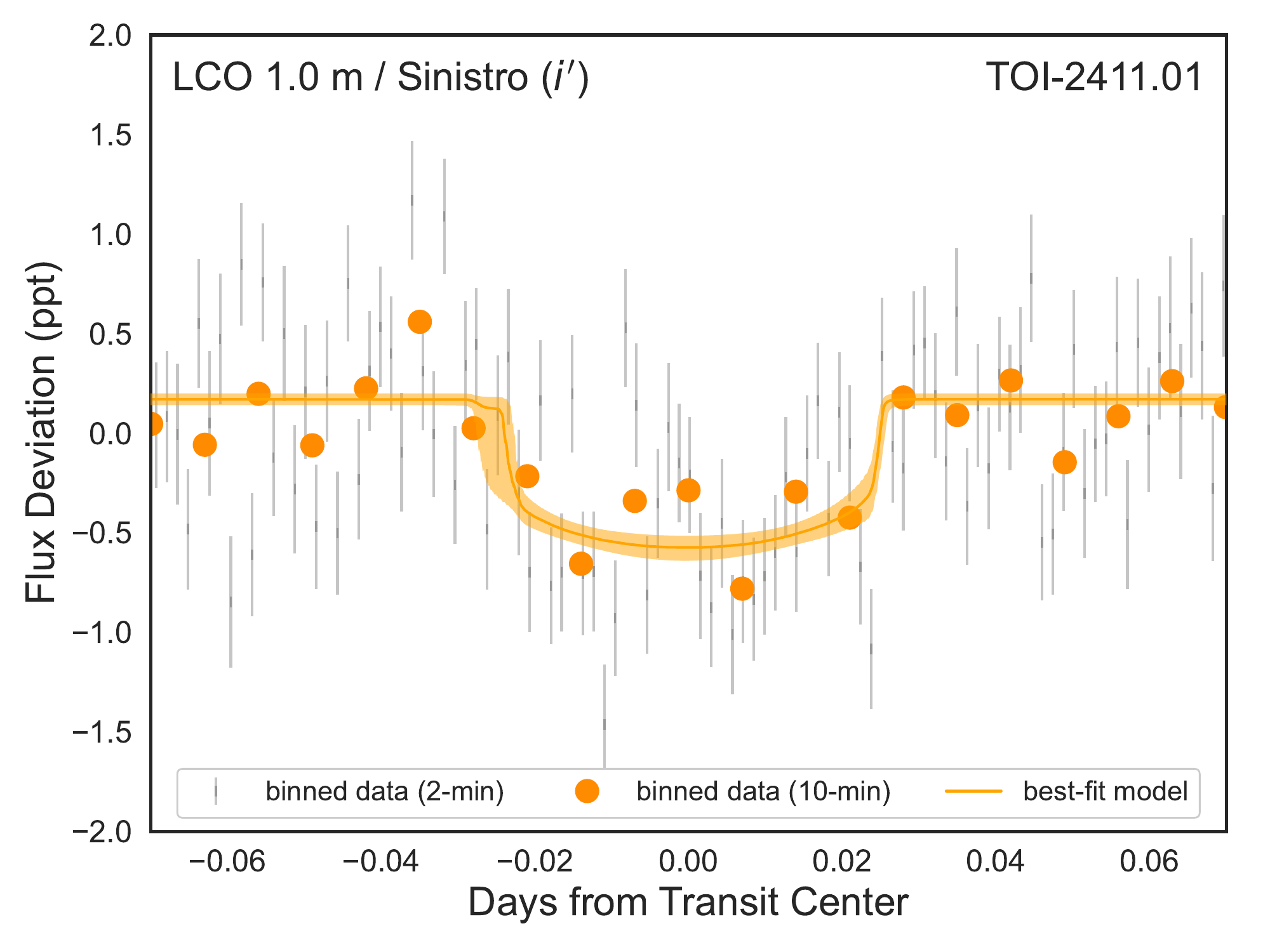}
    \caption{Phase-folded ground-based data and best-fit model of the transit of TOI-2411.01. The data is detrended with a linear model and 3$\sigma$ outliers are removed.}
    \label{fig: 2411-joint}
\end{figure}

Follow-up observations have found no evidence of this TOI being an FP. Time-series photometric follow-up has made several detections of the transit of TOI-2411.01 on TIC 10837041 (shown in Figure \ref{fig: 2411-joint}).

\dave\ is unable to analyze this TOI due to the very low S/N of the data. The SPOC data validation report for this TOI reports no significant centroid offset or any other FP indicators.

The \tri\ analysis of this TOI finds ${\rm FPP} = (1.17 \pm 0.05) \times 10^{-3}$. Because transits have been verified on-target, \tri\ finds ${\rm NFPP} = 0.0$. This FPP is sufficiently low to consider the planet validated. We hereafter refer to this planet as TOI-2411 b. 

We estimate the semiamplitude of the RV signal for this planet to be $K_{\rm RV} = 3.6^{+2.5}_{-1.3}$ m/s, corresponding to $M_{\rm p} = 3.9^{+2.8}_{-1.4} \, M_\oplus$.

\subsection{TOI-2427.01}

TOI-2427.01 is a $1.80 \pm 0.12 \, R_\oplus$ planet candidate with a 1.31 day orbital period orbiting a K dwarf (TIC 142937186) that is 28.5 pc away and has a $V$ magnitude of 10.30. A Lomb-Scargle periodogram of the photometry from each TESS sector finds a maximum peak of 0.05, indicating that the star is quiet. TOI-2427 has been observed in 1 TESS sector (31).

\begin{figure}
    \centering
    \includegraphics[width=0.45\textwidth]{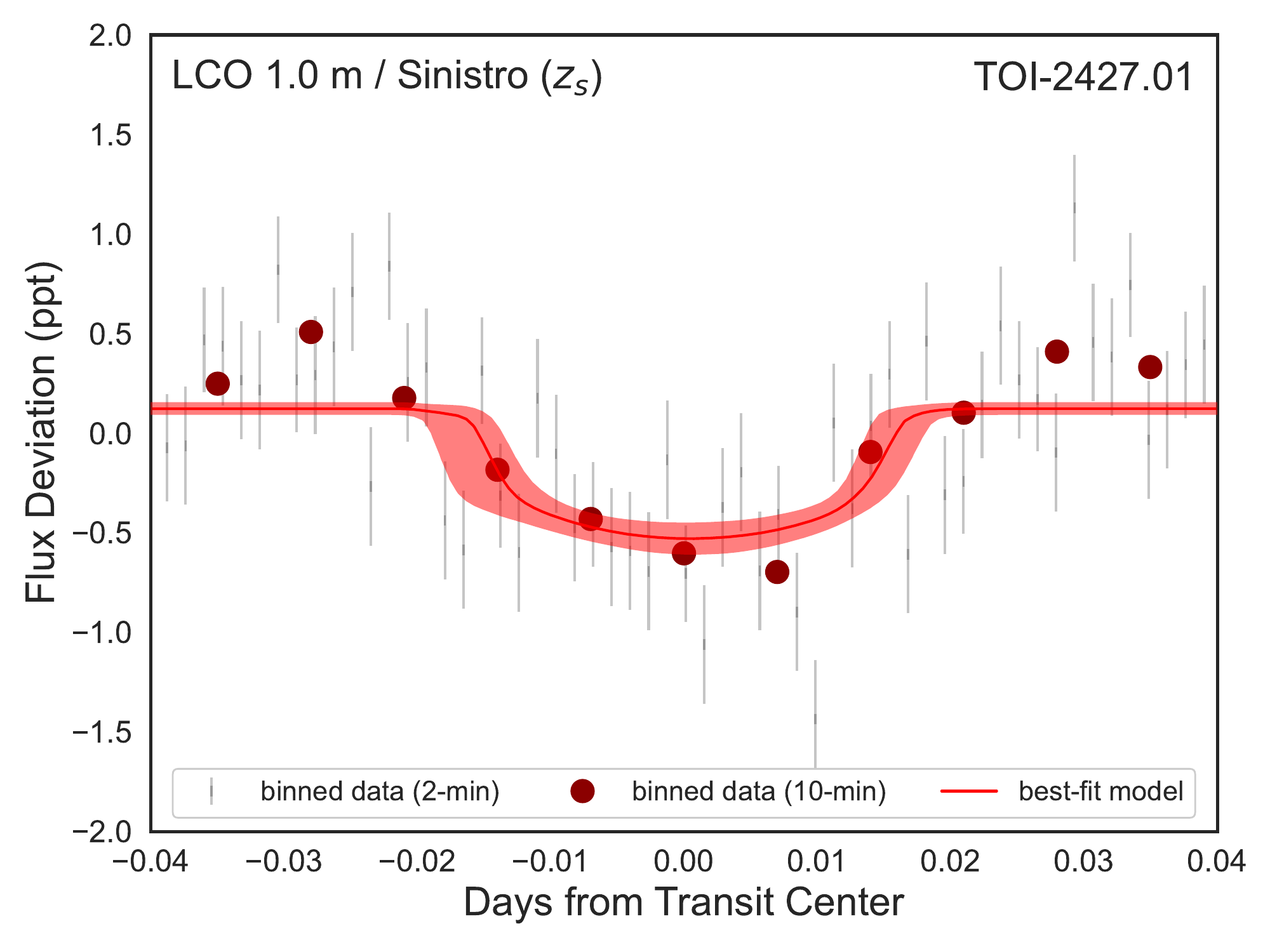}
    \caption{Phase-folded ground-based data and best-fit model of the transit of TOI-2427.01. The data is detrended with a linear model and 3$\sigma$ outliers are removed.}
    \label{fig: 2427-joint}
\end{figure}

Follow-up observations have found no evidence of this TOI being an FP. Time-series photometric follow-up has made several detections of the transit of TOI-2427.01 on TIC 142937186 (shown in Figure \ref{fig: 2427-joint}).

The \dave\ analysis of this TOI finds a potential centroid offset, but finds no other indicators that this TOI is an FP. The SPOC data validation report for this TOI also reports a significant centroid offset. However, given that all neighboring stars have been cleared from being nearby eclipsing binaries, this offset is unlikely to be caused by a FP originating from a nearby star.

The \tri\ analysis of this TOI finds ${\rm FPP} = (7.35 \pm 2.72) \times 10^{-3}$. Because transits have been verified on-target, \tri\ finds ${\rm NFPP} = 0.0$. This FPP is sufficiently low to consider the planet validated. We hereafter refer to this planet TOI-2427 b. 

We estimate the semiamplitude of the RV signal for this planet to be $K_{\rm RV} = 3.2^{+2.4}_{-1.2}$ m/s, corresponding to $M_{\rm p} = 4.1^{+3.1}_{-1.5} \, M_\oplus$.

\subsection{TOI-2445.01}

TOI-2445.01 is a $1.25 \pm 0.08 \, R_\oplus$ planet candidate with a 0.37 day orbital period orbiting a M dwarf (TIC 439867639) that is  48.6 pc away and has a $V$ magnitude of 15.69. A Lomb-Scargle periodogram of the photometry from each TESS sector finds a maximum peak of 0.04, indicating that the star is quiet. TOI-2445 has been observed in 2 TESS sectors (4 and 31).

\begin{figure*}
    \centering
    \includegraphics[width=\textwidth]{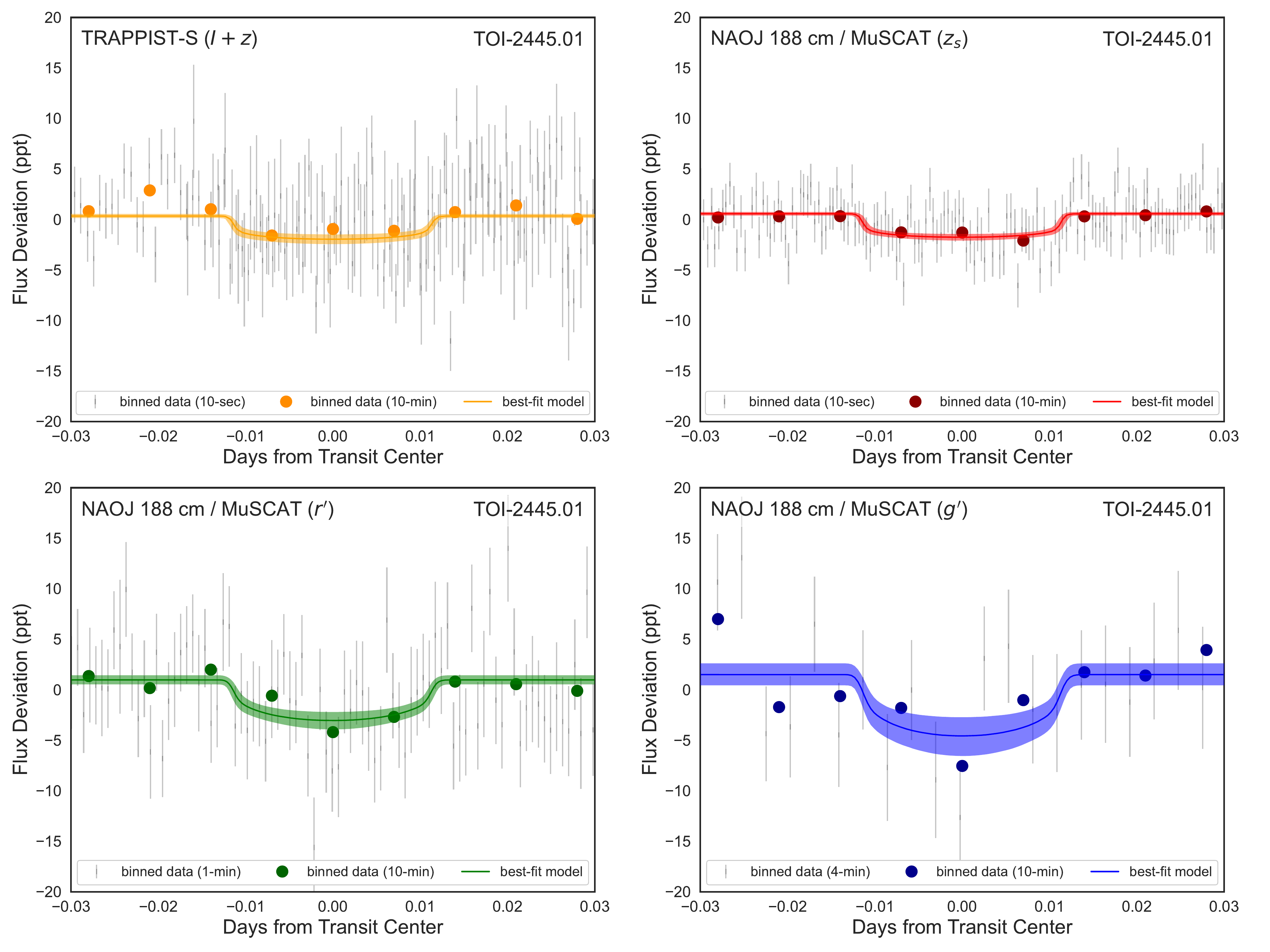}
    \caption{Phase-folded ground-based data and best-fit model of the transit of TOI-2445.01. The data is detrended with a linear model and 3$\sigma$ outliers are removed.}
    \label{fig: 2445-joint}
\end{figure*}

Follow-up observations have found no evidence of this TOI being a FP, although no spectroscopic observations of this TOI have been collected. Time-series photometric follow-up has made several detections of the transit of TOI-2445.01 on TIC 439867639 (shown in Figure \ref{fig: 2445-joint}).

The \dave\ analysis of this TOI finds no strong indicators that the candidate is a FP. However, like TOI-739, the S/N of the per-transit difference images used by \dave\ is very low and the measured centroids are unreliable. No data validation reports have been generated by the SPOC pipeline for this TOI.

The \tri\ analysis of this TOI finds ${\rm FPP} = (1.88 \pm 0.45) \times 10^{-4}$. Because transits have been verified on-target, \tri\ finds ${\rm NFPP} = 0.0$. This FPP is sufficiently low to consider the planet validated. We hereby refer to this planet as TOI-2445 b. 

We estimate the semi-amplitude of the RV signal for this planet to be $K_{\rm RV} = 4.5^{+2.8}_{-1.7}$ m/s, corresponding to $M_{\rm p} = 2.0^{+1.2}_{-0.7} \, M_\oplus$.

\section{Discussion}\label{sec: discussion}

\begin{figure*}[t!]
  \centering
    \includegraphics[width=1.0\textwidth]{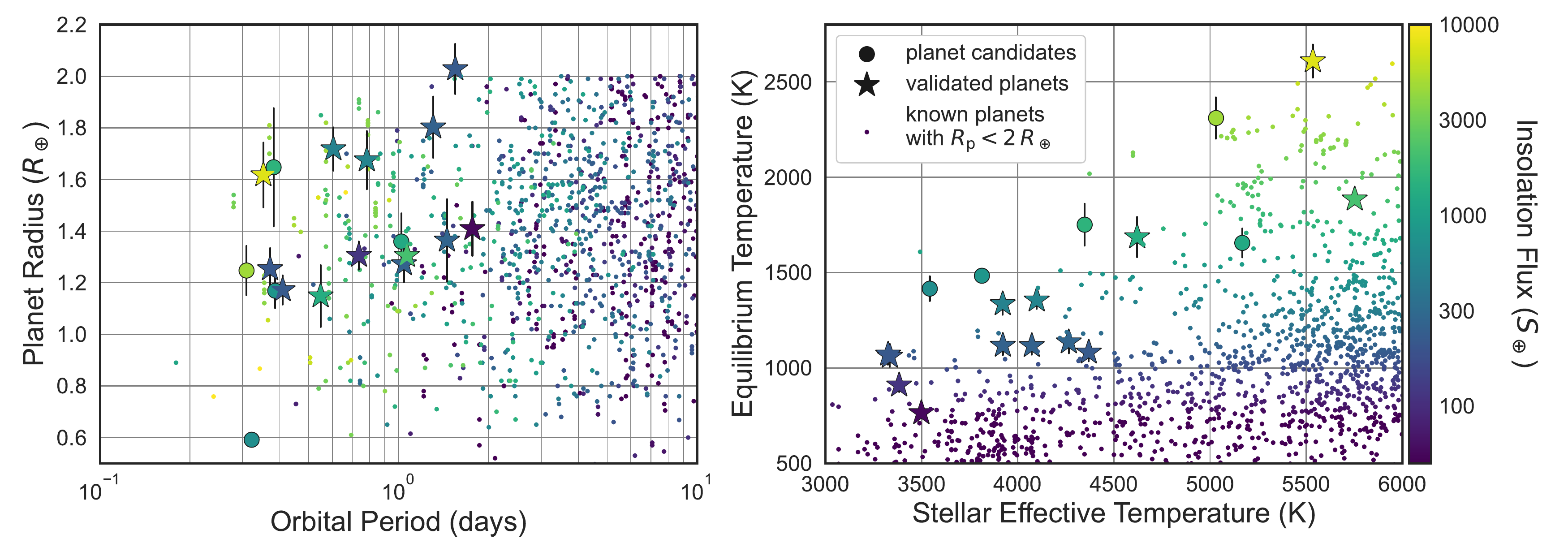}
    \caption{Left: Planet radii and orbital periods of all planet candidates (circles) and validated planets (stars) in this paper, along with all known planets with $R_{\rm p} < 2 \, R_\oplus$ (points). Right: Planet equilibrium temperatures and host star effective temperatures for the same planet candidates, validated planets, and known planets. Color indicates insolation flux. Data for known planets was obtained though the NASA Exoplanet Archive.}
    \label{fig: all}
\end{figure*}

In Section \ref{sec: results}, we scrutinized the available data of 18 potentially terrestrial TESS planet candidates that display promise as subjects of emission spectroscopy observations with JWST. Of these, 13 were validated. In Figure \ref{fig: all}, we show how our targets are distributed in the planet radius -- orbital period plane and the planet equilibrium temperature -- stellar effective temperature plane, with all other known planets with $R_{\rm p} < 2 \, R_\oplus$ included for reference.

The planet candidates and planets analyzed in this paper cover a wide region of parameter space that will allow for studies of hot, potentially terrestrial planets across different environments. For instance, many of the planets validated in this paper are among the hottest known planets with $R_{\rm p} < 2 \, R_\oplus$. For stars with $T_{\rm eff} < 3500$ K, TOI-1442 b and TOI-206 b rank as the fifth and sixth hottest planets, respectively, with $T_{\rm eq} = 1072 \pm 54$ K and $T_{\rm eq} = 910 \pm 36$ K, only being surpassed by GJ 1252 b \citep[$T_{\rm eq} \sim 1089$ K;][]{shporer2020gj}, K2-137 b \citep[$T_{\rm eq} \sim 1608$ K;][]{smith2018k2}, TOI-1634 b \citep[$T_{\rm eq} \sim 1608$ K;][]{cloutier2021toi}, and TOI-1685 b \citep[$T_{\rm eq} \sim 1066$ K;][]{bluhm2021ultra}. For stars with $3500 < T_{\rm eff} < 4000$ K, TOI-1075 b and TOI-833 b are the first and second hottest planets, respectively, with $T_{\rm eq} = 1336 \pm 56$ K and $T_{\rm eq} = 1118 \pm 49$ K, but would be superseded by TOI-2290.01 if found to be a bona fide planet. Lastly, TOI-2260 b is the fourth hottest known planet of this size to orbit any star, with $T_{\rm eq} = 2609 \pm 86$ K, only being surpassed by KOI-55 b \citep[$T_{\rm eq} \sim 8000$ K;][]{charpinet2011compact}, TOI-55 c \citep[$T_{\rm eq} \sim 7000$ K;][]{charpinet2011compact}, and Kepler-1340 b \citep[$T_{\rm eq} \sim 2860$ K;][]{morton2016false}. All of these planets will be valuable for studying the evolution of planets with high equilibrium temperatures, which is a key parameter in core-powered atmospheric mass-loss models for small planets \citep{ginzburg2016super, ginzburg2018core}.

Of our 13 validated planets, 7 (TOI-206 b, TOI-500 b, TOI-1075 b, TOI-1442 b, TOI-2260 b, TOI-2411 b, and TOI-2445 b) are ultra-short-period planets, which are named for their $< 1$ day orbital periods \citep[e.g.,][]{leger2009transiting, batalha2011kepler, sanchis2013transits}. One interesting case is that of TOI-2260 b, whose star we determine to have a metallicity of [Fe/H] = $0.22 \pm 0.06$ dex. While an ultra-short-period planet orbiting such a metal-rich star is not unheard of, other planets of this type tend to orbit stars with lower metallicities \citep{winn2017absence}. Specifically, according to the NASA Exoplanet Archive,\footnote{\url{https://exoplanetarchive.ipac.caltech.edu/} } fewer than $10 \%$ of ultra-short-period planets orbit stars with metallicities greater than 0.2 dex. Further characterization of these planets could be helpful for understanding how these planets form around stars of different metal contents.

For TOI-1860 b and TOI-2260 b, we were able to use the TESS light curves of their host stars to estimate their ages, which we found to be $133 \pm 26$ and $321 \pm 96$ Myr, respectively. These ages make the planets some of the youngest known transiting planets to date. In addition to the recently validated TOI-1807 b, a $\sim 1.82 \, R_\oplus$ planet that was found to have an age of $180 \pm 40$ by \cite{hedges2021toi}, these planets will be important case studies for determining how terrestrial planets evolve in hot environments. Specifically, they will allow us to test two competing theories behind the existence and behavior of the radius gap. Photoevaporative atmospheric mass-loss \citep{jackson2012coronal, lopez2013role, owen2013kepler, jin2014planetary, owen2017evaporation, jin2018compositional} predicts small planets to be stripped of their atmospheres within the first $\sim 100$ Myr of the system lifetime, when the host star is still active enough to produce the high-energy photons responsible for atmospheric escape \citep{ribas2005evolution, jackson2012coronal}. Conversely, core-powered atmospheric mass loss is predicted to occur over a steadier $\sim 1$ Gyr timescale \citep{gupta2019sculpting}. Some studies have explored this distinction by examining how the occurrence rate gap evolves over Gyr timescales \citep{berger2020gaia, david2021evolution, sandoval2021influence}. By characterizing these planets further, either by measuring their masses or observing their emission spectra with JWST, we will be able to determine to what extent these planets have experienced atmospheric mass loss over their short lives. Observations that support the lack of an atmosphere around these planets would provide evidence for the former, while observations that support the existence of atmospheres would provide evidence for the latter.

As was mentioned in Section \ref{TOI-1860}, TOI-1860 is also a Solar twin. With an age of $133 \pm 26$ Myr, this star is the youngest Solar twin with a transiting planet discovered yet. Future studies of this system could shed light on the formation and evolution of planets around Sun-like stars. 

The last notable feature of the targets included in this paper is that they span a wide range of stellar spectral types. It is believed that the radius at which short-period planets transition from having volatile-rich atmospheres to having terrestrial-like or negligible atmospheres depends on the mass of the host star. Specifically, \citep{fulton2018california} found evidence that this transition radius increases with increasing stellar mass. In other words, a $1.6 \, R_\oplus$ planet has a higher probability of having a volatile-rich atmosphere when orbiting a K dwarf than it does when orbiting a G dwarf. Because our sample spans from low-mass M dwarfs to Sun-like stars, acquiring emission spectroscopy observations of our targets would allow for a direct test of this hypothesis.

\movetabledown=18mm
\begin{longrotatetable}
\begin{deluxetable*}{cccccccccl}\label{tab:previous_planets}
\tabletypesize{\footnotesize}
\tablewidth{\textwidth}
 \tablecaption{Confirmed and Validated Planets with $R_{\rm p} < 2 \, R_\oplus$ and ${\rm ESM} > 7.5$}
 \tablehead{
 \colhead{TOI} & \colhead{Alt Name} & \colhead{$K_s$ mag} & \colhead{$T_{\rm eff}$ ($K$)} & \colhead{$P_{\rm orb}$ (days)} & \colhead{$R_{\rm p}$ ($R_\oplus$)} & \colhead{$M_{\rm p}$ ($M_\oplus$)} & \colhead{$T_{\rm eq}$ (K)} & \colhead{ESM} & \colhead{Confirmation/Validation Paper}
 }
 \startdata 
 134.01  & L 168-9 b   & $7.082 \pm 0.031$ & $3800 \pm 70$  & $1.401500 \pm 0.000180$  & $1.39 \pm 0.09$ & $4.60 \pm 0.56$ & $981 \pm 27$   &  $9.9 \pm 1.4$ & \cite{astudillo2020hot} \\
 136.01  & LHS 3844 b  & $9.145 \pm 0.023$ & $3036 \pm 77$  & $0.462929 \pm 0.000002$  & $1.30 \pm 0.02$ &    -            & $805 \pm 27$   &  $28.8 \pm 1.8$ & \cite{vanderspek2019tess} \\
 141.01  & HD 213885 b & $6.419 \pm 0.024$ & $5978 \pm 50$  & $1.008035 \pm 0.000020$  & $1.75 \pm 0.05$ & $8.83 \pm 0.66$ & $2131 \pm 21$  &  $14.1 \pm 0.8$ & \cite{espinoza2020hd} \\
 396.01  & HR 858 c    & $5.149 \pm 0.020$ & $6201 \pm 50$  & $5.972930 \pm 0.000600$  & $1.94 \pm 0.07$ &         -       & $1317 \pm 16$  &  $9.7 \pm 0.7$ & \cite{vanderburg2019tess} \\
 431.02  & HIP 26013 b & $6.723 \pm 0.021$ & $4850 \pm 75$  & $0.490047 \pm 0.000010$  & $1.28 \pm 0.04$ & $3.07 \pm 0.35$ & $1888 \pm 50$  &  $16.0 \pm 1.2$   & Osborn et al. 2021 (submitted) \\
 667.01  & GJ 1132 b   & $8.322 \pm 0.027$ & $3270 \pm 140$ & $1.628931 \pm 0.000027$  & $1.13 \pm 0.06$ & $1.66 \pm 0.23$ & $584 \pm 30$  &  $9.5 \pm 1.4$ & \cite{berta2015rocky}\tablenotemark{a} \\
 732.01  & LTT 3780 b  & $8.204 \pm 0.021$ & $3331 \pm 157$ & $0.768448 \pm 0.000054$  & $1.33 \pm 0.07$ & $2.62 \pm 0.47$ & $892 \pm 44$  &  $13.4 \pm 1.6$ & \cite{cloutier2020pair} \\
 836.02  & HIP 73427 b & $6.804 \pm 0.018$ & $4250 \pm 120$ & $3.816514 \pm 0.000757$  & $1.81 \pm 0.27$ & $5.76 \pm 1.14$ & $834 \pm 47$  &  $8.7 \pm 2.7$ & \cite{teske2020magellan} \\
 1078.01 & GJ 1252 b   & $7.915 \pm 0.023$ & $3458 \pm 140$ & $0.518235 \pm 0.000006$  & $1.19 \pm 0.07$ & $2.09 \pm 0.56$ & $1089 \pm 53$  &  $16.3 \pm 2.2$ & \cite{shporer2020gj} \\
 1416.01 & HIP 70705 b & $7.708 \pm 0.024$ & $4884 \pm 70$  & $1.069763 \pm 0.000005$  & $1.73 \pm 0.05$ & $5.00 \pm 1.10$ & $1514 \pm 24$  &  $11.0 \pm 0.7$  & Deeg et al. (in prep)   \\
 1462.01 & HD 158259 b & $4.965 \pm 0.023$ & $5801 \pm 157$ & $2.178000 \pm 0.000100$  & $1.25 \pm 0.10$\tablenotemark{b} & $2.22 \pm 0.42$ & $1673 \pm 76$  &  $8.4 \pm 1.5$  & \cite{hara2020sophie} \\
 1469.01 & HD 219134 b & $3.261 \pm 0.304$ & $4699 \pm 16$  & $3.093500 \pm 0.000300$  & $1.60 \pm 0.06$ & $4.74 \pm 0.19$ & $1014 \pm 8$  &  $37 \pm 6$ & \cite{motalebi2015harps} \\
 1469.02 & HD 219134 c & $3.261 \pm 0.304$ & $4699 \pm 16$  & $6.764580 \pm 0.000330$  & $1.51 \pm 0.05$ & $4.36 \pm 0.22$ & $782 \pm 6$  &  $18.3 \pm 2.9$ & \cite{gillon2017two} \\
 1773.01 & 55 Cnc e    & $4.015 \pm 0.036$ & $5172 \pm 18$  & $0.736547 \pm 0.000001$  & $1.88 \pm 0.03$ & $7.99 \pm 0.33$ & $1947 \pm 13$  &  $69.9 \pm 2.7$ & \cite{mcarthur2004detection, winn2011super}\tablenotemark{c} \\
 1634.01 & TOI-1634 b  & $8.600 \pm 0.014$ & $3550 \pm 69$  & $0.989343 \pm 0.000015$  & $1.79 \pm 0.08$ & $4.91 \pm 0.69$ & $923 \pm 23$  &  $13.9 \pm 1.3$ & \cite{cloutier2021toi}\tablenotemark{d} \\
 1685.01 & TOI-1685 b  & $8.758 \pm 0.020$ & $3434 \pm 51$  & $0.6691403 \pm 0.000002$ & $1.70 \pm 0.07$ & $3.78 \pm 0.63$ & $1066 \pm 24$  &  $13.8 \pm 1.2$ & \cite{bluhm2021ultra}\tablenotemark{d} \\
 1807.01 & HIP 65469 b & $7.568 \pm 9.995$\tablenotemark{e} & $4757 \pm 50$ & $0.549372 \pm 0.000007$ & $1.82 \pm 0.05$ &   -  & $1730 \pm 28$  &  $22.6 \pm 1.4$ &  \cite{hedges2021toi} \\
 1827.01 & GJ 486 b    & $6.362 \pm 0.018$ & $3340 \pm 54$  & $1.467119 \pm 0.000031$  & $1.31 \pm 0.07$ & $2.82 \pm 0.12$ & $700 \pm 17$  &  $21.5 \pm 2.4$ & \cite{trifonov2021nearby} \\
 2431.01 & HIP 11707 b & $7.554 \pm 0.023$ & $4079 \pm 126$ & $0.224200 \pm 0.000020$  & $1.62 \pm 0.21$ &       -         & $2048 \pm 125$  &  $29 \pm 8$
 & Malavolta et al. (in prep)\tablenotemark{f} \\
-        & HD 3167 b   & $7.066 \pm 0.020$ & $5261 \pm 60$  & $0.959641 \pm 0.000012$  & $1.70 \pm 0.17$ & $5.02 \pm 0.38$ & $1746 \pm 46$  &  $14.0 \pm 2.9$ & \cite{vanderburg2016two, christiansen2017three} \\
  -      & K2-141 b    & $8.401 \pm 0.023$ & $4599 \pm 79$  & $0.280324 \pm 0.000002$  & $1.51 \pm 0.05$ & $5.08 \pm 0.41$ & $2115 \pm 48$  &  $15.0 \pm 1.1$ & \cite{malavolta2018ultra} \\
 -       & GJ 9827 b   & $7.193 \pm 0.024$ & $4340 \pm 47$  & $1.208982 \pm 0.000007$  & $1.58 \pm 0.03$ & $4.91 \pm 0.49$ & $1183 \pm 15$  &  $14.9 \pm 0.6$ & \cite{niraula2017three, rodriguez2018system}\tablenotemark{g}
 \enddata
 \tablenotetext{a}{Listed host star and planet properties from \cite{bonfils2018radial}.}
 \tablenotetext{b}{Planet radius was calculated using the transit depth listed on ExoFOP-TESS, $\delta = 90 \pm 7$ ppm.}
 \tablenotetext{c}{Listed host star and planet properties from \cite{bourrier201855}.}
 \tablenotetext{d}{Also confirmed by \cite{hirano2021two}, which finds the mass of TOI-1634 b to be about twice what is listed here.}
 \tablenotetext{e}{The large uncertainty in this $K_s$ mag is reported by 2MASS. The 2MASS $J$ and $H$ mag are $8.103 \pm 0.023$ and $7.605 \pm 0.018$, respectively.}
 \tablenotetext{f}{Entries for this row are taken from TICv8 and ExoFOP-TESS, as the authors of this paper were unable to share exact figures at the time of writing. The mass of the transiting object has been measured and is consistent with that of a planet (via personal communication).}
 \tablenotetext{g}{Listed host star and planet properties from \cite{rice2019masses}.}
 \vspace{-25pt}
\end{deluxetable*}
\end{longrotatetable}

\begin{figure*}[t!]
  \centering
    \includegraphics[width=\textwidth]{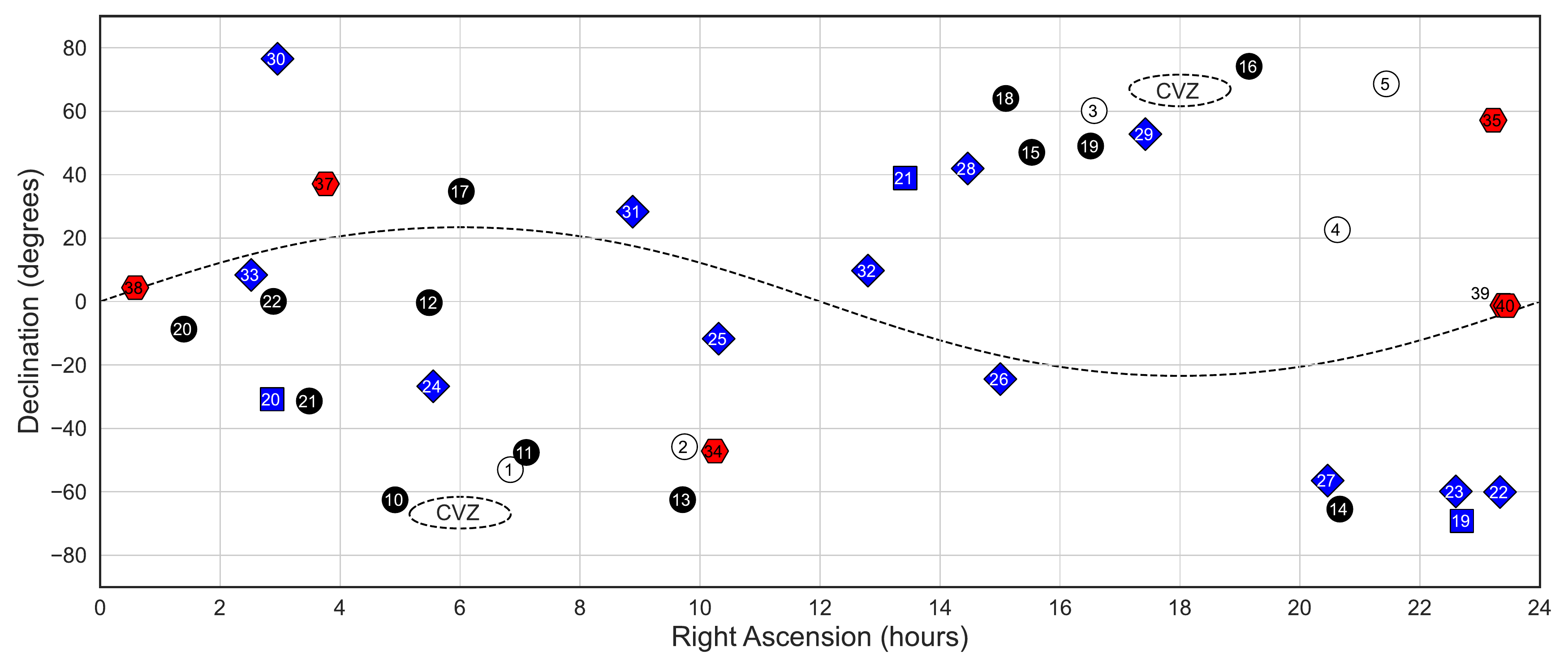}
    \includegraphics[width=\textwidth]{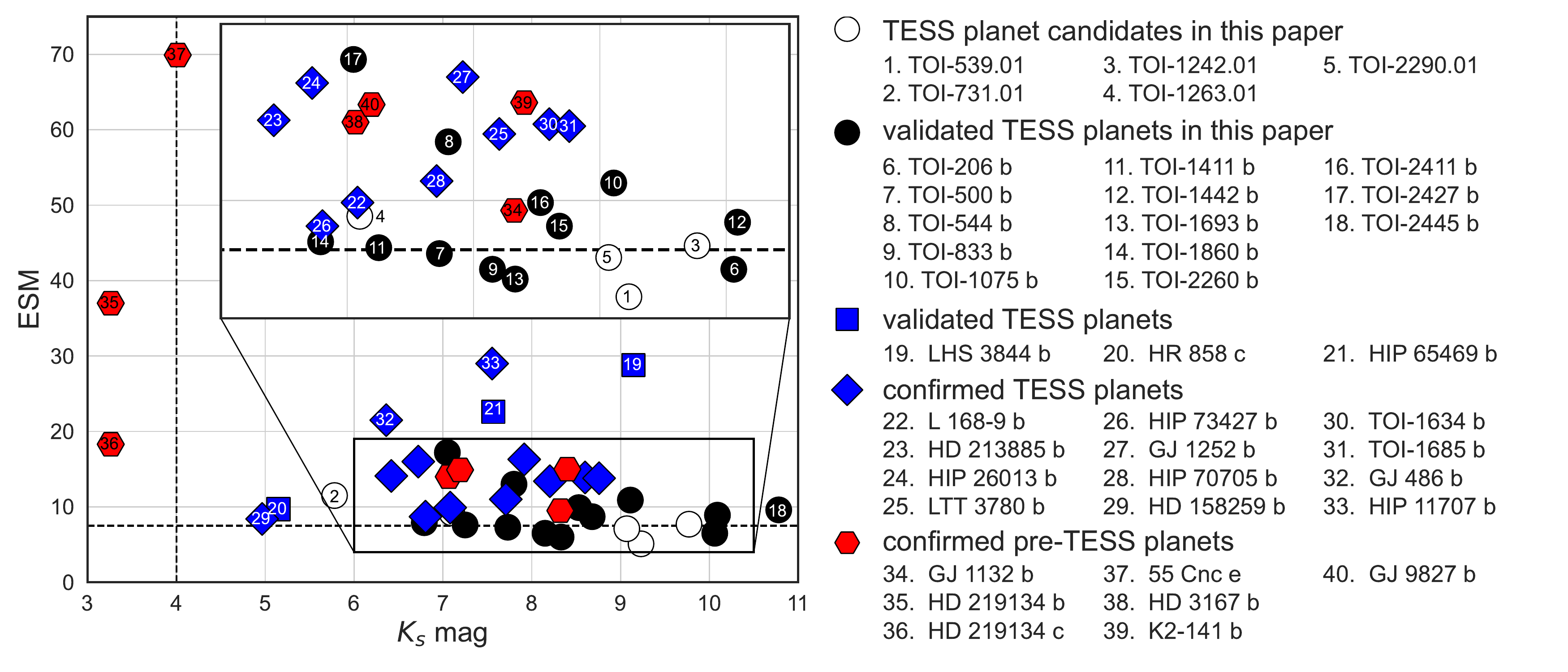}
    \caption{Top: coordinates of the TESS planet candidates in this paper (white circles), validated TESS planets in this paper (black circles), validated TESS planets (blue squares), confirmed TESS planets (blue diamonds), and confirmed pre-TESS planets (red hexagons). The ecliptic plane and ecliptic poles (i.e., the JWST continuous viewing zones) are shown as dashed black lines. Bottom: emission spectroscopy metric versus apparent $K_s$ magnitude for each planet candidate and planet. The dashed lines indicate the minimum values a target should have to be observed with JWST.}
    \label{fig: map}
\end{figure*}

To explore how the TESS mission has thus far increased the number of potentially terrestrial planets amendable to emission spectroscopy observations, we compile a list of all terrestrial planets with ${\rm ESM} > 7.5$ that were confirmed (i.e., have had their masses measured with precise radial velocities) or validated (i.e., have had their planetary natures certified using methods that do not involve a mass measurement) prior to the writing of this paper. Table \ref{tab:previous_planets} shows the host and planet properties of these systems, which were identified using the NASA Exoplanet Archive. Of these, 7 were discovered and confirmed prior to the TESS mission, 12 were discovered by TESS and subsequently confirmed, and 3 were discovered by TESS and subsequently validated. Going by these numbers, TESS has increased the number of potential JWST emission spectroscopy targets from 7 to 22. If we include the planets validated in this paper, this count increases to 35 -- a five-fold increase in the size of the sample available prior to TESS. 

Another aspect one must consider when planning for JWST observations of these targets is their locations in the sky. JWST operates in an ecliptic coordinate framework that makes the telescope capable of observing targets within $5^\circ$ of the north and south ecliptic poles (regions dubbed the ``continuous viewing zones,'' or CVZs) at any time of the year and all other regions of the sky twice per year over time intervals that vary with ecliptic longitude. In other words, targets at or near the ecliptic poles will be observable for longer periods of time than targets near the ecliptic plane. In the top panel of Figure \ref{fig: map}, we show the position of each planet candidate and confirmed/validated planets in our sample. Though no targets lie within the CVZs, several targets (e.g., TOI-206, TOI-500, TOI-539, TOI-1242, TOI-1442, and HD 158259) are only a short distance away. While most of the systems in our sample would make excellent targets for these observations, those close to the CVZs would allow for more flexibility when planning observations. 

Lastly, in addition to ESM, there are other properties of these systems that must be considered when planning for JWST observations. For instance, a star that is too bright in the passband could saturate the instrument in the minimum number of groups (two) required for a JWST observation.\footnote{A ``group'' is JWST terminology for the number of consecutively read frames with no resets.} A vast majority of terrestrial planet emission spectroscopy observations will be conducted using MIRI LRS, a low-resolution spectrograph with a wavelength range of $5-12 \, \mu$m. \texttt{PandExo} \citep{batalha2017pandexo, batalha2019pandexo}, a tool created to calculate the optimal exposure times for exoplanetary JWST observations, estimates the brightest star one can observe with MIRI LRS without saturating to have $K \sim 4$. All of the planet candidates discussed in Section \ref{sec: results} meet this criterion, and all but three previously confirmed/validated planets (HD 219134 b, HD 219134 c, and 55 Cnc e) meet this criterion. This indicates that nearly all planets in our sample will be observable with this instrument. The location of each planet candidate and confirmed/validated planet in ESM -- $K$ mag space is shown in the bottom panel of Figure \ref{fig: map}.

\section{Conclusion}\label{sec: conclusion}

We vet 18 hot TESS planet candidates that are potentially terrestrial ($R_{\rm p} < 2 \, R_\oplus$) and would make good targets for emission spectroscopy observations with JWST (${\rm ESM} \gtrsim 7.5$) using several follow-up observations from the TFOP and analyses performed with \dave\ and \tri. Of these 18, 13 were validated.

The 13 validated planets exist in a diverse set of environments that will allow for differential studies of small planets in and around the $1.5 - 2.0 \, R_\oplus$ radius gap. Some key takeaways about these validated planets are as follows:
\begin{itemize}
    \item Seven of the validated planets (TOI-206 b, TOI-500 b, TOI-1075 b, TOI-1442 b, TOI-2260 b, TOI-2411 b, and TOI-2445 b) are ultra-short-period planets.
    \item TOI-1860 b is a $1.34 \, R_\oplus$ planet orbiting a young ($133 \pm 26$ Myr) solar twin. This is the youngest planetary system discovered around a solar twin to date.
    \item TOI-2260 b is a $1.68 \, R_\oplus$ ultra-short-period planet orbiting a young ($321 \pm 96$ Myr) late G dwarf. With a stellar metallicity of [Fe/H] = $0.22 \pm 0.06$ dex, this star ranks among the most metal-rich to host an ultra-short-period planet. TOI-2260 b has a $T_{\rm eq}$ of $2609 \pm 86$ K and is the fourth hottest planet with $R_{\rm p} < 2 \, R_\oplus$ discovered to date.
\end{itemize}

Lastly, we assemble a list of all other previously discovered transiting planets that met our selection criteria for being ideal JWST emission spectroscopy targets. We discuss the prospects of using JWST to observe each of these known planets, along with the planet candidates and validated planets discussed in this paper.

\acknowledgments

We thank the NASA TESS Guest Investigator Program for supporting this work through grant 80NSSC18K1583 (awarded to C. D. Dressing). S. Giacalone and C. D. Dressing also appreciate and acknowledge support from the Hellman Fellows Fund, the Alfred P. Sloan Foundation, the David and Lucile Packard Foundation, and the NASA Exoplanets Research Program (XRP) through grant 80NSSC20K0250. 

We acknowledge the use of public TESS Alert data from the pipelines at the TESS Science Office and at the TESS Science Processing Operations Center. Resources supporting this work were provided by the NASA High-End Computing (HEC) Program through the NASA Advanced Supercomputing (NAS) Division at Ames Research Center for the production of the SPOC data products.

Data presented herein were obtained at the WIYN Observatory from telescope time allocated to NN-EXPLORE through the scientific partnership of the National Aeronautics and Space Administration, the National Science Foundation, and the National Optical Astronomy Observatory. NESSI was funded by the NASA Exoplanet Exploration Program and the NASA Ames Research Center. NESSI was built at the Ames Research Center by S. B. Howell, Nic Scott, E. P. Horch, and Emmett Quigley. The authors are honored to be permitted to conduct observations on Iolkam Du'ag (Kitt Peak), a mountain within the Tohono O'odham Nation with particular significance to the Tohono O'odham people. 

MEarth is funded by the David and Lucile Packard Fellowship for Science and Engineering, the National Science Foundation under grants AST-0807690, AST-1109468, AST-1004488 (Alan T. Waterman Award) and AST-1616624, and the John Templeton Foundation.  This publication was made possible through the support of a grant from the John Templeton Foundation.  The opinions expressed in this publication are those of the authors and do not necessarily reflect the views of the John Templeton Foundation. 

The authors wish to recognize and acknowledge the very significant cultural role and reverence that the summit of Maunakea has always had within the indigenous Hawaiian community.  We are most fortunate to have the opportunity to conduct observations from this mountain. D.H. acknowledges support from the Alfred P. Sloan Foundation, the National Aeronautics and Space Administration (80NSSC18K1585, 80NSSC19K0379), and the National Science Foundation (AST-1717000). 

Some of the observations in the paper made use of the High-Resolution Imaging instrument(s) ‘Alopeke (and/or Zorro). ‘Alopeke (and/or Zorro) was funded by the NASA Exoplanet Exploration Program and built at the NASA Ames Research Center by S. B. Howell, Nic Scott, E. P. Horch, and Emmett Quigley. Data were reduced using a software pipeline originally written by E. P. Horch and Mark Everett. ‘Alopeke (and/or Zorro) was mounted on the Gemini North (and/or South) telescope of the international Gemini Observatory, a program of NSF’s OIR Lab, which is managed by the Association of Universities for Research in Astronomy (AURA) under a cooperative agreement with the National Science Foundation. on behalf of the Gemini partnership: the National Science Foundation (United States), National Research Council (Canada), Agencia Nacional de Investigación y Desarrollo (Chile), Ministerio de Ciencia, Tecnología e Innovación (Argentina), Ministério da Ciência, Tecnologia, Inovações e Comunicações (Brazil), and Korea Astronomy and Space Science Institute (Republic of Korea). These observations were collected under program GN-2019B-LP-101. Observations acquired with Gemini-S/DSSI were collected as a part of program GS-2018A-Q-202 (PI: J. Winters).

Some of the results in this paper are based on observations made with the Nordic Optical Telescope, operated by the Nordic Optical Telescope Scientific Association at the Observatorio del Roque de los Muchachos, La Palma, Spain, of the Instituto de Astrofisica de Canarias. A.A.B., B.S.S. and I.A.S. acknowledge the support of Ministry of Science and Higher Education of the Russian Federation under the grant 075-15-2020-780 (N13.1902.21.0039). 

This paper is partially based on observations made at the CMO SAI MSU with the support by M.V. Lomonosov Moscow State University Program of Development. 

Based on observations at Cerro Tololo Inter-American Observatory at NSF’s NOIRLab (NOIRLab Prop. IDs 2019A-0294, 2019B-0302, 2020A-0390, 2020B-0262, 2021A-0268; PI: S. Quinn), which is managed by the Association of Universities for Research in Astronomy (AURA) under a cooperative agreement with the National Science Foundation. This research has been supported by RECONS (www.recons.org) members Todd Henry, Hodari James, Leonardo Paredes, and Wei-Chun Jao, who provided data as part of the CHIRON program on the CTIO/SMARTS 1.5m, which is operated as part of the SMARTS Consortium.

The research leading to these results has received funding from the ARC grant for Concerted Research Actions, financed by the Wallonia-Brussels Federation. TRAPPIST is funded by the Belgian Fund for Scientific Research (Fond National de la Recherche Scientifique, FNRS) under the grant PDR T.0120.21, with the participation of the Swiss National Science Fundation (SNF). M. Gillon and E. Jehin are F.R.S.-FNRS Senior Research Associate.

This work is partly supported by JSPS KAKENHI Grant Numbers JP20K14518, JP17H04574, JP18H05439, Grant-in-Aid for JSPS Fellows, Grant Number JP20J21872, JST PRESTO Grant Number JPMJPR1775, JST CREST Grant Number JPMJCR1761, and the Astrobiology Center of National Institutes of Natural Sciences (NINS) (Grant Number AB031010).

This paper is based on observations made with the MuSCAT2 instrument, developed by ABC, at Telescopio Carlos Sánchez operated on the island of Tenerife by the IAC in the Spanish Observatorio del Teide.

This paper is based on observations made with the MuSCAT3 instrument, developed by the Astrobiology Center and under financial supports by JSPS KAKENHI (JP18H05439) and JST PRESTO (JPMJPR1775), at Faulkes Telescope North on Maui, HI, operated by the Las Cumbres Observatory.

This work makes use of observations from the LCOGT network.  

Work by J. N. Winn was supported by the Heising-Simons Foundation. 

We thank Rhodes Hart for his contributions to this paper.

\facilities{TESS, CAO:2.2m (AstraLux), WIYN (NESSI), SOAR (HRCam), Shane (ShARCS), Hale (PHARO), Gemini:Gillett ('Alopeke), Gemini:South (Zorro and DSSI), Keck:II (NIRC2), FLWO:1.5m (TRES), NOT (FIES), CTIO:1.5m (CHIRON), Keck:I (HIRES), MEearth, LCOGT, OMM:1.6 (PESTO), OAO:1.88m (MuSCAT), TRAPPIST, SAAO:0.5m, Sanchez (MuSCAT2).}

\software{\texttt{exoplanet} \citep{exoplanet:exoplanet}, \texttt{lightkurve} \citep{2018ascl.soft12013L}, \dave\ \citep{kostov2019discovery}, \tri\ \citep{2020ascl.soft02004G, giacalone2021triceratops}, \texttt{Tapir} \citep{Jensen:2013}, \texttt{AstroImageJ} \citep{Collins:2017}.}

\bibliography{bibliography}{}
\bibliographystyle{aasjournal}

\suppressAffiliationsfalse
\allauthors

\end{document}